\documentclass[11pt]{article}
\pdfoutput=1

\usepackage[pagebackref,hidelinks]{hyperref}
\usepackage[utf8]{inputenc}
\usepackage[T1]{fontenc}
\usepackage[english]{babel}

\usepackage{
adjustbox,
amscd,
amsmath,
amssymb,
amsthm,
array,
authblk,
bbm,
boldline,
booktabs,
calc,
cite,
cleveref,
color,
colortbl,
diagbox,
enumerate,
epsfig,
fancyvrb,
float,
fp,
graphicx,
keyval,
listings,
longtable,
pdflscape,
makecell,
mathtools,
multicol,
multirow,
pict2e,
pifont,
subfig,
tabularx,
tikz,
tikz-cd,
xcolor,
xspace,
}

\usetikzlibrary{arrows,calc,decorations,decorations.pathreplacing}

\definecolor{sacramentostategreen}{rgb}{0.0, 0.34, 0.25}

\DeclareFontFamily{U}{wncy}{}
    \DeclareFontShape{U}{wncy}{m}{n}{<->wncyr10}{}
    \DeclareSymbolFont{mcy}{U}{wncy}{m}{n}
    \DeclareMathSymbol{\Sh}{\mathord}{mcy}{"58} 

\makeatletter

\renewcommand*{\backref}[1]{}
\renewcommand*{\backrefalt}[4]{({%
    \ifcase #1 Not cited.%
          \or Page~#2.%
          \else Pages #2.%
    \fi%
    })}

\def\CO{{\mathcal{O}}}

\def\bbZ{{\mathbb{Z}}}

\makeatother

\newtheorem{corollary}{Corollary}
\newtheorem{lemma}{Lemma}

\definecolor{promptColor}{rgb}{0.0,0.0,0.589}
\definecolor{brkpromptColor}{rgb}{0.589,0.0,0.0}
\definecolor{gapinputColor}{rgb}{0.589,0.0,0.0}
\definecolor{gapoutputColor}{rgb}{0.0,0.0,0.0}

\sbox0{\small\ttfamily A}
\edef\mybasewidth{\the\wd0 }
\lstnewenvironment{gap}
{\lstset{language=GAP,
	frame=single,
	basicstyle=\ttfamily\small,
	numbers=left,
	numberstyle=\small,
	keywordstyle=\color{red},
	stringstyle=\color{blue},
	commentstyle=\color{green!70!black},
	columns=fullflexible,
	basewidth=\mybasewidth,
	keepspaces=true,
	stepnumber=1,
	showstringspaces=false,
	breaklines=true}}
{}

\newenvironment{gapConsole}{\VerbatimEnvironment \begin{Verbatim}[commandchars=!@&,fontsize=\small,frame=single,xleftmargin = -.3em, xrightmargin=-0.3em,label=Example]}{\end{Verbatim}}

\hypersetup{
    pdftitle={Machine Learning and Algebraic Approaches towards Complete Matter Spectra in 4d F-theory},
    pdfauthor={Martin Bies, Mirjam Cvetic, Ron Donagi, Ling Lin, Muyang Liu, Fabian Ruehle},
    pdfsubject={machine learning, bundle cohomology, F-theory, vector-like spectrum, string phenomenology}
}

\begin{document}

\baselineskip=18pt  
\numberwithin{equation}{section}  



\baselineskip=18pt  
\numberwithin{equation}{section}  



\vspace*{-2cm} 
\begin{flushright}
{\tt UPR-1305-T,  CERN-TH-2020-111}\\
\end{flushright}

\vspace*{0.2cm} 
\begin{center}
{\LARGE Machine Learning and Algebraic Approaches\\towards Complete Matter Spectra in 4d F-theory}

\vspace*{1cm}
{Martin Bies$^{1}$, \, Mirjam Cveti{\v c}$^{2,3,4}$, \, Ron Donagi$^{3,2}$, \\
 Ling Lin$^5$, \, Muyang Liu$^2$, \, Fabian Ruehle$^{5,6}$}

\small
\vspace*{1cm}
{\it $^1$Mathematical Institute, University of Oxford, Woodstock Road, Oxford, OX2 6GG, United Kingdom} 

\bigskip
{\it $^2$Department of Physics and Astronomy, University of Pennsylvania, Philadelphia, PA-19104, USA}

\bigskip
{\it $^3$Department of Mathematics, University of Pennsylvania, Philadelphia, PA-19104, USA}

\bigskip
{\it $^4$Center for Applied Mathematics and Theoretical Physics, University of Maribor,  Maribor, Slovenia}

\bigskip
{\it $^5$CERN Theory Department, CH-1211 Geneva, Switzerland}

\bigskip
{\it $^6$Rudolf Peierls Centre for Theoretical Physics, University of Oxford
Department of Physics,\\Parks Road, Oxford OX1 3PU, United Kingdom} 

%
%
%
\vspace*{0.8cm}
\end{center}
\vspace*{.4cm}
%
\normalsize
\begin{abstract}
\noindent Motivated by engineering vector-like (Higgs) pairs in the spectrum of 4d F-theory compactifications, we combine machine learning and algebraic geometry techniques to analyze line bundle cohomologies on families of holomorphic curves. To quantify jumps of these cohomologies, we first generate 1.8 million pairs of line bundles and curves embedded in $dP_3$, for which we compute the cohomologies. A white-box machine learning approach trained on this data provides intuition for jumps due to curve splittings, which we use to construct additional vector-like Higgs-pairs in an F-Theory toy model. We also find that, in order to explain quantitatively the full dataset, further tools from algebraic geometry, in particular Brill--Noether theory, are required. Using these ingredients, we introduce a diagrammatic way to express cohomology jumps across the parameter space of each family of matter curves, which reflects a stratification of the F-theory complex structure moduli space in terms of the vector-like spectrum. Furthermore, these insights provide an algorithmically efficient way to estimate the possible cohomology dimensions across the entire parameter space.
\end{abstract}


\newpage

\tableofcontents

\section{Introduction}

The spectrum of light chiral particles is a defining feature of any four dimensional quantum field theory.
Their precise number affects aspects such as the moduli space of vacua, or the behavior of the theory under RG flow.
Moreover, they are also of paramount importance to phenomenology, in particular when it comes to models of beyond-the-Standard-Model physics.
Therefore, to be able to draw formal and phenomenological lessons from string theory about 4d field theories, one needs efficient methods to compute the spectrum in compactification scenarios.

From an effective field theory perspective, the chiral excess $\chi({\bf R})$ --- the difference between chiral and anti-chiral modes of the same matter representation ${\bf R}$ --- is a discrete parameter, whereas the individual number of light (anti-)chiral modes depend on continuous mass parameters.
In string theory, this is reflected by the fact that $\chi(\bf R)$ is typically a topologically protected quantity, whereas the (perturbative) mass parameters\footnote{In this work we will neglect moduli stabilization, flux-induced superpotentials and non-perturbative effects.} are captured by continuous deformations, or \emph{moduli}, which for certain values can lead to a pair of chiral and anti-chiral modes --- a \emph{vector-like pair} --- to become massless.

In many string compactification scenarios, we know in principle what the relevant computations are:
massless fields are zero modes of some differential operators on the internal space, and therefore counted by appropriate sheaf cohomologies.
However, oftentimes these computations are so complicated that in practice, they can only be carried out explicitly for toy models, or for specialized values of the deformation parameters.
On the other hand, an exact understanding of how the cohomologies depend on these parameters is necessary for a complete description of the physical interpretation.
The moduli dependence and the possibility of jumps in the massless spectrum have been first discussed in the context of heterotic string theory in~\cite{Berglund:1990rk,Berglund:1990hy,Donagi:2004qk,Donagi:2004ia,Bouchard:2005ag,Bouchard:2006dn}. More recently, the complex structure moduli dependence of the cohomology dimensions has been studied in~\cite{Buchbinder:2019hyb,Buchbinder:2019eal} and~\cite{Gray:2019tzn} in the context of instanton and perturbative superpotential terms, respectively.

In comparison, an analogous analysis in the context of F-theory compactifications \cite{Vafa:1996xn} is largely missing and has only been discussed in part in~\cite{Watari:2016lft}.
The main reason is because, unlike the chiral spectrum which is accessible via intersection theory \cite{Donagi:2008ca,Braun:2011zm,Marsano:2011hv,oai:arXiv.org:1111.1232,Krause:2011xj,Krause:2012yh,Mayrhofer:2013ara,Braun:2013nqa,Cvetic:2013uta,Cvetic:2015txa,Lin:2015qsa,Lin:2016vus,Cvetic:2018ryq,Cvetic:2019gnh}, the vector-like spectrum in F-theory depends on a gauge background, which is encoded in mathematically rather intricate objects such as the intermediate Jacobian and Deligne co\-homology \cite{Curio:1998bva, Donagi:2011jy, Donagi:2011dv, Anderson:2013rka}. Recent progress \cite{Bies:2014sra, Bies:2017fam} has made the spectrum computationally more accessible. Namely, for a four-dimensional ${\cal N}=1$ F-theory compactifications on an elliptically fibered Calabi--Yau fourfold $\pi: Y_4 \rightarrow B_3$ with a given gauge background, the massless spectrum of chiral particles in representation ${\bf R}$ can be counted by certain line bundle cohomologies $h^i(C_{\bf R}, {\cal L}_{\bf R}), i=0,1$ on complex curves $C_{\bf R} \subset B_3$ --- the matter curves --- in the base. Given a compact model with a fixed gauge background, $C_{\bf R}$ and ${\cal L}_{\bf R}$ are specified by global data in terms of polynomials on $B_3$ , whose coefficients are (parts of) the complex structure parameters of $Y_4$. In this case, one can model the line bundle as a coherent sheaf on $B_3$, whose cohomology computation can be systematized in a computer algebra system \cite{Bies:2018uzw}. While this algorithm can be applied to a broad class of global F-theory models, the calculations for almost all phenomenologically interesting examples overburden even super-computers specifically designed for such tasks. The reason is that here, and in fact in many cohomology computations using commutative algebra or computational algebraic geometry, we need to compute Groebner Bases, whose computational complexity scales extremely poorly.

The introduction of ideas from \emph{Big Data} and \emph{machine learning} (ML) to string phenomenology~\cite{He:2017aed,Krefl:2017yox,Ruehle:2017mzq,Carifio:2017bov} provides new perspectives; see~\cite{Ruehle:2020jrk} for an introduction and comprehensive overview. One advantage that a trained algorithm provides is that it recognizes more subtle patterns without the need of a complete, ``microscopic'' understanding of the task. In particular, recent studies suggest that supervised learning can be used to predict line bundle cohomologies in string compactifications~\cite{Ruehle:2017mzq,Klaewer:2018sfl,Brodie:2019dfx}. One may be tempted to apply these techniques, which are mostly motivated by heterotic compactifications, directly to the F-theory. However, there is a significant difference in the way the line bundle data are specified in global heterotic vs.~F-theory models. In heterotic examples, the line bundles are typically given in a ``canonical'' way, namely as an element of the Picard group Pic$(X)$ of the underlying manifold $X$. This was used, e.g., in~\cite{Brodie:2019ozt,Brodie:2019pnz} to derive formulae for line bundle cohomologies in terms of topological indices.

However, in the F-theory setting, there is no straightforward fashion to extract even the structure of the Picard group of $C_{\bf R}$, given its polynomial description.
Likewise, because the same data specifies ${\cal L}_{\bf R}$ essentially as a sum of points $p_i$ on $B_3$ that also lie on $C_{\bf R}$, it is by no means obvious if, say, $p_1 - p_2$ is trivial or not on $C_{\bf R}$. What makes the situation particularly challenging is that, by varying the complex structure parameters, the structure of Pic$(C_{\bf R})$ as well as the points specifying ${\cal L}_{\bf R}$ will change. Together with the fact that we simply do not have a large data set of non-trivial F-theory examples, it is a priori unclear whether we could train an algorithm that reliably predicts the cohomologies for realistic models with arbitrary complex parameters.

Instead, we will use machine learning techniques on less complex examples to gain some intuition for circumstances under which line bundle cohomologies \emph{jump}. Physically, this is already interesting as such a jump can engineer one or possibly more massless vector-like pairs in situations where one generically expects none. Even if the trained algorithm does not perform perfectly, understanding its strategy can provide a guiding principle for the behavior of the vector-like spectrum in non-trivial examples. For this reason, we focus on white-box machine learning techniques, in particular on decision trees.

To fully understand the results of the machine learning, we further employ ``formal'' techniques from algebraic geometry, in the form of \emph{Brill--Noether theory}. This allows to identify ``microscopically'' the sources for jumps in cohomology, either from the curve $C_{\bf R}$ or the line bundle ${\cal L}_{\bf R}$ becoming non-generic. With these insights, we provide an algorithmic way to estimate the admissible numbers of vector-like pairs over the entire parameter space of a matter curve in a global F-theory model with given gauge background. Furthermore, our analysis also reveals a convenient diagrammatic way to encode the \emph{stratification} on the parameter space induced by the number of vector-like pairs. We believe that this is progress towards understanding the full complex structure dependence of the vector-like spectrum in global F-theory models.

The paper is organized as follows. In \cref{sec:MachineLearning} we discuss our machine learning approach. Using the exact methods implemented in \cite{SheafCohomologyOnToricVarieties}, we generate a database \cite{Database} of cohomologies of pullback line bundles on hypersurface curves in $dP_3$.
Interpreting these results with decision trees, we find that curve splittings typically lead to jumps in the vector-like spectrum. In \cref{sec:F-theoryModel}, we demonstrate that such curve splittings provide a practial way to engineer jumps in a global F-theory GUT-model.
To investigate the origin of these jumps, we turn in \cref{sec:DataMeetsAnalysis} to algebraic and analytic techniques.
We find a unified perspective on jumps due to curve splittings and non-generic line bundles described by Brill--Noether theory, and introduce a diagrammatic way to illustrate the natural stratification of the complex structure parameter space in terms of the vector-like spectrum.
In \cref{sec:LocalToGlobalSectionCounting}, we present a refined analysis of jumps due to curve splittings. This rests on a procedure to count the global sections by gluing ``local contributions'' along intersections of curve components, which leads to two interesting results: First, we are able to formulate sufficient conditions for jumps of vector-like spectra. 
Second, we can propose an algorithmic $h^0$ estimate, which relies mostly on topological data, and hence provides a quick, approximative scan of the vector-like spectrum over the entire parameter space of a matter curve. In contrast to currently existing exact methods, such as \cite{SheafCohomologyOnToricVarieties}, our implementation \cite{H0Approximator} has a much lower demand of computational resources and run times.

\section{Machine Learning} \label{sec:MachineLearning}

\subsection{Introduction to Decision Trees}
\label{sec:IntroDT}
We are interested in tuning complex structure moduli to engineer jumps in the dimensions of sheaf cohomologies over complex curves. It is a priori not clear how to efficiently identify these subloci in complex structure moduli space. In order to state (at least) necessary conditions for jumps to occur, we address the problem using ML. Since we are interested in interpreting the results of the ML algorithm, we resort to white-box models, in particular to binary decision trees.

In more detail, we use binary decision trees as classifiers in supervised machine learning, following the notation and conventions of~\cite{Ruehle:2020jrk}. Supervised learning means that we have a set of inputs $x_i^\mu$ (called features) together with associated labels\footnote{In general, there could be more than one label for each feature vector; however, for the cases studied in this paper, the label corresponds to a class the input belongs to, labeled by an integer.} $y_i$, where $i=1,\ldots,N$ counts the feature-label-pairs, and $\mu=1,\ldots,F$ counts the $F$ features of each input. This set of feature-label combinations is now divided into a train set and a test set (typically around 90 percent of the pairs are assigned to the train set and 10 percent to the test set). Using the train set, an algorithm is trained to learn a map from the features to the labels. The training consists of adjusting parameters of the algorithm to optimize the map. This is typically done by minimizing the loss, which is a measure for how well the algorithm reproduces the labels. Once training ends, the algorithm is tested on the test set. This is necessary in order to see how well it performs on (hitherto unseen) data. If the test set have been chosen generically enough, performance on the test set will serve as an indicator for how well the trained algorithm will perform.

After this general discussion, let us describe these steps in the context of binary decision trees. Trees are data structures that appear abundantly in computer science. They can be thought of as acyclic, directed, connected graphs with a unique root vertex (in trees, vertices are called nodes). In binary trees, each node has either zero or exactly two vertices, each of which is connected to a unique node. These two subnodes are called child nodes, and the original node is called parent node. A node with no children is called a leaf node.

A decision tree expects numerical features $x_i^{(0)}$. It then introduces boolean splitting criteria of the type $x_i^{(0)}\leq \kappa_i$ for some constant $\kappa_i\in\mathbbm{R}$. All data that satisfy this criterion are assigned to one child node, while data that does not satisfy the criterion is assigned to the other child node. The tree is now built recursively by splitting each child node according to some other feature $x_j^{(0)}\leq \kappa_j$, etc. This procedure segments feature space (which is in our case $\mathbbm{R}^N$) along hyperplanes $x_i=\kappa_i$ with the goal to find regions such that all inputs in that region belong to the same class.

At each node, it is checked how many of the data carry which label. For single membership classification problems, which is what we will be using, the labels are just the different classes which the input feature vector belongs to. A typical loss function is the Gini impurity of a node, which measures how ``impure'' the data at that node actually is, i.e., how many features with different classes are in the region in feature space corresponding to this node. Denoting the set of features in the region of node $a$ by $N_a$, we find for $K$ classes the fraction of elements that belong to a class $y_k\in K$ via
\begin{align}
p_{a,k} = \frac{1}{|N_a|}\sum_{i\in N_a}\delta_{i,k}\,.
\end{align}
The Gini impurity $G_a$ at node $a$ can then be written as
\begin{align}
G_a = \sum_{k=1}^K p_{a,k}(1-p_{a,k})\,.
\end{align}
In particular, if all elements of $N_a$ belong to the same class, $G_a=0$. In such a case, the node is turned into a leaf, since no further splits are necessary.

The decision tree is now trained by starting from the root node and trying to split by any of the $F$ features. For $\kappa_i$, one tries all\footnote{In case of many different values for a feature, this might be unfeasible, in which case a number of equally spaced values are tried for $\kappa_i$.} intermediate values between consecutive values of feature $i$. The solution that leads to the lowest Gini impurity at the child nodes is accepted, and the procedure is repeated for the two child nodes and the remaining features, etc.

In cases where the map from the input to the labels is not one-to-many, one can eventually reach a perfect classification, if need be with a single element in each region. Typically, this is undesired and hence one stops splitting a node if there are less than some fixed number of elements in its corresponding region. Turning this around, if the minimal number at which a node is split is set to 2, and if the tree does not find a solution where all leaves have Gini impurity zero, this means that the map defined by the input-label-pairs is many-to-one, i.e., even all features combined are not sufficient to distinguish between the class labels.

\subsection[Divisors and line bundles of \texorpdfstring{$dP_3$}{dP3}]{Divisors and line bundles on \boldmath{$dP_3$}}

While in the general F-theoretic setup, matters curves $C_{\bf R}$ are a priori defined on a threefold ${\cal B}_3$, in most models there is a distinguished surface $S \subset {\cal B}_3$ that is wrapped by the 7-branes supporting a non-abelian gauge theory, in which the matter curve sits.
A part of the complex structure moduli then parametrizes deformations of the curve inside $S$, which will in general affect the vector-like spectrum.
These deformations can be described by pulling back all defining polynomials on ${\cal B}_3$ onto $S$, and then simply consider the coefficients of these in terms of the homogeneous coordinates on $S$.

For our data collection, we will mimic such a ``pulled back'' description by focusing on curves embedded inside the del Pezzo surface $dP_3$.
One advantage of this choice is that $dP_3$ has a toric description in terms of a reflexive polygon, which simplifies many computations.
Another one is that it fits the setup for section \ref{sec:F-theoryModel}, where we consider an F-theory toy model with non-abelian gauge degrees of freedom localized precisely on a $dP_3$ surface.

To set the notation, we denote the toric coordinates of $dP_3$ by $x_i, i=1,...,6$.
They are graded by homogeneous scalings with associated divisor classes, which are summarized in the following table:
\begin{align}
\label{eq:dP3_Grading}
\begin{array}{c|cccccc}
\toprule
 & x_1 & x_2 & x_3 & x_4 & x_5 & x_6 \\
 \midrule
H     & 1  &   &   &  1 &  1 &   \\
E_1 & -1 & 1 &   & -1 &    &   \\
E_2 & -1 &   & 1 &    & -1 &   \\
E_3 &    &   &   & -1 & -1 & 1 \\
 \bottomrule
\end{array}
\end{align}
The columns give the divisor classes of the coordinate's vanishing loci. E.g., $[\{x_1\}] = H - E_1 - E_2$.
The Stanley--Reisner ideal is
\begin{align}
I_{\text{SR}} = \langle 
x_3 x_6,
x_2 x_6,
x_1 x_6,
x_4 x_5,
x_2 x_5,
x_1 x_5,
x_3 x_4,
x_1 x_4,
x_2 x_3 
\rangle\,,
\end{align}
and the anti-canonical class is $-K_{dP_3} = \sum_{i} [\{x_i\}] = 3H - E_1 - E_2 - E_3$.
The independent intersection numbers are
\begin{align}\label{eq:intersection_numbers_dP3}
	H^2 = 1 \, , \quad E_i \cdot E_j = - \delta_{ij} \, , \quad H \cdot E_i = 0 \, .
\end{align}
In order to simplify the notation, we introduce the short-hand notation $(a;b,c,d)$ with $a,b,c,d \in \bbZ$ for a divisor $D = a H + b E_1 + c E_2 + d E_3$.

We then define curves $C$ inside $dP_3$ via $C = \{ P = 0\} \equiv V( P )$ with
\begin{align}
P &= \sum_{i} c_i m_i(x_1,\ldots,x_6)\,,
\end{align}
where $m_i(x_1,\ldots,x_6)$ are monomials of appropriate multi-degree under the grading in~\eqref{eq:dP3_Grading}.
Importantly, the coefficients $c_i$ parametrize the shape of the curve and thus model (parts of) the complex structure parameters of a global F-theory compactification.
The (arithmetic) genus of the curve depends only on the divisor class $[C]$ of the curve (equivalently, the multi-degree of the monomials in $P$) and is given via adjunction formula as
\begin{align}
	g = 1 + \frac12 [C] \cdot ([C] + K ) \,.
\end{align}

Next, we also need to specify a line bundle ${\cal L}$ on $C$. Again, instead of focusing on the most general setup, where ${\cal L}$ is directly specified by a set of points on $C$, we consider the slightly simpler cases where ${\cal L}$ is a pullback of a line bundle $L = {\cal O}_{dP_3} ( D )$ on $dP_3$:
\begin{align}
\mathcal{L} = \left. \mathcal{O}_{dP_3} (  D  ) \right|_{dP_3} \, .
\end{align}
One can think of the points then as the (weighted) intersections $\{a_i \, p_i\}$ between $C$ and a generic representative in the class $D$.
Note that in this case, another representative of $D$, intersecting $C$ at $\{b_j \, p'_j\}$, necessarily must give the same divisor on $C$, i.e., $\{a_i \, p_i\} \sim \{b_j \, p'_j\}$ are linearly equivalent on $C$.
However, in general we cannot say anything about linear equivalences among any two of the points.
Therefore, we expect, and also will find, that even for pullback line bundles, there can be special divisor alignments, i.e., $p_1$ and $p_2$, say, move into special positions, when we deform $C$, thus leading to jumps in the cohomology.

\subsection{Generating the data set}\label{subsec:generate_data}
We generate training data by picking 6 different curve classes $[C]$ with genus $1\leq g \leq 6$. 
For each class we consider several line bundles $L$ on $dP_3$ and compute (using techniques from~\cite{Bies:2018uzw}) the cohomologies $h^i(C({\bf c}), L|_{C({\bf c})} )$, where we vary the curve $C({\bf c})$ by considering all possible combinations of $c_i\in \{0,1\}, i = 1, \ldots, d$ for the coefficients.\footnote{We exclude the case where all $c_i=0$.}
This way, we calculate cohomologies of $L$ pulled back to $2^d-1$ genus $g$ curves in the class $[C]$.
While this seems to be a very limited choice, it nevertheless reveals enough structures to correlate jumps in cohomology with degenerations of the geometry.
On the other hand, it also introduces some bias in the data.
For example, a common way the curve degenerates is if all monomials in the defining polynomial share a common variable; this happens frequently if many $c_i$ are set to 0.
However, for certain polynomials, restricting $c_i \in \{0,1\}$ misses out possible factorizations, where factors are not just a single variable.
We will see later that we can easily generalize the interpretation based on our data with algebraic methods to these cases as well.

For this data set, we then compute/collect the following features for each choice of line bundle $L$ on each curve $C$ with coefficients $c_i$:
\begin{enumerate}[F1)]
 \item The coefficients $c_i$ that define the curve.
 \item The genus of the curve.
 \item The number of global sections of the line bundle.\footnote{The dimension of $H^1(C,L)$ can then be computed from the index which is topological and does not depend on $c_i$.}
 \item Are the curves smooth?\label{itm:smooth}
 \item The number of components the curve splits into. \label{itm:nosplitcomponents}
 \item Are the splits smooth?\label{itm:splitssmooth}
 \item Are the splits reduced?\label{itm:splitsreduced}
 \item The genera of the split components.\label{itm:splitsgenera}
 \item The intersection numbers among the split components.\label{itm:intnums}
\end{enumerate}
Note that all of this data is numerical (the true/false features are encoded as 1/0). We aggregate the features~F\ref{itm:smooth}-F\ref{itm:intnums} into a single feature called the \textit{split type}. We want to consider two curves as identical if their features~F\ref{itm:smooth}-F\ref{itm:intnums} are identical (up to relabeling the individual components). In order to check this, we would in principle have to check all permutations of all split components and see whether any of them have the same data. Since this becomes prohibitively expensive, we perform the following necessary checks:
\begin{itemize}
\item Are the data~F\ref{itm:smooth} and~F\ref{itm:nosplitcomponents} identical for the two curves?
\item Are the data~F\ref{itm:splitssmooth}-F\ref{itm:splitsgenera} identical as sets for the two curves? This can be checked by ordering the tuples and comparing them, which is much faster than checking actual permutations.
\item Is the determinant of the intersection matrix in~F\ref{itm:intnums} identical for the two curves? Note that the determinant is permutation invariant. However, at that point we do not check whether the permutations that make all sets match are actually the same. 
\end{itemize}
Curves which are identical under these checks are assigned the same integer that encodes the split type.

Equipped with this data, we generate four different data sets which we use to train the decision trees and compare the results. In the first, we use the coefficients $c_i$ as features and assign a label of 0 if the cohomology dimension of $H^0(C({\bf c}),\mathcal{L})$ has the generic (i.e., the lowest) value and a label of 1 if there is a jump. Note that at this point, we only classify the curve according to whether a jump occurs, but not according to how large the jump is. For the second data set, we use the same labels, while the features are taken to be the topological intersection numbers between the curve components and the line bundle divisors. For the third data set we use the split type as explained above. Finally, for the fourth data set, we use both the split type and the topological intersection numbers between the curve components and the line bundle divisor as features. In addition, we perform a train:test split of 90:10 for all four data sets.

\subsection{Decision Trees to learn cohomology jumps}
\begin{figure}[t]
\includegraphics[width=\textwidth]{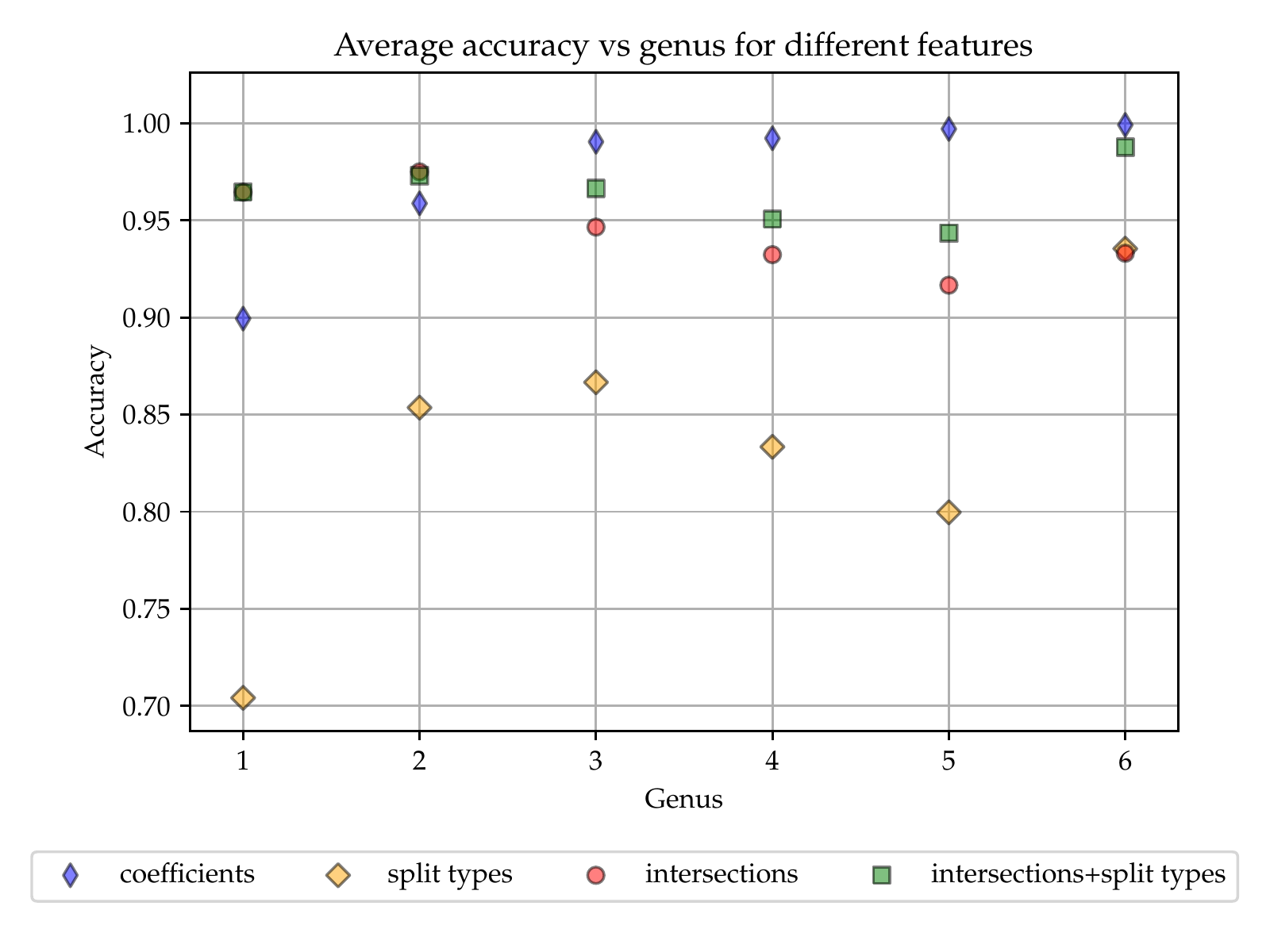}
\caption{Average accuracy on the test set as a function of the genera of the curves for different features.}
\label{fig:AccVSgenera}
\end{figure}

Training the decision trees only takes a few seconds on a modern desktop computer. We train a separate decision tree for each line bundle and each of the four data sets. It is instructive to compare the performance of all four training sets on both the train and the test set. 

The results for the accuracy of the trained trees on the test set are summarized in Figure~\ref{fig:AccVSgenera}. One notices that the accuracy of all data sets improves with the genus of the curve. This is due to the fact that the size of the data set grows with the genus: While the genus 0 curve we are considering has only 7 coefficients $c_i$ and hence only $2^7-1=128$ data points per line bundle, the genus 6 curve has $2^{18}-1=262143$ data points.

For the blue data points, which uses the coefficients $c_i$ as labels, we find that the decision tree performs best. This is to be expected, since these are the finest feature set, i.e., the one with the most information, out of the four feature sets we studied. Indeed, the trees reach an accuracy of essentially 1 as soon as the training set becomes large enough (there are 3685 points in the training set for genus 3). For the other three data sets, we see that they perform worse, but still reaches high accuracies. Using just the split type as a feature, for the larger genus cases where enough data is available, we reach accuracies around 80 to 85 percent. Using the intersection numbers, accuracies around 94 percent are obtained. Lastly, combining the split type and the intersection numbers, improves the results obtained when either is used individually, to an accuracy of around 97 percent. This means that the two features contain different types of information which the three can use in order to improve its prediction when given access to both.

One can learn more information about the data by also analyzing the performance on the training set, as explained in Section~\ref{sec:IntroDT}. Indeed, we find that, when not imposing constraints on the tree, the accuracy on the train set when using the coefficients as features is always 100 percent. This is not surprising, since the coefficients uniquely identify each case and hence the tree can learn a sequence of splits that puts each data point in the correct leaf node (if necessary, this leaf might only contain this single data point). For the other data sets, we find that the performance on the test set is already below 100 percent. Hence, the features are not enough to decide whether a jump in cohomology occurs, not even in principle.

\begin{landscape}
\begin{figure}[t]
\centering
\includegraphics[width=\linewidth]{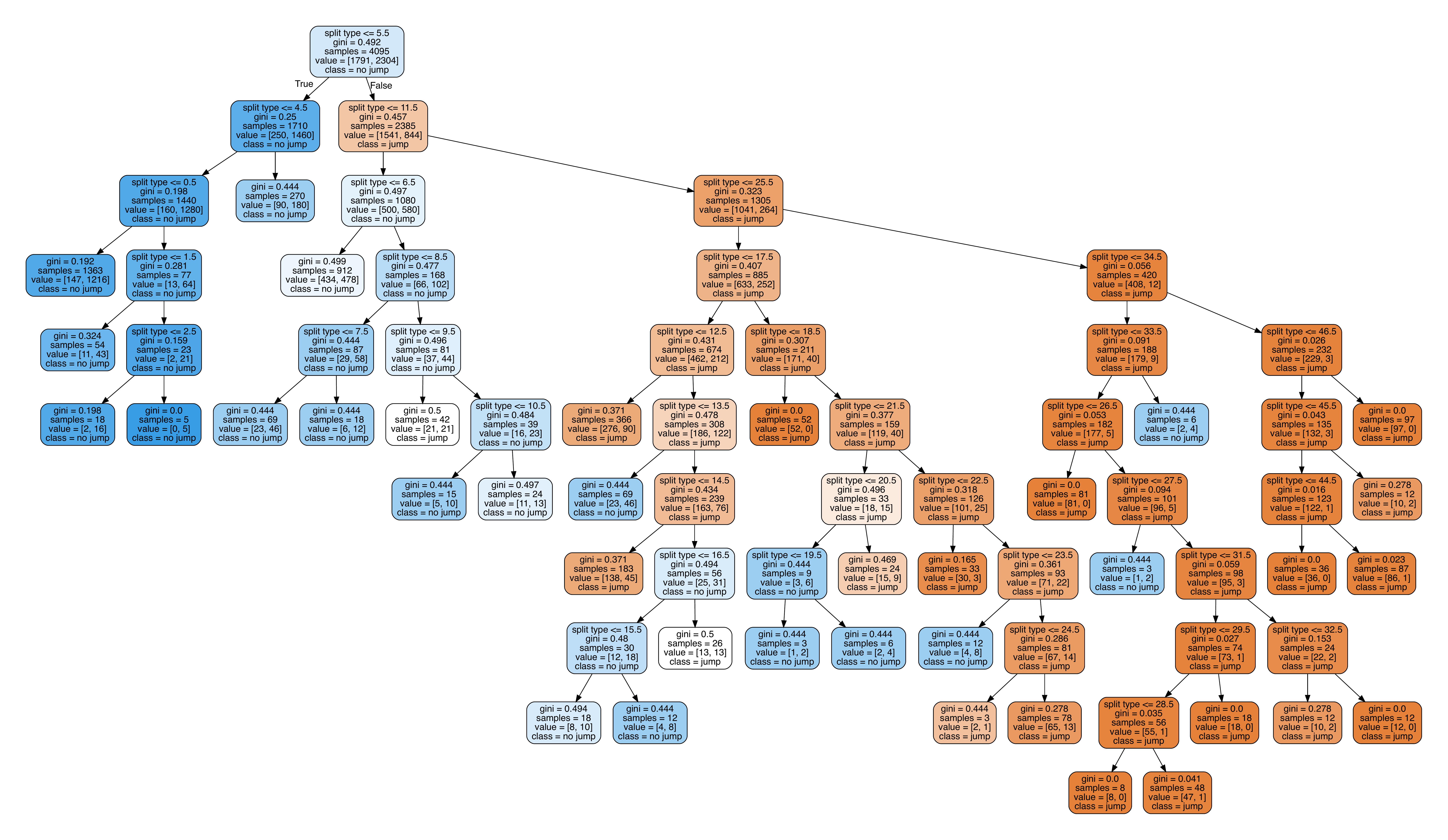}
\caption{Trained decision tree that classifies the presence of cohomology jumps based on split types for the genus three curve example.}
\label{fig:DecisionTree}
\end{figure}
\end{landscape}

Let us illustrate this by looking at the decision tree trained on the full data set for a genus three curve $D_C=(4;-1,-1,-1)$ inside $dP_3$ with line bundle $D_L=(1, 2, -2, -1)$, cf.~Section~\ref{sec:Genus3Details}. We give the full decision tree in Figure~\ref{fig:DecisionTree}. Looking at the root node, we see that for this bundle, there are 4095 different data points (``samples''). Out of these, 1791 exhibit a cohomology jump for this line bundle, while 2304 do not. The tree assigns a class label to this (non-leaf) node based on the majority, which is ``no jump''. However, there are almost as many data points with a jump as there are data points without, which is why the uncertainty is high. This is encoded in the light blue color: the more certain a node predicts no jump, the darker blue it is colored. Similarly, the more certain there is a jump, the darker orange it is.

Recall that integers labelling the split type (based on the features~F\ref{itm:smooth}-F\ref{itm:intnums}) are by construction small if the number of components the curve splits into is small. Hence, small split types correspond to irreducible curves, or curves with only few split components. 
We expect such curves being close to generic (in a sense that will be made mathematically more precise in Section~\ref{sec:DataMeetsAnalysis}), hence the cohomologies should also take generic values.

\pagebreak[4]

Indeed, we observe that the first split is performed according to whether or not the split type is smaller than $5.5$. This first split already gives a good indicator in the sense that out of the 1710 training data points that have a split type of 5 or smaller, 85 percent actually do not have a jump in their cohomologies. This also illustrates that decision trees can be used for feature selection: important features that are good indicators for the classes tend to be used for splitting higher up in the tree, while more unimportant features are used further down the tree (or not at all, if they do not have any predictive power for the class membership). Now, in our case, we only have a single feature, but it is a composite feature of several quantities. The fact that the first split does not occur  around the median (which would be 27) but at much smaller value indicates that the number of split components is a good criterion to distinguish jumps.

While the split types are integers, the tree always chooses half-integer decision boundaries. The reason is that the tree does not know that the feature only takes integer values. Hence, splitting in the middle between the feature values that appear in the train set will allow the most slack in either direction when the tree is presented with unseen data. 

By focusing on the leaf nodes, we can also see that the tree is not classifying the data perfectly, not even the training data. Indeed, many nodes have a non-zero Gini impurity, i.e., both curves with and without jumps share the same split type associated with this leaf node. Looking for example at the bottom right leaf node, we see that three curves have the same split type (with value 48). However, two of these have a jump while one does not. This means that the topological data~F\ref{itm:smooth}-F\ref{itm:intnums} used to construct the split type is not enough to decide whether or not a cohomology jump occurs.

\subsection{Interpretation of results}

\subsubsection{Jumps from curve splittings}

We have seen that the decision tree trained on a combination of split types and intersection numbers performs very well. Moreover, the tree trained with just the split types splits on small split types first. This suggests that there is a tight correlation between changes in the topology of the curve and jumps in the line bundle cohomology.
In particular, the data set has an abundance of cases with jumps where the curve $C$ splits off one or more rigid components:
For 78 (about 95\%) of the 82 pairs of geometries $D_C$ and line bundles $D_L$ considered in our database, we find that we can split off a rigid component $E$, i.e., $C \rightarrow \tilde{C} \cup E$, such that
\begin{align}
h^0_{\text{min}} ( \tilde{C} \cup E, L|_{\tilde{C} \cup E}  ) > h^0_{\text{min}} ( C, L|_{C} ) \, .
\end{align}
Put differently, for almost all pairs $(D_C, D_L)$ in our database, there exists a rigid divisor such that splitting off this rigid divisor from the curve $C$ leads to a jump in the number of global sections on that curve. At the same time, for a given combination $(D_C, D_L)$, we observe a jump of $h^0_\text{min}$ only for a subset of all possible splits $C \rightarrow \tilde{C} \cup E$, suggesting that $E$ and $D_L$ must have some correlation in order for the cohomology to enhance. We list the details of these splittings and jumps in \cref{sec:DetailsRigidDivisorJumps}.

It is obvious that the jumps stemming from rigid component splittings can be associated with the curve $C$ becoming non-generic. While per se not unexpected, the machine learning process reveals --- without explicitly ``knowing'' algebraic geometry --- these features.

It is important in this context to address the bias in the data coming from considering only values of $\{0,1\}$ for the coefficients.
Namely, within the data, we only observe jumps associated with splittings of rigid components.
Naively, one might conclude that rigidity of a split component is a necessary condition.
However, as we already stressed in the beginning of \cref{subsec:generate_data}, setting enough coefficients to 0 usually factors out one of the homogeneous coordinates $x_i$.
The corresponding curve splitting then always involves the toric divisor $V(x_i)$ which on a $dP_3$ is rigid for any $i=1,...,6$.
Therefore, the strong correlation between a rigid component and a jump is likely due to the bias in the data.

Indeed, we will find in sections \ref{sec:DataMeetsAnalysis} and \ref{sec:LocalToGlobalSectionCounting} with insights from algebraic geometry, that the main source for cohomology jumps in cases of curve splittings is actually insensitive to components being rigid.
We will also supplement a concrete example in \cref{subsubsec:JumpNonRigid} where we find a jump from non-rigid curve splittings.
Furthermore, we will combine these arguments with the intuition about curve splittings we gained through the data to phrase a sufficient condition for a jump in cohomology to occur in terms of topological data only. We will discuss this idea in \cref{sec:LocalToGlobalSectionCounting}.

\subsubsection{Unpredicted jumps}

The fact that the decision tree cannot predict all jumps hints towards sources for additional sections (and hence cohomology jumps) beyond curve splitting. Within the data set, we observe that in rare occasions, the curve remains smooth despite a deformation which induces a jump. 

For illustration purposes, consider again the genus three curve with the line bundle discussed above.
Generically, this genus $3$ curve is cut out by the polynomial
\begin{align}
\begin{split}
P ( \mathbf{c} ) &= c_1 x_1^3 x_2^3 x_3^2 x_4 + c_2 x_1^2 x_2^3 x_3 x_4^2 x_6 + c_3 x_1 x_2^3 x_4^3 x_6^2 + c_4 x_1^3 x_2^2 x_3^3 x_5 + c_5 x_1^2 x_2^2 x_3^2 x_4 x_5 x_6 \\
                 &\qquad + c_6 x_1 x_2^2 x_3 x_4^2 x_5 x_6^2 + c_7 x_2^2 x_4^3 x_5 x_6^3 + c_8 x_1^2 x_2 x_3^3 x_5^2 x_6 + c_9 x_1 x_2 x_3^2 x_4 x_5^2 x_6^2 \\
                 &\qquad + c_{10} x_2 x_3 x_4^2 x_5^2 x_6^3 + c_{11} x_1 x_3^3 x_5^3 x_6^2 + c_{12} x_3^2 x_4 x_5^3 x_6^3 \, .
\end{split}
\end{align}
The pullback of $\CO_{dP_3} (D_L)$ onto $C$ defines a line bundle ${\cal L}$ of degree $d = 3$. 
By Riemann--Roch we have $\chi ( {\cal L} ) = h^0 - h^1 = 1$.

In our database, we have computed the number of global sections for this line bundle for coefficient choices $\mathbf{c} \in \{ 0, 1 \}^{12} - \mathbf{0}$. For these 4095 curves, we find
\begin{itemize}
\item $h^0 = 1$: 2304 (56.3\%) \,,
\item $h^0 = 2$: 1664 (40.6\%) \,,
\item $h^0 = 3$: 127 (3.1\%) \,.
\end{itemize}
Our database indicates that a jump to $h^0 = 3$ occurs whenever $c_1 = c_2 = c_3 = c_{11} = c_{12} = 0$. 
This corresponds to a splitting
\begin{align}
C = V( x_2 ) \cup V( x_5 ) \cup V( \left. P \right|_{c_1 = c_2 = c_3 = c_{11} = c_{12} = 0} ) \, . \label{equ:MaximalSplitting}
\end{align}

The majority of the cases with $h^0 = 2$ are where either $V(x_2)$ or $V(x_5)$ splits off, each being a rigid $\mathbb{P}^1$.
This is in line with the above observation.
However, we also have instances (about 9\% of all curves with $h^0 = 2$) where the curve remains smooth and irreducible.
Despite having $h^0 = 2$, the split type features cannot distinguish these cases from the generic setup with $h^0 = 1$, thus leading to an imperfect performance of the decision tree.

While we will come back to a detailed discussion of this phenomenon and the associated algebraic description in terms of Brill--Noether theory in \cref{subsec:BNJumps}, it is evident that these cases of jumps are associated to the line bundle ${\cal L}$ on $C$ becoming non-generic. Moreover, we also observe that such Brill--Noether-type jumps can sometimes produce values of $h^0$ that cannot be obtained by splittings off rigid curve components. This becomes particularly important in F-theory models, as we will discuss now.

\section{Application: F-theory model building} \label{sec:F-theoryModel}

In the previous section, we have used machine learning techniques to gain some intuition on how line bundle cohomologies jump under complex structure deformations.
While we will discuss the underlying ``precise'' description of these various sources of jumps in the next section, we would like to show that these ``rules of thumb'' inferred from the withe-box machine learning results can be applied directly in string phenomenology. To this end, we consider an F-theory toy model and exemplify how curve splittings help ``controlling'' the number of vector-like pairs.\footnote{For the purpose of this work, and in particular this section, we will only focus on the matter curves and their embeddings into the ``GUT''-surface that supports the non-abelian gauge symmetry.
We refer the interested reader to recent reviews \cite{Weigand:2018rez,Cvetic:2018bni} for detailed introduction to F-theory.}

Let us first summarize the relevant features of the model, whose explicit construction is detailed in \cite{Bies:2018uzw}.
The model has an $SU(5)$ gauge symmetry localized on a $dP_3$ surface inside the compact base threefold ${\cal B}_3$, which itself is a smooth hypersurface inside a toric variety.
There are matter states in the representations ${\bf 10}_1$, ${\bf 5}_3$ and ${\bf 5}_{-2}$, where the subscript denote the charges under an additional $U(1)$ gauge symmetry.
Each representation ${\bf R}$ resides on a curve $C_{\bf R}$ inside the $dP_3$ surface.
One can find a globally consistent vertical $G_4$-flux configuration that induces the chiral spectrum
\begin{align}\label{eq:chiral_indices}
  \chi({\bf 10}_1) = 3 \, , \quad \chi({\bf 5}_3) = 15 \, , \quad \chi({\bf 5}_{-2}) = -18 \, .
\end{align}
In the following, we will analyze in detail the vector-like spectrum in this setup.

\subsubsection*{Geometry of curves}

In the global geometry, the matter curves $C_{\bf R}$ are complete intersections involving the $dP_3$ surface and another divisor on the base ${\cal B}_3$.
As discussed in \cite{Bies:2018uzw}, a generic choice of the complex structure parameters for the elliptic fourfold also induces a generic curve $C_{\bf R}$ on $dP_3$.
In other words, we can parametrize them in terms of global sections of ${\cal O}_{dP_3}([C_{\bf R}])$, where $[C_{\bf R}]$ denotes the divisor class of the curve inside $dP_3$.

Furthermore, the data defining the zero mode spectrum in a global F-theory model can be extracted from the $G_4$-configuration and packaged into a line bundle (or, more generally, a coherent sheaf) for each curve $C_{\bf R}$ \cite{Bies:2014sra,Bies:2017fam}.
For the case at hand, the flux inducing the chiral spectrum \eqref{eq:chiral_indices} induces line bundles which are pullbacks of various bundles on $dP_3$ to the curves \cite{Bies:2018uzw}.

Using the same notation as in the previous section\footnote{Divisor classes $aH + b E_1 + c E_2 + d E_3$ are denoted by $(a;b,c,d)$.}, the curves with their genus and their corresponding zero-modes counting bundles are:
\begin{align}\label{tab:data_curves__in_GUT}
\begin{tabular}{ccc|cc}
\toprule
curve & class & genus & bundle & $h^i$ \\
\midrule
$C_{\mathbf{10}_1}$ & $(4;-1,-1,-2)$ & 2 & $ \mathcal{O}_{dP_3} \left( 1;-1,-1,1 \right) $ & $(3,0)$ \\
$C_{\mathbf{5}_3}$ & $(10;-3,-3,-4)$ & 24 & $ \mathcal{O}_{dP_3} \left( 5;-4,-4,3 \right) $ & $(15 + n, n)$ \\
$C_{\mathbf{5}_{-2}}$ & $(17;-5,-5,-7)$ & 79 & $ \mathcal{O}_{dP_3} \left( 6;0,0,-6 \right) $ & $(7,25)$ \\
\bottomrule
\end{tabular}
\end{align}
Note that the cohomologies on $C_{{\bf 10}_1}$ and $C_{{\bf 5}_{-2}}$ are fixed by the exactness of the corresponding Koszul resolutions, and hence there are no complex-structure-dependent jumps possible.\footnote{This can change if we modify the flux by, e.g., horizontal pieces. However, for the purpose of this work, we focus on jumps induced by geometric changes.}
For the representation ${\bf 5}_3$, no such arguments apply, and thus we expect the number $n$ of light vector-like pairs to vary.

The curve $C_{{\bf 5}_3} = \{a_{3,2} = 0\}$ is the vanishing locus of a polynomial with class $(10; -3, -3, -4)$, whose explicit expression in the parametrization of the toric $dP_3$ coordinates $x_i$ are given in appendix~\ref{sec:Tools}, cf.~\eqref{eq:explicit_form_a32}. 
With the curve having genus 24, it would be almost impossible to perform a scan by varying all the complex structure parameters (\eqref{eq:explicit_form_a32} has 44 coefficients), as we did previously for the low genus cases.
However, the intuition we gained from the low genus examples will help us to ``control'' $n$ --- that is, to efficiently find suitable geometries realizing the desired vector-like spectrum.

\subsection{Engineering jumps in cohomology} \label{subsec:JumpsInF-theoryExample}

What we have learned from the machine learning results is that the line bundle cohomology is more likely to jump if the curve in question is reducible.
Though we have already emphasized that rigidity of the components is not necessary, the abundance of toric coordinates makes it handy to factor out various different curves which in this case happen to be rigid.
For the purpose of finding \emph{a} concrete realization of a particular jump in the vector-like spectrum, these rigid factors turn out to be sufficient.

We thus modify the coefficients of the defining polynomial $a_{3,2}$ in~\eqref{eq:explicit_form_a32} such that individual toric coordinates $x_i$ of $dP_3$ factor out.
Of course, not every such factorization will lead to a jump: the rigid component must in some way receive a ``non-trivial contribution'', i.e., intersection, from the divisor $D_L$ defining the line bundle.
The intuitions we gained from the previous section is that a \emph{negative} intersection of $D_L$ with $V(x_i)$ will lead to a jump.
It is then intuitive to assume that the more rigid components splits off, the higher the jumps tend to be.
With this intuition, we now proceed to engineer step-wise jumps of the vector-like spectrum.

Using the linear relations \eqref{eq:dP3_Grading} and intersection numbers \eqref{eq:intersection_numbers_dP3}, we easily verify the divisor defining the line bundle, $D_L = 5 H - 4 E_1 - 4 E_2 + 3 E_3$, has only negative intersections with $[x_1]$ and $[x_6]$.
Inspecting~\eqref{eq:explicit_form_a32}, one finds that if we set
\begin{align}
    \begin{split}
      c_{40}=c_{41}=c_{42}=c_{43}=c_{44} = 0 \, ,
    \end{split}
\end{align}
the polynomial factors as $a_{3,2} = x_6 \, R_2$, where $R_2$ is an irreducible polynomial in the class $(10;-3,-3,-5)$. And indeed, a computer-assisted computation with methods from \cite{Bies:2018uzw} reveals that for this curve $C_2 = \{x_6 \, R_2 = 0\}$, we have
\begin{align}
    h^i( C_2, \left. \mathcal{O}_{dP_3} (5;-4,-4,3) \right|_{C_2} ) = (17,2) \, ,
\end{align}

We can factor out another factor $x_6$ from $R_2$ by setting 
\begin{align}
  \begin{split}
    c_{34}=c_{35}=c_{36}=c_{37}=c_{38}=c_{39}=c_{40}=c_{41}=c_{42}=c_{43}=c_{44} = 0 \, ,
  \end{split}
\end{align}
yielding $C_{{\bf 5}_3} \rightarrow C_3 = \{ x_6^2 \, R_3 =0 \}$, with $R_3$ an irreducible polynomial of class $(10; -3, -3, -6)$.
In this case, we find a jump by three,
\begin{align}
  h^i( C_3 , \left. \mathcal{O}_{dP_3} ((5;-4,-4,3) \right|_{C_3} ) = (18,3) \, .
\end{align}

To achieve a jump by four, we factorize $C_{{\bf 5}_3} \rightarrow C_4 = \{ x_1 \, x_6 \, R_4 =0 \}$, with $[R_4] = (9;-2,-2,-5)$, with the following choice of complex structure: 
\begin{align}
  \begin{split}
    c_1=c_2=c_3=c_4=c_5=c_{40}=c_{41}=c_{42}=c_{43}=c_{44} = 0 \, .
  \end{split}
\end{align}
Then we find
\begin{align}
  h^i( C_4 , \left. \mathcal{O}_{dP_3} ((5;-4,-4,3) \right|_{C_4} ) = (19,4) \, .
\end{align}

Lastly, we also easily construct a model with five vector-like pairs, by setting
\begin{align}
  \begin{split}
    &c_1=c_2=c_3=c_4=c_5=c_{34}=c_{35}=c_{36}=c_{37}=0\\
    &c_{38}=c_{39}=c_{40}=c_{41}=c_{42}=c_{43}=c_{44} = 0\,.
  \end{split}
\end{align}
On this sublocus in complex structure moduli space, the matter curve factorizes as $C_{{\bf 5}_3} \rightarrow C_5 = \{ x_1 \, x_6^2 \, R_5 =0 \}$, with $[R_5] = (9;-2,-2,-6)$.
In this case we have
\begin{align}
  h^i( C_5 , \left. \mathcal{O}_{dP_3} ((5;-4,-4,3) \right|_{C_5} ) = (20,5) \, .
\end{align}

\subsection{Single vector-like pair from Brill--Noether theory}

The above examples demonstrate how the machine learning intuition led us to a step-wise increase in the number of vector-like pairs by suitable tuning of the complex structure parameters.
These jumps occur because the matter curve in question splits into several components.
However, such splittings induce a jump from zero vector-like pairs to at least two (or three, or four, or five).
If we are interested in models with a single vector-like pair --- such as for the Higgs field in MSSM realizations --- then we need to look for other effects than curve splitting. 

As we have seen earlier, such effects are related to the cases not predicted by the trained decision tree.
Here, the jumps in cohomology are not due to the curve becoming non-generic, but rather the line bundle.
In fact, Brill--Noether theory (to be discussed in the next section, see also \cref{subsec:BNTheory}) tells us that for the matter curve $C_{{\bf 5}_3}$ of genus 24, we expect that a scenario with a single vector-like pair --- i.e., one having $h^i = (16,1)$ --- to occur on a subvariety of dimension $\rho = g - h^0 \cdot h^1 = 8$ of the space Jac$(C_{{\bf 5}_3})$ which parametrizes the line bundles on $C_{{\bf 5}_3}$.
Note that the same formula would yield $\rho = -10$ for jumps by two, and hence no such jumps can occur \emph{for a generic} $C_{{\bf 5}_3}$. This agrees with the above instances, as each of those requires the curve to become non-generic.

Because of this, engineering the jump by 1 becomes more challenging, and in particular requires additional tools from algebraic geometry. We defer the details of the relevant computations to \cref{subsec:Tools} and simply remark here that the necessary tuning is 
\begin{align}
  \begin{split}
  &c_1=c_2=c_3 = c_4= c_5 =c_7=c_8=c_9 = c_{10}=c_{35}=c_{36}=c_{37}=c_{38}=1 \, , \\
  &c_{40}=c_{41}=c_{42}=c_{43}=c_{44}=1, \quad c_{11}=c_{34}=-1,\quad   c_6 = c_{39}=2 \, .
\end{split}
\end{align}
One can easily verify that the polynomial $a_{3,2}$ in~\eqref{eq:explicit_form_a32} does not factorize in this case, and that the curve $C_{{\bf 5}_3}$ remains smooth. Therefore, the enhancement in cohomology in this case is indeed of Brill--Noether type.

\section{Cohomology jumps throughout the moduli space} \label{sec:DataMeetsAnalysis}

To put the intuition we gained from machine learning onto more solid grounds, we now apply tools from algebraic geometry to develop a more complete, ``microscopic'' understanding for the various sources of jumps we encountered in our data.
As we will see, the resulting insights lead to a diagrammatic representation of a \emph{stratification} of the complex structure moduli space of F-theory compactifications induced by vector-like spectra.

As we have alluded to in \cref{sec:MachineLearning}, based on our database we can essentially distinguish two types of jumps:
\begin{enumerate}
	\item Jumps due to a non-generic line bundle.
	\item Jumps due to a non-generic curve.
\end{enumerate}
This shows that our samplings are very atypical. Namely, true jump loci have lower dimensionality than the full set of parameters. Therefore, jump loci form sets of measure $0$ and should never be encountered by a genuinely random sample.

It is central to our discussion that algebraic geomemtry can bound from below the ‘size’ of such jump loci. In particular, this is true for jumps due to non-generic line bundles. Such jumps have been analyzed since 1874 in the context of \emph{Brill--Noether theory}\footnote{The physics community may find it entertaining to learn that this theory is named after Max Noether, the father of Emmy Noether.} \cite{Brill1874}. Given a generic curve $C_g$ of genus $g$ and an integer $d$, Brill--Noether theory provides an integer $\rho ( r, g, d )$ which measures how likely it is that a line bundle ${\cal L}_d$ of degree $d$ on $C_g$ has $r+1$ independent non-trivial global sections, i.e., has $h^0(C_g, {\cal L}_d) = r+1$.

To formulate this more precisely, first recall that the Jacobian $\mathrm{Jac} ( C_g )$ of the curve $C_g$ is isomorphic to $\mathbb{C}^g / \Lambda$ where $\Lambda$ is the full-dimensional period lattice of $C_g$. By the Abel--Jacobi map, equivalence classes of line bundles of degree $d$ form a copy of the Jacobian $\mathrm{Jac} ( C_g )$. Let us focus on the subset of the Jacobian formed by all equivalence classes of line bundles of degree $d$ which admit exactly $r+1$ global sections. Then a lower bound on the dimension of this space is given by the integer
\begin{align}
         \rho( r, g, d ) = g - \left( r + 1 \right) \cdot \left( r+1 - ( d - g + 1 ) \right) \equiv g - n^0 \cdot n^1 \, .
\end{align}
In the last equality we use the intuitive notation $n^0 = r+1$. Furthermore, we have used that by the Riemann--Roch theorem, $n^1 \equiv n^0 - ( d - g + 1 )$ is equal to $h^1(C_g, {\cal L}_d)$ if $h^0(C_g, {\cal L}_d) = n^0$. Further details on Brill--Noether theory can be found in appendix~\cref{subsec:BNTheory}, and a more complete presentation is given in~\cite{mumford1975curves,Griffiths:433962}.

An important result follows from \cite{griffiths1980}: If the curve is generic, then lines bundles of degree $d$ only admit numbers $r+1$ of global sections for which $\rho( r, g, d )$ is non-negative. Put differently, there are no line bundles on generic curves with $r+1$ global sections with $\rho( r, g, d ) < 0$. Furthermore, the value of $\rho$ gives a very clear notion of the likelihood to have $r+1$ sections in terms of a dimension on the ``moduli'' space of line bundles.

Let us demonstrate this for a line bundle $\mathcal{L}$ of degree $d = 2$ on a curve $C_g$ of genus $g = 3$. By general theory, the number of section of this line bundle cannot exceed its degree. Hence, it has $0$, $1$ or $2$ sections. With this information, let us compute $\rho( r, d, g )$:
\begin{equation}
\begin{array}{cc|c}
\toprule
r & h^i & \rho( r, d, g ) \\
\midrule
-1     & (0,0) & 3 \\
0     & (1,1) & 2 \\
1     & (2,2) & -1\\
\bottomrule
\end{array}
\end{equation}
From this we learn, that most line bundles $\mathcal{L}$ of degree $d = 2$ on a genus $g = 3$ curve $C_3$ satisfy $h^0 \left( C_3, \mathcal{L} \right) = 0$. Since for these bundles $\rho$ matches the dimension of the Jacobian of $C_3$, we can say that these line bundles are associated to generic points of the Jacobian. Furthermore, we learn that there are line bundles with $h^0 \left( C_3, \mathcal{L} \right) = 1$. However, these are special in the sense that they are associated to a codimension-1 locus in the Jacobian $\mathrm{Jac} ( C_3 )$.

Finally, $\rho = -1$ for $r = 1$ begs for an explanation. This explanation follows from work of Griffiths and Harris \cite{griffiths1980}:
\begin{center}
On \textbf{generic} curves, $\mathrm{dim} ( G^{r+1}_d ) = \rho \left( r, d, g \right)$.
\end{center}
So in particular, on generic curves it holds $G^{r+1}_d = \emptyset$ if and only if $\rho \left( r, d, g \right) < 0$. Consequently, we conclude from \cref{equ:BNSimpleExample2}, that on generic genus $g = 3$ curve, there is no line bundle $\mathcal{L}$ of degree $2$ such that $h^0 ( C_3, \mathcal{L} ) = 2$.

Note however, that this does not rule out the possibility that non-generic curves may host such line bundles. In the case at hand, it follows from the theorem of Clifford \cite{griffiths1980} that hyperelliptic curves $H_3$ of genus $g = 3$ admit line bundles $\mathcal{L}$ of degree $d = 2$ and $h^0 ( H_3, \mathcal{L} ) = 2$. Note that hyperelliptic curves of genus $g > 2$ are non-generic. Hence, this points us to jumps of the vector-like spectrum, which originate from non-generic deformations of the curve.

Let us give another such example, which illustrates a jump on a singular curve. To this end, let us consider a line bundle $\mathcal{L}$ of degree $d = 5$ on a genus $g = 2$ curve. Then $\chi ( \mathcal{L} ) = 4$ and $h^0 ( C_2, \mathcal{L} ) \in \{ 4,5 \}$. Let us compute $\rho( r, d, g )$ for these two values of global sections:
\begin{equation}
\begin{tabular}{cc|c}
\toprule
$r$ & $h^i( C_2, \mathcal{L} )$ & $\rho( r, d, g )$ \\
\midrule
3 & (4,0) & $2$ \\
4 & (5,1) & $-3$ \\
\bottomrule
\end{tabular} \label{equ:BNSimpleExample}
\end{equation}
Thus, on a smooth curve of genus $g = 2$, any line bundle of degree $d = 5$ has $4$ global sections. Even more, since the degree $d$ is in the stable range, we find $4$ global sections for this line bundle on every smooth curve of genus $g = 2$ --- generic or not. Hence, $5$ sections can only be realized on a singular curve.

This can be achieved by choosing the curve parameters (which model the complex structure moduli of global F-theory models) such that the curve becomes reducible, and factors into various components which intersect transversely in a number of points. A way to construct global sections on such curves is then as follows: First, consider each component individually and identify which sections they support. Then, by demanding that these sections agree at the intersection points, we glue these local sections to global sections.
We will return to this gluing procedure in more detail in \cref{sec:LocalToGlobalSectionCounting}.

In this section, we will take a closer look at the interplay of jumps that occur due to non-genericity both of the line bundle and the curve. In particular, since in global F-theory models, both the bundle and the curve depend on the complex structure parameters of the elliptic fibration in the same fashion (namely through the coefficients of its defining polynomials), they should be treated on the same footing, which we can summarize diagrammatically. The following analysis requires, at a technical level, a working understanding of the Koszul resolution of a pullback bundle, its associated long exact sequence in sheaf cohomology, inferring the maps in this long exact sequence from \v{C}ech ochomology as well as a basic understanding of on-reduced curves. For convenience of the reader, further details are provided in \cref{subsec:Tools}.

\subsection{Jumps from curve splittings}\label{subsec:jumps_from_splitting}

We first analyze examples with jumps from curve splittings. We will see that rigidity of the components that split off play no role in the section counting. The reason why we found in earlier chapters that rigid divisors split off is due to our special choice of setting all coefficients in the polynomial that specify the curve in $dP_3$ to either zero or one.

\subsubsection{Example: one additional section} \label{subsec:OneAdditionalSection}

\paragraph{Setup}

Let us return to the example of a line bundle on a genus 2 curve discussed above. In more detail, the curve and line bundle are given by
\begin{align}
D_C = ( 4 ; -1, -2, -1 ) \, , \qquad  D_L = ( 3; -3, -1, -2 ) \, .
\end{align}
The curve $C \left( \mathbf{c} \right) = V \left( P( \mathbf{c} ) \right)$ is defined by a polynomial $P( \mathbf{c} ) \in H^0 \left( dP_3, \mathcal{O}_{dP3} ( D_C ) \right) \cong \mathbb{C}^{10}$ with
\begin{align}
\begin{split}
P( \mathbf{c} ) &= 
c_1 x_1^3 x_2^3 x_3 x_4
+ c_2 x_1^2 x_2^3 x_4^2 x_6
+ c_3 x_1^3 x_2^2 x_3^2 x_5
+ c_4 x_1^2 x_2^2 x_3 x_4 x_5 x_6
+ c_5 x_1 x_2^2 x_4^2 x_5 x_6^2 \\
& + c_6 x_1^2 x_2 x_3^2 x_5^2 x_6
+ c_7 x_1 x_2 x_3 x_4 x_5^2 x_6^2
+ c_8 x_2 x_4^2 x_5^2 x_6^3
+ c_9 x_1 x_3^2 x_5^3 x_6^2
+ c_{10} x_3 x_4 x_5^3 x_6^3 \, ,
\end{split} \label{equ:PForFirstExample}
\end{align}
where the coefficients $\mathbf{c} \in \mathbb{C}^{10}$ form the parameter space of this genus $g = 2$ setup.
The line bundle $\mathcal{L} ( \mathbf{c} ) = \left. \mathcal{O}_{dP3}( D_L ) \right|_{C ( \mathbf{c} )}$ satisfies $\mathrm{deg} ( \mathcal{L} ( \mathbf{c} ) ) = 5$. Hence, on smooth curves, the theorem of Riemann--Roch tells us
\begin{align}
\chi ( \mathcal{L} ( \mathbf{c} ) ) = \mathrm{deg} ( \mathcal{L} ( \mathbf{c} ) ) - g + 1 = 5 - 2 + 1 = 4 \, .
\end{align}
Moreover, since $\mathrm{deg} ( \mathcal{L} ( \mathbf{c} ) ) = 5 > 2g - 2$, we know that for smooth curves $h^1( C ( \mathbf{c} ), \mathcal{L} ( \mathbf{c} ) ) = 0$. Hence, $h^0 ( \mathcal{L} ( \mathbf{c} ) ) = 5$ is only possible on non-smooth curves.

\paragraph{Comparison with database}

In our database, we have considered choices of parameters $\mathbf{c} \in \{ -1,0,1 \}^{10} - \mathbf{0}$. On about 96\% of these 59048 curves, $\mathcal{L} ( \mathbf{c} )$ has 4 sections. This fits with the above picture, that generically we expect 4 sections. However, we also find 2186 curves for which $\mathcal{L} ( \mathbf{c} )$ has 5 sections. Those curves satisfy $c_3 = c_6 = c_9 = 0$, which means that $C( \mathbf{c} ) = V( x_4 ) \cup B$, where
\begin{align}
\begin{split}
B &= V( c_1 x_1^3 x_2^3 x_3 + c_2 x_1^2 x_2^3 x_4 x_6 + c_4 x_1^2 x_2^2 x_3 x_5 x_6 + c_5 x_1 x_2^2 x_4 x_5 x_6^2 \\
  & \qquad \qquad \qquad \qquad \qquad \qquad + c_7 x_1 x_2 x_3 x_5^2 x_6^2 + c_8 x_2 x_4 x_5^2 x_6^3 + c_{10} x_3 x_5^3 x_6^3 )
\end{split}
\end{align}
is a genus-0 curve with $V( x_4 ) \cdot B = 3$. We will now argue that $\mathcal{L} ( \mathbf{c} )$ admits 5 sections if and only if $C( \mathbf{c} )$ decomposes in this way.

\paragraph{Classification of jump geometries}

To this end, we consider the Koszul resolution
\begin{align}
0 &\to \mathcal{O}_{dP_3} \left( D_L - D_C \right) \xrightarrow{\alpha} \mathcal{O}_{dP_3} \left( D_L \right) \to \mathcal{L} ( \mathbf{c} ) \to 0 \, .
\end{align}
Its associated long exact sequence in sheaf cohomology takes the form
\begin{equation}
\begin{tikzpicture}[scale=2, baseline=(current  bounding  box.center)]
          \matrix(m)[matrix of math nodes,column sep=15pt,row sep=15pt]{
                0 & 0
                  & H^0 \left( dP_3, D_L \right) \cong \mathbb{C}^1
                  & H^0 \left( C ( \mathbf{c} ), \mathcal{L} ( \mathbf{c} ) \right) \\
                  & H^1 \left( dP_3, D_L - D_C \right) \cong \mathbb{C}^4
                  & H^1 \left( dP_3, D_L \right) \cong \mathbb{C}^1
                  & H^1 \left( C ( \mathbf{c} ), \mathcal{L} ( \mathbf{c} ) \right) \\
                  & 0
                  & 0
                  & 0 & 0 \\
                     };
               \draw[->,font=\scriptsize,every node/.style={above},rounded corners]
                     (m-1-1) edge (m-1-2)
                     (m-1-2) edge (m-1-3)
                     (m-1-3) edge (m-1-4)
                     (m-1-4.east) --+(5pt,0)|-+(0,-7.5pt)-|([xshift=-5pt]m-2-2.west)--(m-2-2.west);
                \draw[->,font=\scriptsize,every node/.style={above},rounded corners]
                     (m-3-2) edge (m-3-3)
                     (m-3-3) edge (m-3-4)
                     (m-3-4) edge (m-3-5);
                \draw[->] (m-2-2) edge node [above] {$\varphi$} (m-2-3);
                \draw[->] (m-2-3) edge node [above] {} (m-2-4);
                \draw[->,font=\scriptsize,every node/.style={above},rounded corners] (m-2-4.east) --+(5pt,0)|-+(0,-7.5pt)-|([xshift=-5pt]m-3-2.west)-- (m-3-2.west);
\end{tikzpicture}
\end{equation}
The exactness of this sequence implies that
\begin{align}
\begin{split}
h^0 ( C ( \mathbf{c} ), \mathcal{L} ( \mathbf{c} ) ) = 5 - \mathrm{dim} \left( \mathrm{im} \varphi \right) = 5 - \mathrm{rk} \left( M_\varphi \right) \, , \label{equ:H1ForFirstExample}
\end{split}
\end{align}
where $M_\varphi = \left( c_3, c_6, c_9, 0 \right)$. We explain the construction of the mapping matrix $M_\varphi$ in more detail in \cref{subsec:Tools}.

Obviously, $M_\varphi$ has rank $1$ iff $(c_3,c_6,c_9) \neq \mathbf{0}$ and its rank vanishes iff $(c_3, c_6, c_9) = \mathbf{0}$. This immediately leads to the following classification of curve geometries:
\begin{align}
\begin{tabular}{ccc}
\toprule
$\mathrm{rk} \left( M_\varphi \right)$ & explicit condition & curve splitting \\
\midrule
1 & $( c_3, c_6, c_9 ) \neq \mathbf{0}$ & $C$ \\
0 & $( c_3, c_6, c_9 ) = \mathbf{0}$ & $V( x_4 ) \cup B$ \\
\bottomrule
\end{tabular}
\end{align}
showing that we obtain one additional vector-like pair if and only if the curve factors as $V( x_4 ) \cup B$.
We illustrate this result in the following diagram:
\begin{equation}
\begin{tikzpicture}[baseline=(current bounding box.center)]

          \def\w{2};
          \def\h{2};

          \node (A) at (0,0) [draw, blue] {$C$};
          \node (E) at (0,-1*\h) [draw, red] {$V( x_4 ) \cup B$};
          
          \node (L1) at (-1.8*\w,0) {$(h^0, \rho ) = (4,1)$};
          \node (L3) at (-1.8*\w,-1*\h) {$(h^0, \rho ) = (5,-3)$};
          
          \draw[-latex,thick] (A) --node[left] {} (E);
          
\end{tikzpicture} \label{equ:StratDiagram1}
\end{equation}
In this diagram, the $a^\text{th}$ node represents a family $\mathcal{F}_a$ of curves, for which we give the generic element in this family. 

For example, the family $\mathcal{F}_1$ of curves at the first node is defined by the condition $( c_3, c_6, c_9 ) \neq \mathbf{0}$ and has the curve $C$ as its generic element,  which is a smooth, irreducible curve of genus $g = 2$. Note that (non-generic) members of $\mathcal{F}_1$ can also be singular curves with several components. For example, the curve $V( x_1^3 x_2^2 x_3^2 x_5 )$ is defined by the condition that all $c_i$ but $c_3$ vanish. This curve is clearly singular and has several connected components. Recall that $\mathcal{F}_1$ is the family of curves on which the line bundle in question admits four global sections. Hence, the statement is that even on such a very singular curve, the bundle in question admits exactly four sections.

This feature changes exactly on the family of curves $\mathcal{F}_2$, which are defined by $( c_3, c_6, c_9 ) \equiv \mathbf{0}$. Its generic element is a curve of the form $V( x_4 ) \cup B$, where $B$ is a smooth genus $g = 0$ curve touching $V( x_4 )$ in 3 distinct points. 
We can also view ${\cal F}_1 = \{ {\bf c} \, | \, (c_3, c_6, c_9) \neq 0 \}$ and ${\cal F}_2 = \{ {\bf c} \, | \, (c_3, c_6, c_9) = 0 \}$ as subspaces of the parameter space $\mathbb{C}^{10} \ni {\bf c}$.
In this case it is trivial to see that
\begin{align}
\mathcal{F}_1 \cap \mathcal{F}_2 = \emptyset \, , \qquad \mathcal{F}_2 \subset \overline{\mathcal{F}_1} \, ,
\end{align}
where $\overline{\mathcal{F}_1}$ the closure with respect to the standard topology on $\mathbb{C}^{10}$.
We will come back to this property shortly.

\subsubsection{An \texorpdfstring{\boldmath{$h^0$}}{h0}-gap} \label{subsubsec:GapExample}

Whilst factoring-off curve components typically increases the number of global sections, this effect need not necessarily generate exactly one additional section, as we have already seen above. Rather, it can force multiple additional sections to appear simultaneously.
An example of this sort is
\begin{align}
D_C = ( 3 ; -1, -1, -1 ) \, , \qquad D_L = ( 1; -1, -3, -1 ) \, .
\end{align}
In this case, $C( \mathbf{c} ) = V( P( \mathbf{c} ) )$ is a genus $1$ curve defined by
\begin{align}
\begin{split}
P( \mathbf{c} ) &= c_1 x_1^2 x_2^2 x_3 x_4 + c_2 x_1^2 x_2 x_3^2 x_5 + c_3 x_1 x_2^2 x_4^2 x_6 + c_4 x_1 x_2 x_3 x_4 x_5 x_6 \\
  &\qquad \qquad + c_5 x_1 x_3^2 x_5^2 x_6 + c_6 x_2 x_4^2 x_5 x_6^2 + c_7 x_3 x_4 x_5^2 x_6^2 \, .
\end{split}
\end{align}
Moreover, $\mathcal{L}$ is a line bundle of degree $d = -2$. Hence, its degree is in the stable regime and on any smooth curve we find $h^0( C, \mathcal{L} ) = 0$. Still, as demonstrated in \cref{subsec:OneAdditionalSection}, non-smooth curves can admit higher numbers of global sections. Here, we will argue, that even on singular curve, the pullback line bundle $\mathcal{L}$ can never have exactly one section.

To see this, let us look at the long exact sequence in sheaf cohomology associated to the Koszul resolution of the setup:
\begin{equation}
\begin{tikzpicture}[scale=2, baseline=(current  bounding  box.center)]
          \matrix(m)[matrix of math nodes,column sep=15pt,row sep=15pt]{
                0 & 0
                  & 0
                  & H^0 \left( D_C, \mathcal{L} \right) \\
                  & H^1 \left( dP_3, D_L - D_C \right) \cong \mathbb{C}^{3}
                  & H^1 \left( dP_3, D_L \right) \cong \mathbb{C}^{5}
                  & H^1 \left( C, \mathcal{L} \right) \\
                  & 0
                  & 0
                  & 0 & 0 \\
                     };
               \draw[->,font=\scriptsize,every node/.style={above},rounded corners]
                     (m-1-1) edge (m-1-2)
                     (m-1-2) edge (m-1-3)
                     (m-1-3) edge (m-1-4)
                     (m-1-4.east) --+(5pt,0)|-+(0,-7.5pt)-|([xshift=-5pt]m-2-2.west)--(m-2-2.west);
                \draw[->,font=\scriptsize,every node/.style={above},rounded corners]
                     (m-3-2) edge (m-3-3)
                     (m-3-3) edge (m-3-4)
                     (m-3-4) edge (m-3-5);
                \draw[->] (m-2-2) edge node [above] {$\varphi$} (m-2-3);
                \draw[->] (m-2-3) edge node [above] {} (m-2-4);
                \draw[->,font=\scriptsize,every node/.style={above},rounded corners] (m-2-4.east) --+(5pt,0)|-+(0,-7.5pt)-|([xshift=-5pt]m-3-2.west)-- (m-3-2.west);
\end{tikzpicture}
\end{equation}
The exactness of this sequence implies $h^0 ( C, \mathcal{L} ) = 3 - \mathrm{dim} \left( \mathrm{im} \varphi \right) = 3 - \mathrm{rk} \left( M_\varphi \right)$ with
\begin{align}
\begin{split}
\qquad M_\varphi = 
\begin{psmallmatrix}
c_6 & 0 & 0 \\
c_7 & c_6 & 0 \\
c_3 & 0 & 0 \\
c_1 & 0 & c_3 \\
c_4 & c_3 & c_6 \\
\end{psmallmatrix} \, . \label{equ:MVarphiEample5}
\end{split}
\end{align}
Consequently, the statement that on the curves in class $D_C$ the pullback of $D_L$ never has exactly one section is equivalent to saying that $M_\varphi$ never has rank $2$. We see this by studying the four non-trivial and independent $3 \times 3$-minors of $M_\varphi$:
\begin{align}
m_1 = c_6^2 c_3 \, , \qquad m_2 = c_6^3 \, \qquad m_3 = c_3^3 \, , \qquad m_4 = c_6 c_4 c_3 - c_7 c_3^2 - c_6^2 c_1 \, .
\end{align}
Now, $\mathrm{rk} ( M_\varphi ) < 3$ requires $m_1 = m_2 = m_3 = m_4 = 0$. This is equivalent to $c_3 = c_6 = 0$ and
\begin{align}
\left. M_\varphi \right|_{c_3 = c_6 = 0} = 
\begin{psmallmatrix}
0 & 0 & 0 \\
c_7 & 0 & 0 \\
0 & 0 & 0 \\
c_1 & 0 & 0 \\
c_4 & 0 & 0 \\
\end{psmallmatrix} \, , \label{equ:MVarphiEample5-II}
\end{align}
which can have at most rank 1. More generally, we can classify the rank of $M_\varphi$ and thereby summarize the curve geometry as follows:
\begin{align}
\begin{tabular}{ccc}
\toprule
$\mathrm{rk} ( M_\varphi )$ & explicit condition (${\cal F}_i$) & splitting of curve \\
\midrule
3 & $c_3, c_{6} \neq 0$ & $C$ \\
\midrule
1 & $c_3 = c_{6} = 0$ & $E_2 \cup B$ \\
\midrule
0 & $c_1 = c_3 = c_4 = c_6 = c_7 = 0$ & $E_6 \cup E_4 \cup E_2^{(2)} \cup A$ \\
\bottomrule
\end{tabular}
\end{align}
Observe again that within the parameter space of ${\bf c}$, we have 
\begin{align}
	{\cal F}_i \cap {\cal F}_j = \emptyset \, , \quad \overline{{\cal F}_1} \supset {\cal F}_2 \, , \quad \overline{{\cal F}_2} \supset {\cal F}_3 \, .
\end{align}
The corresponding diagram is
\begin{equation}
\begin{tikzpicture}[baseline=(current bounding box.center)]

          \def\w{4};
          \def\h{2};

          \node (A) at (0,0) [draw, blue] {$C$};
          \node (E) at (0,-1*\h) [draw, red] {$E_2 \cup B$};
          \node (G) at (0,-2*\h) [draw, red] {$E_6 \cup E_4 \cup E_2^{(2)} \cup A$};
          
          \node (L1) at (-1.8*\w,0) {${\cal F}_1 : \, (h^0, \rho ) = (0,1)$};
          \node (L3) at (-1.8*\w,-1*\h) {${\cal F}_2 : \, (h^0, \rho ) = (2,-7)$};
          \node (L4) at (-1.8*\w,-2*\h) {${\cal F}_3 : \, (h^0, \rho ) = (3,-14)$};
          
          \draw[-latex,thick] (A) --node[left] {} (E);
          \draw[-latex,thick] (E) --node[left] {} (G);
          
\end{tikzpicture}\label{fig:stratdiagram2}
\end{equation}

\subsubsection{Jump from non-rigid curve splitting} \label{subsubsec:JumpNonRigid}

We now address the bias in our data, and provide a concrete example of jumps from curve splitting where none of the components are rigid.
To this end, we consider $D_C = (2;-1,-1,0)$ and $D_L = (-2,0,4,0)$. This curve is thus given by
\begin{align}
P = c_1 x_4 x_5 x_6^2 + c_2 x_1 x_3 x_5 x_6 + c_3 x_1 x_2 x_4 x_6 + c_4 x_1^2 x_2 x_3 \, .
\end{align}
For generic coefficients $c_i$, the curve $C$ is a smooth curve of genus $g = 0$ and $\mathcal{L}$ has degree $d = 0$. Hence we conclude $h^0( C, \mathcal{L} ) = 1$.

To understand jumps at special coefficients, we employ the Koszul resolution and find $h^0 \left( C ( \mathbf{c} ), \mathcal{L} ( \mathbf{c} ) \right) = 7 - \mathrm{rk} ( M )$ where
\begin{align}
M = \left( \begin{array}{ccccccc} 0 & 0 & 0 & c_1 & 0 & c_3 & 0 \\
                         c_4 & 0 & 0 & 0 & c_1 & c_2 & c_3 \\
                         c_3 & 0 & 0 & 0 & 0 & c_1 & 0 \\
                         0 & c_1 & 0 & c_2 & c_3 & c_4 & 0 \\
                         0 & 0 & c_1 & c_3 & 0 & 0 & 0 \\
                         0 & 0 & 0 & 0 & c_2 & 0 & c_4
\end{array} \right) \, .
\end{align}
The rank drops of this matrix include both cases of rigid and non-rigid splittings. Explicitly, let us set $A_i = V( x_i )$, which are rigid components. 
Moreover, we also have the following possible genus $g = 0$ components which are non-rigid:
\begin{align}
\begin{split}
D_1 &= V( c_2 x_3 x_5 x_6 + c_3 x_2 x_4 x_6 + c_4 x_1 x_2 x_3 ) \, , \\
D_2 &= V( c_3 x_4 x_6 + c_4 x_1 x_3 ) \, , \\
D_3 &= V( c_4 x_1 x_2 + c_2 x_5 x_6 ) \, , \\
D_4 &= V( c_2 x_1 x_3 + c_1 x_4 x_6 ) \, , \\
D_5 &= V( c_3 x_1 x_2 + c_1 x_5 x_6 ) \, .
\end{split} \label{equ:ComponentsNonRigidSplit}
\end{align}
With these, we can then summarize the rank drops as follows:
\begin{align}
\begin{tabular}{ccc}
\toprule
$\mathrm{rk} ( M )$ & explicit condition & curve splitting \\
\midrule
6 & generic & $C$ \\
\midrule
5 & $c_1 = 0$ & $A_1 \cup D_1$\\
5 & $c_1 c_4 = c_2 c_3$ & $D_2 \cup D_3$ \\
\midrule
3 & $c_1 = c_3 = 0$ & $A_1 \cup A_3 \cup D_3$ \\
\bottomrule
\end{tabular}
\end{align}
The corresponding diagram is of the form
\begin{equation}
\begin{tikzpicture}[baseline=(current bounding box.center)]
          
          \def\w{3.5};
          \def\h{2};
          
          \node (A) at (0,0) [draw, blue] {$C$};
          \node (B) at (-.6*\w,-\h) [draw, red] {$A_1 \cup D_1$};
          \node (C) at (.6*\w,-\h) [draw, red] {$D_2 \cup D_3$};
          \node (E) at (0,-2*\h) [draw, red] {$A_1 \cup A_3 \cup D_3$};
          
          \node (L1) at (-1.8*\w,0) {${\cal F}_1 : \, (h^0, \rho ) = (1,0)$};
          \node (L2) at (-1.8*\w,-\h) {${\cal F}_2 : \, (h^0, \rho ) = (2,-2)$};
          \node (L3) at (-1.8*\w,-2*\h) {${\cal F}_4 : \, (h^0, \rho ) = (4,-12)$};
          
          \draw[-latex,thick] (A) --node[left] {} (B);
          \draw[-latex,thick] (A) --node[left] {} (C);
          
          \draw[-latex,thick] (B) --node[left] {} (E);
          \draw[-latex,thick] (C) --node[left] {} (E);
          
\end{tikzpicture}\label{fig:diagram_non-rigid_split}
\end{equation}
Similar to our discussion in \cref{subsubsec:GapExample}, there is a gap at $h^0 = 3$. 
Crucially, since $D_2$ and $D_3$ are non-rigid, the deformation $C \to D_2 \cup D_3$ provides an explicit example of a jump associated to curve splitting with no rigid components.

\subsection{Jumps from non-generic line bundles}\label{subsec:BNJumps}

We now turn to jumps due to special alignments of the points that define a line bundle divisor. These phenomena are described by Brill--Noether theory.

\subsubsection{Additional section due to special divisors} \label{subsubsec:BN1}

Let us consider the pair
\begin{align}
D_C = ( 4; -1, -1, -1 ) \, , \qquad D_L = ( 1; 2, -2, -1 ) \, .
\end{align}
This genus $g = 3$ curve $C( \mathbf{c} ) = V( P( \mathbf{c} ) )$ is defined by
\begin{align}
\begin{split}
P ( \mathbf{c} ) &= c_1 x_1^3 x_2^3 x_3^2 x_4 + c_2 x_1^2 x_2^3 x_3 x_4^2 x_6 + c_3 x_1 x_2^3 x_4^3 x_6^2 + c_4 x_1^3 x_2^2 x_3^3 x_5 + c_5 x_1^2 x_2^2 x_3^2 x_4 x_5 x_6 \\
                 &\qquad + c_6 x_1 x_2^2 x_3 x_4^2 x_5 x_6^2 + c_7 x_2^2 x_4^3 x_5 x_6^3 + c_8 x_1^2 x_2 x_3^3 x_5^2 x_6 + c_9 x_1 x_2 x_3^2 x_4 x_5^2 x_6^2 \\
                 &\qquad + c_{10} x_2 x_3 x_4^2 x_5^2 x_6^3 + c_{11} x_1 x_3^3 x_5^3 x_6^2 + c_{12} x_3^2 x_4 x_5^3 x_6^3 \, . \label{equ:DefiningEquation}
\end{split}
\end{align}
Brill--Noether theory implies
\begin{align}
\begin{tabular}{cc|ccc|ccc}
\toprule
curve & $g$ & $\mathcal{L}$ & $\chi$ & $d$ & \multicolumn{3}{c}{BN-theory} \\
\midrule
\multirow{4}{*}{$C = V( P )$} & \multirow{4}{*}{3} & \multirow{4}{*}{$\left. \mathcal{O}_{dP3}( D_L ) \right|_{C}$} & \multirow{4}{*}{1} & \multirow{4}{*}{3} & $h^0$ & $h^1$ & $\rho$ \\
                  &    &                                                                                  &    &   & 1 & 0 & 3 \\
                  &    &                                                                                  &    &   & 2 & 1 & 1 \\
                  &    &                                                                                  &    &   & 3 & 2 & $-3$ \\
\bottomrule
\end{tabular}
\end{align}
Hence, a jump on the generic curve --- a Brill--Noether jump --- to $h^0 ( C( \mathbf{c} ), \mathcal{L} ( \mathbf{c} ) ) = 2$ is possible. 
To explicitly construct such curves, we again inspect the long exact sequence, associated to the Koszul resolution of $\mathcal{L} ( \mathbf{c} )$, which is given by
\begin{equation}
\begin{tikzpicture}[scale=2, baseline=(current  bounding  box.center)]
          \matrix(m)[matrix of math nodes,column sep=15pt,row sep=15pt]{
                0 & 0
                  & 0
                  & H^0 \left( D_C, \mathcal{L} \right) \\
                  & H^1 \left( dP_3, D_L - D_C \right) \cong \mathbb{C}^3
                  & H^1 \left( dP_3, D_L \right) \cong \mathbb{C}^2
                  & H^1 \left( C, \mathcal{L} \right) \\
                  & 0
                  & 0
                  & 0 & 0 \, . \\
                     };
               \draw[->,font=\scriptsize,every node/.style={above},rounded corners]
                     (m-1-1) edge (m-1-2)
                     (m-1-2) edge (m-1-3)
                     (m-1-3) edge (m-1-4)
                     (m-1-4.east) --+(5pt,0)|-+(0,-7.5pt)-|([xshift=-5pt]m-2-2.west)--(m-2-2.west);
                \draw[->,font=\scriptsize,every node/.style={above},rounded corners]
                     (m-3-2) edge (m-3-3)
                     (m-3-3) edge (m-3-4)
                     (m-3-4) edge (m-3-5);
                \draw[->] (m-2-2) edge node [above] {$\varphi$} (m-2-3);
                \draw[->] (m-2-3) edge node [above] {} (m-2-4);
                \draw[->,font=\scriptsize,every node/.style={above},rounded corners] (m-2-4.east) --+(5pt,0)|-+(0,-7.5pt)-|([xshift=-5pt]m-3-2.west)-- (m-3-2.west);
\end{tikzpicture}
\end{equation}
From the exactness of this sequence, we learn that $h^0 ( C( \mathbf{c} ), \mathcal{L} ( \mathbf{c} ) ) = 3 - \mathrm{rk} \left( M_\varphi \right)$ with
\begin{align}
M_\varphi = \begin{psmallmatrix} c_3 & c_2 & c_1 \\ 0 & c_{12} & c_{11} \end{psmallmatrix} \, . \label{equ:MatrixExample3}
\end{align}
We set $\mathbb{P}^1_a = V( x_2 )$, $\mathbb{P}^1_b = V( x_5 )$. 
Then the possible $h^0$ jumps are classified as
\begin{align}
\begin{tabular}{ccc}
\toprule
$\mathrm{rk} ( M_\varphi )$ & explicit condition & curve splitting \\
\midrule
2 & $\left( c_3 c_{11}, c_3 c_{12}, c_2 c_{11} - c_1 c_{12} \right) \neq \mathbf{0}$ & $C^1$ \\
\midrule
1 & $c_3 = 0$, $c_2 c_{11} - c_1 c_{12} = 0$ & $C^2$ \\
1 & $c_1 = c_2 = c_3 = 0$ & $B_2 \cup \mathbb{P}^1_b$ \\
1 & $c_{11} = c_{12} = 0$ & $\mathbb{P}^1_a \cup B_1$ \\
\midrule
0 & $c_1 = c_2 = c_3 = c_{11} = c_{12} = 0$ & $\mathbb{P}^1_a \cup A \cup \mathbb{P}^1_b$ \\
\bottomrule
\end{tabular}
\end{align}
The corresponding diagram is of the form
\begin{equation}
\begin{tikzpicture}[baseline=(current bounding box.center)]
          
          \def\w{4};
          \def\h{2};
          
          \node (A) at (0,0) [draw, blue] {$C^1$};
          \node (B) at (0,-\h) [draw, blue] {$C^2$};
          \node (C) at (-\w,-\h) [draw, red] {$\mathbb{P}^1_a \cup B_2$};
          \node (D) at (\w,-\h) [draw, red] {$B_1 \cup \mathbb{P}^1_b$};
          \node (E) at (0,-2*\h) [draw, red] {$\mathbb{P}^1_a \cup A \cup \mathbb{P}^1_b$};
          
          \node (L1) at (-1.8*\w,0) {${\cal F}_1 : \, (h^0, \rho ) = (1,3)$};
          \node (L2) at (-1.8*\w,-\h) {${\cal F}_2 : \, (h^0, \rho ) = (2,1)$};
          \node (L3) at (-1.8*\w,-2*\h) {${\cal F}_3 : \, (h^0, \rho ) = (3,-3)$};
          
          \draw[-latex,thick] (A) --node[left] {} (B);
          
          \draw[-latex,thick] (A) --node[left] {} (C);
          \draw[-latex,thick] (A) --node[left] {} (D);
          
          \draw[-latex,thick] (B) --node[left] {} (E);
          \draw[-latex,thick] (C) --node[left] {} (E);
          \draw[-latex,thick] (D) --node[left] {} (E);
          
\end{tikzpicture} \label{equ:DecayChannelExample3}
\end{equation}
The change of coefficients
\begin{align}
\mathbf{c} = (1, 1, 1, 1, 1, 1, 1, 1, 1, 1, 1, 1) \quad \to \quad \mathbf{c} = (1, 1, 0, 1, 1, 1, 1, 1, 1, 1, 1, 1)
\end{align}
leads to a transition $C^1 \to C^2$ of smooth, irreducible curves. Since the topology of the curve does not change for this choice of parameters, such a transition cannot be detected from the topological data which we used for our machine learning. Therefore, such transitions are the major source of error in our decision trees.

On smooth curves $C^i$, the nature of the jump $C^1 \to C^2$ can be analyzed by using Serre duality:
\begin{align}
\begin{split}
h^1 \left( C, \mathcal{O}_C \left( \left. D_L \right|_{C} \right) \right) > 0 &\quad \Leftrightarrow \quad h^0 \left( C, \mathcal{O}_C \left( K_C - \left. D_L \right|_{C} \right) \right) > 0 \\
                                     &\quad \Leftrightarrow \quad K_C - \left. D_L \right|_{C} \text{ effective} \\
                                     &\quad \Leftrightarrow \quad \exists p \in C \colon K_C - p \sim \left. D_L \right|_{C} \, .
\end{split}
\end{align}
Hence, the origin of this jump is that $K_C$ and the line bundle divisor differ, modulo linear equivalence, only by a point on $C$. Such a divisor is known as a \emph{special divisor}. Loosely speaking, we may thus say that the origin of this one additional sections is that the points, which define the line bundle on the curve, move into a special alignment.

Note that also in this case, the diagram \eqref{equ:DecayChannelExample3} encodes a hierarchy $\overline{{\cal F}_1} \supset {\cal F}_2$, $\overline{{\cal F}_2} \supset {\cal F}_3$.
This is a generic feature of the parameter space and reflects a \emph{stratification} induced by the vector-like spectrum.

\subsection[\texorpdfstring{$h^0$}{h0}-stratification of the parameter space]{\boldmath{$h^0$}-stratification of the parameter space} \label{subsubsec:JumpSpecialDivisorOnCurveComponent}

A stratification of a topological space $X$ is a decomposition $X = \bigcup_i {\cal F}_i$ into locally closed subspaces ${\cal F}_i$ such that
\begin{enumerate}
	\item ${\cal F}_i \cap {\cal F}_j = \emptyset$ if $i \neq j$,
	\item if ${\cal F}_i \cap \overline{{\cal F}_j} \neq \emptyset$, then ${\cal F}_i \subset \overline{{\cal F}_j}$.
\end{enumerate}
Intuitively speaking, a feature associated to a subspace ${\cal F}_i$ --- a so-called stratum --- becomes ``less likely'' with increasing codimension of ${\cal F}_i$, and being contained in (the closure of) a higher dimensional stratum ${\cal F}_j$ implies a ``specialization'' of the feature when going from ${\cal F}_i$ to ${\cal F}_j$ with $j>i$.
The second defining property has a convenient diagrammatic representation: Let the strata ${\cal F}_i$ form vertices of a graph, then there is a directed edge going from $j$ to $i$ if ${\cal F}_i \subset \overline{{\cal F}_j}$.
This is precisely the structure of the diagrams \eqref{equ:StratDiagram1}, \eqref{fig:stratdiagram2}, \eqref{fig:diagram_non-rigid_split}, and \eqref{equ:DecayChannelExample3}.
Here, the stratified $X$ is the parameter space $\{ \bf c\}$ associated with a pair $(D_C, D_L)$, and the strata are defined by the value of $h^0(C ({\bf c}), {\cal L}({\bf c}))$ in the notation of the previous subsections. Hence, we call these diagrams \emph{$h^0$-stratification}, or in short, stratification diagrams.

Note that Brill--Noether theory basically provides an analog description of the moduli space of line bundles / divisors on a smooth curve.
In particular, it provides lower bounds on the dimension of the strata in terms of $\rho$. For F-theory models, where also deformations of the curve's topology become relevant, we see that the stratification by $h^0$ can be extended to the enlarged moduli space.

We observe that in this generalized setting, a stratum associated to a certain value of $h^0$ can consist of several disjoint subfamilies of different dimensions. In the example \eqref{equ:DecayChannelExample3}, the stratum ${\cal F}_2$ associated with $h^0 = 2$ decomposes as ${\cal F}_2 = {\cal F}_2^{(a)} \cup {\cal F}_2^{(s)} \cup {\cal F}_2^{(b)}$ with
\begin{align}
\begin{split}
	{\cal F}_2^{(a)} & = \{ {\bf c} \, | \, c_{11} =  c_{12} = 0 \, , c_1 \neq 0, c_2 \neq 0, c_3 \neq 0 \} \, ,\\
	{\cal F}_2^{(b)} & = \{ {\bf c} \, | \, c_1 = c_2 = c_3 = 0, c_{11} \neq 0 \neq c_{12} \} \, , \\
	{\cal F}_2^{(s)} & = \{ {\bf c} \, | \, c_3 = 0 = c_2 c_{11} - c_1 c_{12} \, , c_1 \neq 0 \neq c_2 \, , c_{11} \neq 0 \neq c_{12} \} \, .
\end{split}
\end{align}
It is easy to see that each of these components also satisfies the axioms for strata (since they satisfy ${\cal F}_2^{(x)} \cap \overline{\cal F}_2^{(y)} = \emptyset$ for $x \neq y$).
Furthermore, their closure contains the common stratum ${\cal F}_3 = \{ {\bf c} \, | \, c_1 = ... = c_{12} =0 \}$ of higher codimension with $h^0 = 3$, as can be seen from the arrows connecting the three subfamilies of the stratum $\mathcal{F}_2$ to $\mathcal{F}_3$ in~\eqref{equ:DecayChannelExample3}.

In general, a stratification diagram can be roughly divided into three regions.
At low values of $h^0$, jumps typically occur for divisor alignment, i.e., are allowed by Brill--Noether theory on a smooth curve.
To get to high $h^0$, i.e., many vector-like pairs, the curve typically needs to factorize into many components.
In the middle regime, we can have a mixture, meaning in particular that a jump occurs due to divisor alignment on a split component.

\begin{figure}[tb]
\centering{
\begin{tikzpicture}[baseline=(current bounding box.center)]
    
    \def\w{2.4};
    \def\h{1.5};
    \def\t{1.5};
    
    \node (A) at (0.75*\w,0) [draw] {$\textcolor{blue}{C^0}$};
    
    \node (B1) at (0.75*\w,-\h) [draw] {$\textcolor{blue}{C^1_1}$};
    
    \node (C1) at (0.75*\w,-2*\h) [draw] {$\textcolor{blue}{C^2_1}$};
    
    \node (D1) at (0.75*\w,-3*\h) [draw] {$\textcolor{red}{A_3} \cup \textcolor{blue}{D_1}$};
    
    \node (E1) at (0*\w,-4*\h) [draw] {$\textcolor{red}{A_3} \cup \textcolor{blue}{D_2}$};
    \node (E2) at (1.5*\w,-4*\h) [draw] {$\textcolor{red}{A_3 \cup A_4 \cup D_3}$};
    
    \node (F1) at (0*\w,-5*\h) [draw] {$\textcolor{red}{A_3^{(2)} \cup D_4}$};
    
    \node (G2) at (0*\w,-6*\h) [draw] {$\textcolor{red}{A_3^{(3)} \cup A_5 \cup D_5}$};
    \node (G1) at (1.5*\w,-6*\h) [draw] {$\textcolor{red}{A_3^{(2)} \cup A_4 \cup D_6}$};
    
    \node (H1) at (0.75*\w,-7*\h) [draw] {$\textcolor{red}{A_3^{(3)} \cup A_4 \cup A_5 \cup D_7}$};
    
    \node (L1) at (-\t*\w,0) {$h^0 = 0$};
    \node (L2) at (-\t*\w,-\h) {$h^0 = 1$};
    \node (L3) at (-\t*\w,-2*\h) {$h^0 = 2$};
    \node (L4) at (-\t*\w,-3*\h) {$h^0 = 3$};
    \node (L5) at (-\t*\w,-4*\h) {$h^0 = 4$};
    \node (L6) at (-\t*\w,-5*\h) {$h^0 = 5$};
    \node (L7) at (-\t*\w,-6*\h) {$h^0 = 6$};
    \node (L8) at (-\t*\w,-7*\h) {$h^0 = 7$};
    
    \draw[-latex,thick] (A) --node[left] {} (B1);
    \draw[-latex,thick] (B1) --node[left] {} (C1);
    \draw[-latex,thick] (C1) --node[left] {} (D1);
    \draw[-latex,thick] (D1) --node[left] {} (E1);
    \draw[-latex,thick] (D1) --node[left] {} (E2);
    \draw[-latex,thick] (E1) --node[left] {} (F1);
    \draw[-latex,thick] (E2) --node[left] {} (G1);
    \draw[-latex,thick] (F1) --node[left] {} (G1);
    \draw[-latex,thick] (F1) --node[left] {} (G2);
    \draw[-latex,thick] (G1) --node[left] {} (H1);
    \draw[-latex,thick] (G2) --node[left] {} (H1);
    
    \draw[dashed] (-2.0*\w,-2.5*\h)--(2.5*\w,-2.5*\h);
    \draw[dashed] (-2.0*\w,-4.5*\h) to [out=0,in=180] (0,-4.5*\h) to [out=0,in=180] (\w,-3.5*\h) to [out = 0, in = 180] (2.5*\w,-3.5*\h);
    
    \draw [decorate,decoration={brace,amplitude=10pt},xshift=-0.5cm,yshift=0pt] (-2.0*\w,-2.5*\h) -- (-2.0*\w,0.5*\h) node [black,midway,xshift=-0.6cm,rotate=90] {BN-jumps};
    \draw [decorate,decoration={brace,amplitude=10pt},xshift=-0.5cm,yshift=0pt] (-2.0*\w,-4.5*\h) -- (-2.0*\w,-2.5*\h) node [black,midway,xshift=-1.1cm,rotate=90] {BN-jumps \&};
    \node (mark) at (-2.0*\w,-3.5*\h) [xshift = -1.1cm, rotate = 90] {curve splittings};
    
    \draw [decorate,decoration={brace,amplitude=10pt},xshift=-0.5cm,yshift=0pt] (-2.0*\w,-7.5*\h) -- (-2.0*\w,-4.5*\h) node [black,midway,xshift=-0.6cm,rotate=90] {curve splittings};
    
\end{tikzpicture}}
\caption{The stratification diagram for $D_C = ( 5 ; -1, -1, -2 )$, $D_L = ( 1; 1, -4, 1 )$.}
\label{fig:BigDecayChannel}
\end{figure}
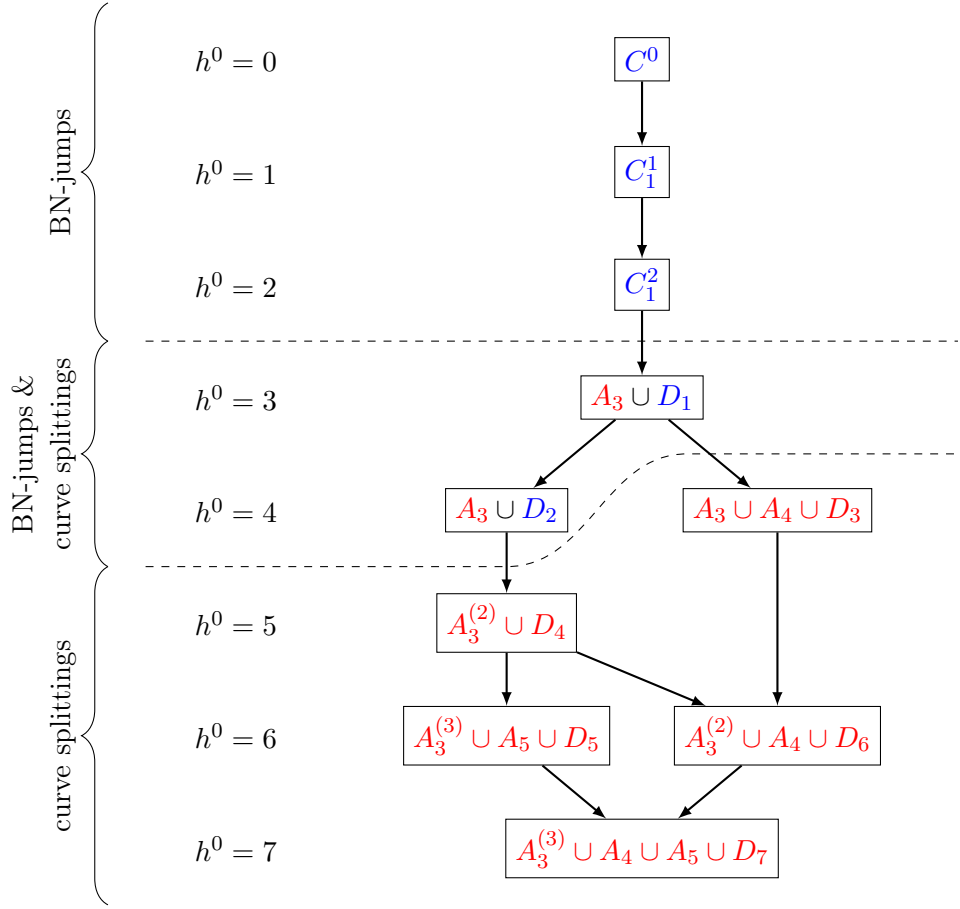

To illustrate such a ``typical'' case, consider
\begin{align}
D_C = ( 5 ; -1, -1, -2 ) \, , \qquad D_L = ( 1; 1, -4, 1 ) \, .
\end{align}
This genus $g = 5$ curve is given by $C( \mathbf{c} ) = V( P ( \mathbf{c} ) )$ with
\begin{align}
\begin{split}
P &:= c_1 x_1^3 x_2^4 x_3^2 x_4^2
+ c_2 x_1^2 x_2^4 x_3 x_4^3 x_6
+ c_3 x_1 x_2^4 x_4^4 x_6^2
+ c_4 x_1^3 x_2^3 x_3^3 x_4 x_5
+ c_5 x_1^2 x_2^3 x_3^2 x_4^2 x_5 x_6 \\
& \qquad + c_6 x_1 x_2^3 x_3 x_4^3 x_5 x_6^2
+ c_7 x_2^3 x_4^4 x_5 x_6^3
+ c_8 x_1^3 x_2^2 x_3^4 x_5^2
+ c_9 x_1^2 x_2^2 x_3^3 x_4 x_5^2 x_6 \\
& \qquad + c_{10} x_1 x_2^2 x_3^2 x_4^2 x_5^2 x_6^2
+ c_{11} x_2^2 x_3 x_4^3 x_5^2 x_6^3
+ c_{12} x_1^2 x_2 x_3^4 x_5^3 x_6
+ c_{13} x_1 x_2 x_3^3 x_4 x_5^3 x_6^2 \\
& \qquad + c_{14} x_2 x_3^2 x_4^2 x_5^3 x_6^3
+ c_{15} x_1 x_3^4 x_5^4 x_6^2
+ c_{16} x_3^3 x_4 x_5^4 x_6^3 \, .
\end{split}
\end{align}
From Brill--Noether theory, we then find
\begin{align}
\begin{tabular}{cc|ccc|ccc}
\toprule
curve & $g$ & $\mathcal{L}$ & $\chi$ & $d$ & \multicolumn{3}{c}{BN-theory} \\
\midrule
\multirow{4}{*}{$C = V( P )$} & \multirow{4}{*}{5} & \multirow{4}{*}{$\left. \mathcal{O}_{dP3}( D_L ) \right|_{C}$} & \multirow{4}{*}{0} & \multirow{4}{*}{4} & $h^0$ & $h^1$ & $\rho$ \\
                  &    &                                                                                  &    &   & 0 & 0 & 5 \\
                  &    &                                                                                  &    &   & 1 & 1 & 4 \\
                  &    &                                                                                  &    &   & 2 & 2 & 1 \\
\bottomrule
\end{tabular}
\end{align}
The stratification of curve geometries follows from the long exact sequence
\begin{equation}
\begin{tikzpicture}[scale=2, baseline=(current  bounding  box.center)]
          \matrix(m)[matrix of math nodes,column sep=15pt,row sep=15pt]{
                0 & 0
                  & 0
                  & H^0 \left( C ( \mathbf{c} ), \mathcal{L} ( \mathbf{c} ) \right) \\
                  & H^1 \left( dP_3, D_L - D_C \right) \cong \mathbb{C}^7
                  & H^1 \left( dP_3, D_L \right) \cong \mathbb{C}^7
                  & H^1 \left( C ( \mathbf{c} ), \mathcal{L} ( \mathbf{c} ) \right) \\
                  & 0
                  & 0
                  & 0 & 0 \\
                     };
               \draw[->,font=\scriptsize,every node/.style={above},rounded corners]
                     (m-1-1) edge (m-1-2)
                     (m-1-2) edge (m-1-3)
                     (m-1-3) edge (m-1-4)
                     (m-1-4.east) --+(5pt,0)|-+(0,-7.5pt)-|([xshift=-5pt]m-2-2.west)--(m-2-2.west);
                \draw[->,font=\scriptsize,every node/.style={above},rounded corners]
                     (m-3-2) edge (m-3-3)
                     (m-3-3) edge (m-3-4)
                     (m-3-4) edge (m-3-5);
                \draw[->,font=\scriptsize] (m-2-2) edge node [above] {$\varphi$} (m-2-3);
                \draw[->,font=\scriptsize] (m-2-3) edge node [above] {} (m-2-4);
                \draw[->,font=\scriptsize,every node/.style={above},rounded corners] (m-2-4.east) --+(5pt,0)|-+(0,-7.5pt)-|([xshift=-5pt]m-3-2.west)-- (m-3-2.west);
\end{tikzpicture}
\end{equation}
Consequently $h^0 \left( C( \mathbf{c} ), \mathcal{L}( \mathbf{c} ) \right) = 7 - \mathrm{rk} ( M_\varphi )$ and we find
\begin{align}
M_\varphi =
\left(
\begin{array}{ccccccc}
 c_{15} & c_{11} & c_{7} & 0 & 0 & 0 & 0 \\
 0 & c_{10} & c_{6} & c_{3} & c_{11} & c_{7} & 0 \\
 c_{12} & c_{6} & c_{3} & 0 & c_{7} & 0 & 0 \\
 0 & c_{5} & c_{2} & 0 & c_{6} & c_{3} & c_{7} \\
 c_{8} & c_{2} & 0 & 0 & c_{3} & 0 & 0 \\
 0 & c_{14} & c_{11} & c_{7} & 0 & 0 & 0 \\
 0 & c_{1} & 0 & 0 & c_{2} & 0 & c_{3} \\
\end{array}
\right) \, .
\end{align}
We list the curve strata in \cref{tab:CurveStrataBigDiagram} and display the corresponding stratification diagram in \cref{fig:BigDecayChannel}.

Of particular interest is the transition $A_3 \cup D_1 \to A_3 \cup D_2$. The former curve admits $3$, the latter $4$ sections. This change in the number of sections is due to a Brill--Noether jump on the curve components $D_i$:
\begin{center}
\begin{tabular}{c|ccc|ccc}
\toprule
curve & class & genus & $d$ & $h^0$ & $h^1$ & $\rho$ \\
\midrule
\multirow{2}{*}{$D_i$} & \multirow{2}{*}{$(5,-1,-2,-2)$} & \multirow{2}{*}{4} & \multirow{2}{*}{0} & 0 & 3 & 4 \\
      &                &   &   & 1 & 4 & 0 \\
\bottomrule
\end{tabular}
\end{center}
Hence, provided that the line bundle divisor is chosen such that $K_{D_i} - \left. D_L \right|_{D_i}$ is effective, we find an additional section on $D_i$, due to a Brill--Noether effect. More explicitly, in the case at hand this condition states that the line bundle divisor is linearly equivalent to the trivial divisor, i.e. $\left. D_L \right|_{D_i} \sim \emptyset$. This condition is satisfied on $D_2$ but not on $D_1$. For this reason we find one additional section on $A_3 \cup D_2$.

\begin{table}[tb]
\centering{
\begin{tabular}{ccc}
\toprule
$\mathrm{rk} ( M_\varphi )$ & explicit condition & curve splitting \\
\midrule
7 & $\mathrm{det} ( M_\varphi ) \neq 0$ & $\textcolor{blue}{C^0}$ \\
6 & $\mathrm{det} ( M_\varphi ) = 0$ & $\textcolor{blue}{C^1_1}$ \\
\midrule
\multirow{2}{*}{5} & $c_3 c_7 c_{12} = c_{15} c_3^2 + c_8 c_7^2$, \quad $c_{11} c_3^2 = c_3 c_6 c_7 - c_2 c_7^2$ & \multirow{2}{*}{$\textcolor{blue}{C^2_1}$} \\
              & $c_1 c_7^3 + c_{10} c_3^2 c_7 = c_{14} c_3^3 + c_3 c_5 c_7^2$ \\
\midrule
4 & $c_{3}=c_{7}=0$ & $A_3 \cup \textcolor{blue}{D_1}$ \\
\midrule
\multirow{2}{*}{3} & $c_3 = c_7 = 0$ \quad $c_{11} c_8 = c_{15} c_2$ \quad $c_{11} c_{12} = c_{15} c_6$ & \multirow{2}{*}{$A_3 \cup \textcolor{blue}{D_2}$} \\
                & $c_{11} c_2 c_5 = c_{14} c_2^2 + c_1 c_{11} c_6$ \quad $c_{10} c_{11} c_2 = c_1 c_{11}^2 + c_{14} c_2 c_6$ \\
\midrule
3 & $c_3=c_7=c_{8}=c_{12}=c_{15}=0$ & $A_3\cup A_4 \cup D_3$ \\
2 & $c_2 = c_3 = c_6 = c_7 = c_{11} = 0$ & $A_3^{(2)} \cup D_4$ \\
1 & $c_1=c_2=c_{3}=c_5=c_{6}=c_7=c_{10}=c_{11}=c_{14}=0$ & $A^{(3)}_3 \cup A_5\cup D_5$ \\
1 & $c_2=c_{3}=c_{6}=c_7=c_8=c_{11}=c_{12}=c_{15}=0$ & $A^{(2)}_3 \cup A_4\cup D_6$ \\
\midrule
\multirow{2}{*}{0} & $c_1=c_2=c_{3}=c_5=c_{6}=c_7=0$ & \multirow{2}{*}{$A^{(3)}_3 \cup A_4 \cup A_5 \cup D_7$} \\
                   & $c_8=c_{10}=c_{11}=c_{12}=c_{14}=c_{15}=0$ \\
\bottomrule
\end{tabular}}
\caption{The curve strata for $D_C = ( 5 ; -1, -1, -2 )$ and $D_L = ( 1; 1, -4, 1 )$.}
\label{tab:CurveStrataBigDiagram}
\end{table}

\section{Local to global section counting} \label{sec:LocalToGlobalSectionCounting}

In this section, we provide an in-depth analysis of the procedure of gluing local sections on reducible curves. As a result, we can place a lower bound on the number of global sections. We find sufficient topological conditions for a jump of $h^0$ to occur. This further allows us to formulate an algorithm to estimate the possible numbers of vector-like pairs on the moduli space of F-theory compactifications.

\subsection{Gluing local sections to global sections} \label{subsec:Gluing}

\subsubsection{Trivial boundary conditions} \label{subsubsec:TrivialBCExample}

Let us start by looking at a simple example. To this end, we go back to the geometry discussed in \cref{subsubsec:GapExample}, i.e.
\begin{align}
D_C = ( 3 ; -1, -1, -1 ) \, , \qquad D_L = ( 1; -1, -3, -1 ) \, .
\end{align}
Recall that in this case, $C( \mathbf{c} ) = V( P( \mathbf{c} ) )$ is a genus $1$ curve defined by
\begin{align}
\begin{split}
P( \mathbf{c} ) &= c_1 x_1^2 x_2^2 x_3 x_4 + c_2 x_1^2 x_2 x_3^2 x_5 + c_3 x_1 x_2^2 x_4^2 x_6 + c_4 x_1 x_2 x_3 x_4 x_5 x_6 \\
  &\qquad \qquad + c_5 x_1 x_3^2 x_5^2 x_6 + c_6 x_2 x_4^2 x_5 x_6^2 + c_7 x_3 x_4 x_5^2 x_6^2 \, .
\end{split}
\end{align}
We found that for $c_1 = c_3 = c_4 = c_6 = c_7 = 0$ we have 3 global sections. Furthermore, we have already seen that for this choice of parameters, the curve has 4 components
\begin{align}
C ( \mathbf{c} ) = E_6 \cup E_4 \cup E_2^{(2)} \cup A \, .
\end{align}
These components have the following properties:
\begin{align}
\begin{tabular}{c|cccc|c}
\toprule
curve component & equation & class & $g$ & $d$ & $h^0( C_i, D_L )$ \\
\midrule
$A$ & $V( c_2 x_1 x_2 + c_5 x_5 x_6 )$ & $(1;0,-1,0)$ & 0 & $-2$ & 0 \\
$E_4$ & $V( x_1 )$ & $(1;-1,-1,0)$ & 0 & $-3$ & 0 \\
$E_6$ & $V( x_5 )$ & $(1;0,-1,-1)$ & 0 & $-3$ & 0 \\
$E_2^{(2)}$ & $V( x_3^2 )$ & $(0;0,2,0)$ & $-2$ & 6 & 9 \\
\bottomrule
\end{tabular}
\end{align}
In the last column we give the number of sections of the restriction of the bundle $\mathcal{O}_{dP_3}( D_L )$ to these curve components. We will refer to these sections in the following as the \emph{local sections}.

\begin{figure}[tb]
\begin{center}
\begin{tikzpicture}

\def\r{1.8};

\draw[blue, thick] (0,-2*\r) circle[radius=\r];
\draw[blue, thick] (-2*\r,0) circle[radius=\r];
\draw[blue, thick] (2*\r,0) circle[radius=\r];
\draw[red, thick] (0,0) circle[radius=\r];

\fill (-\r,0) circle[radius=2.5pt];
\fill (0,-\r) circle[radius=2.5pt];
\fill (\r,0) circle[radius=2.5pt];

\node at (-\r,0) [right] {$2 \times$};
\node at (0,-\r) [above] {$2 \times$};
\node at (\r,0) [right] {$2 \times$};

\node[red] at (0,1.25*\r) {$V( x_3^2 )$};
\node[blue] at (-2*\r,1.25*\r) {$V( x_1 )$};
\node[blue] at (2*\r,1.25*\r) {$A$};
\node[blue] at (1.4*\r,-2*\r) {$V( x_5 )$};

\node at (-2*\r,0.25*\r) {$D_L \cdot V( x_1 ) = -3$};
\node at (-2*\r,-0.25*\r) {$h^0 = 0$};

\node at (2*\r,0.25*\r) {$D_L \cdot A = -2$};
\node at (2*\r,-0.25*\r) {$h^0 = 0$};

\node at (0,0.25*\r) {$D_L \cdot V( x_3^2 ) = 6$};
\node at (0,-0.25*\r) {$h^0 = 9$};

\node at (0,-1.75*\r) {$D_L \cdot V( x_5 ) = -3$};
\node at (0,-2.25*\r) {$h^0 = 0$};

\end{tikzpicture}
\end{center}
\caption{The 9 local sections on $A$ lead to $9 - 3 \times 2 = 3$ global sections.}
\label{Fig:LocalToGlobalCounting}
\end{figure}
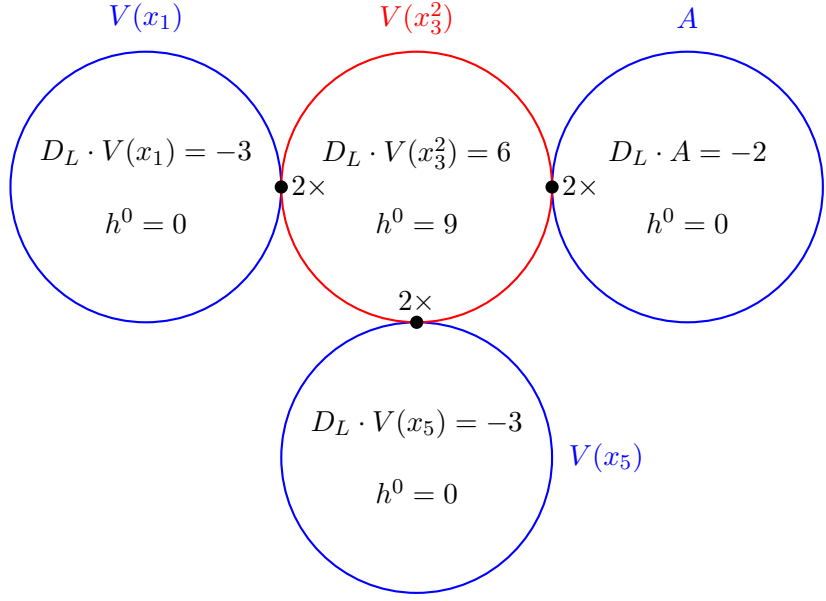

We display this geometry in \cref{Fig:LocalToGlobalCounting}. Our task is to glue the \emph{local sections} to global sections on the curve $C = E_6 \cup E_4 \cup E_2^{(2)} \cup A\,$. To this end, we work out the sections explicitly and then subject them to boundary conditions at the intersection points of the different curve components. 

For the components $A$, $E_4$ and $E_6$ we already know that the only allowed local section vanishes identically. On $E_2^{(2)}$ however, the situation is a bit more involved since $E_2^{(2)}$ is a non-reduced curve. As a set, $E_2^{(2)}$ is the locus $V( x_3 )$. Using the scaling relations of $dP_3$, we can then set $x_2 = x_4 = x_6 = 1$ and thereby identify $(x_1, x_5)$ as coordinates of $E_2^{(2)}$. Note, however, that since $E_2^{(2)}$ is a non-reduced curve, the polynomial $x_3$ is a non-trivial function on this curve component. These observations allow us to conclude
\begin{align}
H^0 \left( E_2^{(2)}, \left. \mathcal{O}_{dP_3}( D_L ) \right|_{E_2^{(2)}} \right) \cong P_3( x_1, x_5 ) \oplus x_3 \cdot P_4 ( x_1, x_5 ) \, ,
\end{align}
where $P_i( x_1, x_5 )$ is the space of polynomials of degree $i$ in $x_1$ and $x_5$. Upon homogenization with $x_2$, $x_4$, $x_6$, we can then write
\begin{align}
\begin{split}
H^0 \left( E_2^{(2)}, \left. \mathcal{O}_{dP_3}( D_L ) \right|_{E_2^{(2)}} \right) &\cong \mathrm{Span}_{\mathbb{C}} \left\{ \frac{x_5^3}{x_2^3 x_4^2}, \frac{x_1 x_5^2}{x_2^2 x_4^2 x_6}, \frac{x_1^2 x_5}{x_2 x_4^2 x_6^2}, \frac{x_1^3}{x_4^2 x_6^3} \right\} \\
&\qquad \oplus x_3 \cdot \mathrm{Span}_{\mathbb{C}} \left\{ \frac{x_5^4}{x_2^4 x_4^3}, \frac{x_5^3 x_1}{x_2^3 x_4^3 x_6}, \frac{x_1^2 x_5^2}{x_2^2 x_4^3 x_6^2}, \frac{x_1^3 x_5}{x_2 x_4^3 x_6^3}, \frac{x_1^4}{x_4^3 x_6^4} \right\} \, .
\end{split}
\end{align}
From this, we learn that the only sections on $V( x_3^2 )$, which vanish at $V( x_1 )$, $V( x_5 )$ and $V( c_2 x_1 x_2 + c_5 x_5 x_6 )$, are linear combinations of the following three sections:
\begin{align}
s_1 &= c_5 \frac{x_1 x_5^2}{x_2^2 x_4^2 x_6} + c_2 \frac{x_1^2 x_5}{x_2 x_4^2 x_6^2} = \frac{x_1 x_5 \left( c_2 x_1 x_2 + c_5 x_5 x_6 \right)}{x_2^2 x_4^2 x_6^2} \, , \\
s_2 &= c_5 \frac{x_5^3 x_1}{x_2^3 x_4^3 x_6} + c_2 \frac{x_1^2 x_5^2}{x_2^2 x_4^3 x_6^2} = \frac{x_1 x_5^2 \left( c_2 x_1 x_2 + c_5 x_5 x_6 \right)}{x_2^3 x_4^3 x_6^2} \, , \\
s_3 &= c_5 \frac{x_1^2 x_5^2}{x_2^2 x_4^3 x_6^2} + c_2 \frac{x_1^3 x_5}{x_2 x_4^3 x_6^3} = \frac{x_1^2 x_5 \left( c_2 x_1 x_2 + c_5 x_5 x_6 \right)}{x_2^2 x_4^3 x_6^3} \, .
\end{align}
Consequently, by extending these sections by zero outside of $V( x_3^2 )$, we obtain 3 global sections.

\subsubsection{Non-trivial boundary conditions} \label{subsubsec:NonTrivialBC}

\begin{figure}[tb]
\begin{center}
\begin{tikzpicture}

\def\o{1.5};

\draw[blue] (-3*\o,0) circle (\o);
\draw[blue] (3*\o,0) circle (\o);

\draw[thick,red] (-2*\o,0) to [out=90,in=90] (0,0.5*\o)
                           to [out=-90,in=90] (-0.5*\o,0)
                           to [out=-90,in=90] (0,-0.5*\o)
                           to [out=-90,in=-90] (-2*\o,0);

\draw[thick,red] (2*\o,0) to [out=90,in=90] (0,0.5*\o)
                           to [out=-90,in=90] (0.5*\o,0)
                           to [out=-90,in=90] (0,-0.5*\o)
                           to [out=-90,in=-90] (2*\o,0);

\fill (0,0.5*\o) circle[radius=2.5pt];
\fill (0,-0.5*\o) circle[radius=2.5pt];
\fill (-2*\o,0) circle[radius=2.5pt];
\fill (2*\o,0) circle[radius=2.5pt];

\node at (-2*\o,0) [left] {$2 \times$};
\node at (0,0.5*\o) [right] {$2 \times$};
\node at (0,-0.5*\o) [right] {$2 \times$};
\node at (2*\o,0) [left] {$2 \times$};

\node at (-3*\o,-1.25*\o) {$\textcolor{blue}{E_3}$};
\node at (-1*\o,-1.25*\o) {$\textcolor{red}{E_5^{(2)}}$};
\node at (1*\o,-1.25*\o) {$\textcolor{red}{E_1^{(2)}}$};
\node at (3*\o,-1.25*\o) {$\textcolor{blue}{E_4}$};

\node at (-3*\o,0) {$h^0 = 2$};
\node at (-1*\o,0) {$h^0 = 1$};
\node at (1*\o,0) {$h^0 = 5$};
\node at (3*\o,0) {$h^0 = 0$};

\end{tikzpicture}
\end{center}
\caption{A non-trivial gluing example which gives no global sections.}
\label{Fig:NonTrivialGluing}
\end{figure}
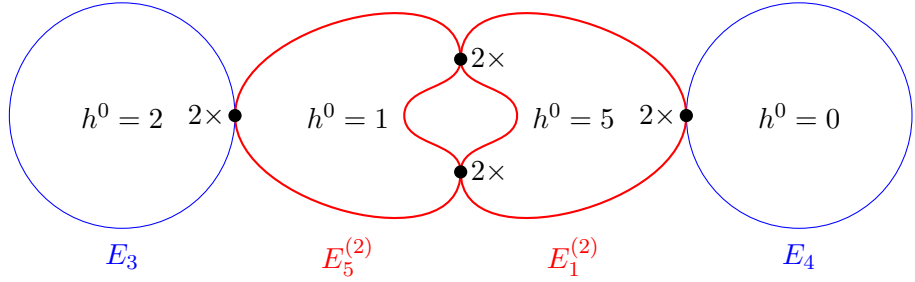

Let us consider $D_C = (3,-1,-1,-1)$ and $D_L = (5;-4,-4,3)$. We pick special values for the parameters such that $C = V( x_1 x_2^2 x_4^2 x_6 )$. The curve thus factors into four components, as displayed in \cref{Fig:NonTrivialGluing}. These components have the following properties:
\begin{align}
\begin{array}{cccccccc}
\toprule
\text{curve} & \text{class} & \text{eqn.} & d & g & h^0 & \text{basis of sections} \\
\midrule
E_3 & (0;0,0,1) & V( x_6 ) & 1 & 0 & 2 & \frac{x_4}{x_3^3}, \frac{x_5}{x_2 x_3^2}\\
E_5^{(2)} & (2;-2,0,-2) & V( x_4^2 ) & -2 & -2 & 1 & \frac{x_4}{x_3^3} \\
E_1^{(2)} & (0;2,0,0) & V( x_2^2 ) & 2 & -2 & 5 & \frac{x_1}{x_3^2 x_6}, \frac{x_4}{x_3^3}, \frac{x_2 x_4^2}{x_3^4 x_5}, \frac{x_2 x_1 x_4}{x_3^3 x_5 x_6}, \frac{x_2 x_1^2}{x_3^2 x_5 x_6^2} \\
E_4 & (1;-1,-1,0) & V( x_1 ) & -3 & 0 & 0 & 0 \\
\bottomrule
\end{array}
\end{align}
We have also listed bases for the sections on the individual curve components. By starting in $E_3$, we see that there is a unique section which extends to $E_5^{(2)}$ and then to $E_1^{(2)}$ -- this section is $\frac{x_4}{x_3^3}$. However, this section fails to vanish on $V( x_1 )$. Consequently, this geometry only admits the global section which is identically zero.

\subsubsection{From trivial to non-trivial boundary conditions} \label{subsubsec:NonTrivialBC2}

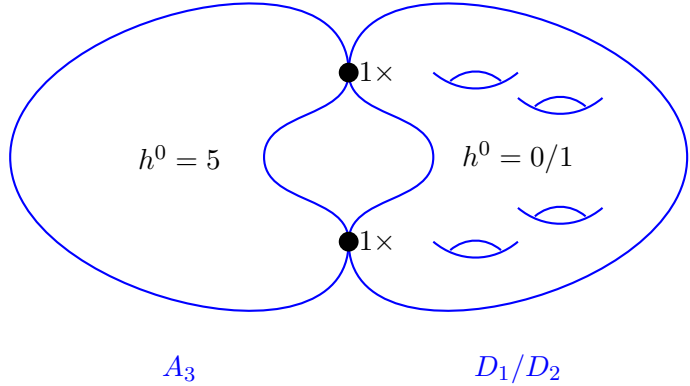
\begin{figure}[tb]
\begin{center}
\begin{tikzpicture}[scale = 1.5]

\def\o{1.5};

\draw[thick,blue] (-2*\o,0) to [out=90,in=90] (0,0.5*\o)
                           to [out=-90,in=90] (-0.5*\o,0)
                           to [out=-90,in=90] (0,-0.5*\o)
                           to [out=-90,in=-90] (-2*\o,0);

\draw[thick,blue] (2*\o,0) to [out=90,in=90] (0,0.5*\o)
                           to [out=-90,in=90] (0.5*\o,0)
                           to [out=-90,in=90] (0,-0.5*\o)
                           to [out=-90,in=-90] (2*\o,0);

\fill (0,0.5*\o) circle[radius=2.5pt];
\fill (0,-0.5*\o) circle[radius=2.5pt];

\node at (0,0.5*\o) [right] {$1 \times$};
\node at (0,-0.5*\o) [right] {$1 \times$};

\node at (-1*\o,-1.25*\o) {$\textcolor{blue}{A_3}$};
\node at (1*\o,-1.25*\o) {$\textcolor{blue}{D_1/D_2}$};

\node at (-1*\o,0) {$h^0 = 5$};
\node at (+1*\o,0) {$h^0 = 0/1$};

\draw[thick,blue] (0.5*\o,-0.5*\o) to [out=-40,in=-140] (1.0*\o,-0.5*\o);
\draw[thick,blue] (0.6*\o,-0.55*\o) to [out=40,in=140] (0.9*\o,-0.55*\o);

\draw[thick,blue] (1.0*\o,-0.3*\o) to [out=-40,in=-140] (1.5*\o,-0.3*\o);
\draw[thick,blue] (1.1*\o,-0.35*\o) to [out=40,in=140] (1.4*\o,-0.35*\o);

\draw[thick,blue] (0.5*\o,0.5*\o) to [out=-40,in=-140] (1.0*\o,0.5*\o);
\draw[thick,blue] (0.6*\o,0.45*\o) to [out=40,in=140] (0.9*\o,0.45*\o);

\draw[thick,blue] (1.0*\o,0.35*\o) to [out=-40,in=-140] (1.5*\o,0.35*\o);
\draw[thick,blue] (1.1*\o,0.3*\o) to [out=40,in=140] (1.4*\o,0.3*\o);

\end{tikzpicture}
\end{center}
\caption{A Brill--Noether jump $D_1 \to D_2$ generates one additional global section.}
\label{Fig:NonTrivialGluing2}
\end{figure}

We have seen an interesting geometric transition when we discussed $D_C = (5,-1,-1,-2)$ and $D_L = (1;1,-4,1)$ in \cref{subsubsec:JumpSpecialDivisorOnCurveComponent}. Namely, the transition
\begin{align}
A_3 \cup D_1 \to A_3 \cup D_2
\end{align}
enforces a Brill--Noether jump on $D_2$. Whilst $D_1$ only supports the trivial section, $D_2$ supports a one-dimensional space of non-trivial sections. As a consequence, $A_3 \cup D_2$ admits one additional section as compared to $A_3 \cup D_1$. Let us investigate this finding in more detail. We depict this geometry in \cref{Fig:NonTrivialGluing2} and recall the following information:
\begin{center}
\begin{tabular}{ccccc}
\toprule
curve & class & degree & genus & $h^0$ \\
\midrule
$A_3$ & $(0;0,1,0)$ & 4 & 0 & 5 \\
$D_1$ & $(5,-1,-2,-2)$ & 0 & 4 & 0 \\
$D_2$ & $(5,-1,-2,-2)$ & 0 & 4 & 1 \\
\bottomrule
\end{tabular}
\end{center}
To simplify our analysis, let us work with a particular class of curves $D_1$ and $D_2$, for which the transition $D_1 \to D_2$ is particularly simple:
\begin{align}
\begin{split}
D_1 &= V \left( c_{12} x_1^2 x_2 x_3^3 x_5^3 x_6 +c_{13} x_1 x_2 x_3^2 x_4 x_5^3 x_6^2 +c_{16} x_3^2 x_4 x_5^4 x_6^3 +c_{4} x_1^3 x_2^3 x_3^2 x_4 x_5 \right. \\
    &\qquad \quad +c_{9} x_1^2 x_2^2 x_3^2 x_4 x_5^2 x_6 +x_1^3 x_2^4 x_3 x_4^2 +x_1^3 x_2^2 x_3^3 x_5^2 +x_1^2 x_2^4 x_4^3x_6 \\
    &\qquad \quad -x_1^2 x_2^3 x_3 x_4^2 x_5 x_6 -x_1 x_2^2 x_3 x_4^2 x_5^2 x_6^2 -x_1 x_3^3 x_5^4 x_6^2 -x_2^2 x_4^3 x_5^2 x_6^3 \\
    &\qquad \quad \left. +x_2 x_3 x_4^2 x_5^3 x_6^3 \right) \, , \\
D_2 &= \left. D_1 \right|_{c_{12} = 0} \, .
\end{split}
\end{align}

Next, we turn to the sections on $A_3 \cong \mathbb{P}^1$. We note that the homogeneous coordinates are $[x_1:x_5]$. Hence, the line bundle sections at hand are of the form ($\lambda = x_2 x_6^{-1}$):
\begin{align}
H^0 \left( A_3, \left. \mathcal{L} \right|_{A_3} \right) = \frac{1}{x_4^3} \cdot \mathrm{Span}_{\mathbb{C}} \left\{ x_1^4 \cdot \lambda^2,~x_1^3 x_5 \cdot \lambda, x_1^2 x_5^2,~x_1 x_5^3 \cdot \lambda^{-1},~x_5^4 \cdot \lambda^{-2} \right\} \, .
\end{align}
At $x_3 = 0$, we may set $x_2 = x_4 = x_6 = 1$ by the scaling relations of $dP_3$. In terms of these inhomogeneous coordinates, we find
\begin{align}
A_3 \cap D_i = V( x_3, x_1 - x_5 ) \cup V( x_3, x_1 + x_5 ) \, .
\end{align}
That all said, we can discuss the global sections on $A_3 \cup D_1$ and $A_3 \cup D_2$:
\begin{itemize}
 \item On $D_1$, the only supported section vanishes identically. Hence, we may only consider sections on $A_3$, which vanish at $A_3 \cap D_1$. It is not too hard to see that the space of 
      these sections is generated by
      \begin{align}
      s_1 = - x_1^4 + x_5^4 \, , \qquad s_2 = - x_1^3 x_5 + x_1 x_5^3 \, , \qquad s_3 = - x_1^4 + x_1^2 x_5^2 \, .
      \end{align}
 \item On $D_2$ however, the line bundle divisor is special. In fact, since it is a divisor of degree zero, this divisor must be the trivial divisor. Consequently, the sections on $D_2$ are the constant ones. It is not too hard to see that the sections on $A_3$, which have value $1$ at the intersection points $A_3 \cap D_2$, are generated by
      \begin{align}
      t_1 = x_1^4 \, , \qquad t_2 = t_1 + s_1 \, , \qquad t_3 = t_1 + s_2 \, , \qquad t_4 = t_1 + s_3 \, .
      \end{align}
\end{itemize}
This explains the one additional section on $A_3 \cap D_2$ as opposed to $A_3 \cap D_1$.

\subsubsection{Overcounting boundary conditions} \label{subsubsec:OvercountingBCs}

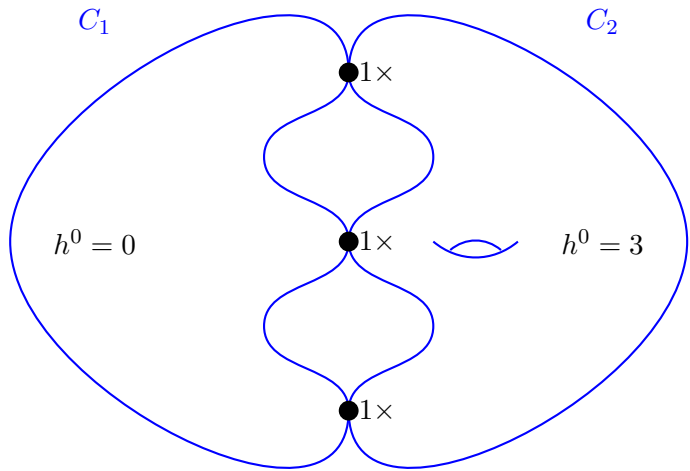
\begin{figure}[tb]
\centering{
\begin{tikzpicture}[scale = 1.5]

\def\o{1.5};

\draw[thick,blue] (-2*\o,-0.5*\o) to [out=90,in=90] (0,0.5*\o)
                           to [out=-90,in=90] (-0.5*\o,0)
                           to [out=-90,in=90] (0,-0.5*\o)
                           to [out=-90,in=90] (-0.5*\o,-\o)
                           to [out=-90,in=90] (0,-1.5*\o)
                           to [out=-90,in=-90] (-2*\o,-0.5*\o);

\draw[thick,blue] (2*\o,-0.5*\o) to [out=90,in=90] (0,0.5*\o)
                           to [out=-90,in=90] (0.5*\o,0)
                           to [out=-90,in=90] (0,-0.5*\o)
                           to [out=-90,in=90] (0.5*\o,-\o)
                           to [out=-90,in=90] (0,-1.5*\o)
                           to [out=-90,in=-90] (2*\o,-0.5*\o);

\fill (0,0.5*\o) circle[radius=2.5pt];
\fill (0,-0.5*\o) circle[radius=2.5pt];
\fill (0,-1.5*\o) circle[radius=2.5pt];

\node at (0,0.5*\o) [right] {$1 \times$};
\node at (0,-0.5*\o) [right] {$1 \times$};
\node at (0,-1.5*\o) [right] {$1 \times$};

\node at (-1.5*\o,0.8*\o) {$\textcolor{blue}{C_1}$};
\node at (1.5*\o,0.8*\o) {$\textcolor{blue}{C_2}$};

\node at (-1.5*\o,-0.5*\o) {$h^0 = 0$};
\node at (+1.5*\o,-0.5*\o) {$h^0 = 3$};

\draw[thick,blue] (0.5*\o,-0.5*\o) to [out=-40,in=-140] (1.0*\o,-0.5*\o);
\draw[thick,blue] (0.6*\o,-0.55*\o) to [out=40,in=140] (0.9*\o,-0.55*\o);

\end{tikzpicture}}
\caption{Naively, we expect $3 - 3 = 0$ global sections. However, one section on $C_2$ automatically vanishes at $C_1 \cap C_2$, leading to $h^0 \left( C_1 \cup C_2, \mathcal{L} \right) = 1$.}
\label{Fig:OverestimateBCs}
\end{figure}

As a final example, let us look at $D_C = (4;-1,-1,-1)$ and $D_L = (1,1,-3,0)$. Let us deform the curve $C$ such that it is given by
\begin{align}
P = x_1 \cdot Q \, , \qquad Q = x_1^2 x_2^2 x_3^3 x_5 + x_2^3 x_4^3 x_6^2 + x_3^3 x_5^3 x_6^2 \, .
\end{align}
We display this curve geometry in \cref{Fig:OverestimateBCs}. The two curve components have the following properties:
\begin{center}
\begin{tabular}{cccccc}
\toprule
component & equation & class & $g$ & $d$ & $h^0$ \\
\midrule
$C_1$ & $V( x_1 )$ & $(1;-1,-1,0)$ & 0 & $-1$ & 0 \\
$C_2$ & $V( Q )$ & $(3;0,0,-1)$ & 1 & 3 & 3 \\
\bottomrule
\end{tabular}
\end{center}
Up to canonical isomorphism (induced from the connection homomorphism), we find a basis of the sections on $C_2$ as
\begin{align}
\mathcal{B} = \left\{ \frac{1}{x_2 x_3^3 x_4^2 x_6}, \frac{x_5}{x_2^2 x_3^2 x_4^3 x_6}, \frac{x_1}{x_2 x_3^2 x_4^3 x_6^2} \right\} \, .
\end{align}
From this we can see that the third section automatically vanishes at the intersection $C_1 \cap C_2$, whilst the other two sections do not vanish there. Consequently, and in agreement with the computational results by \emph{gap}, we find $h^0 \left( C_1 \cup C_2, \mathcal{L} \right) = 1$.

Importantly, a naive guess cannot predict this number. In this case, we would have counted as follows: 3 sections on $C_2$ subject to vanishing conditions at the 3 intersection points $C_1 \cap C_2$ should leave us only with the trivial section. Hence, in this example, a naive counting fails. Such phenomena were originally studied more generally in \cite{Bacharach1886, Cayley_Arthur_On_1889} --- see also \cite{Eisenbud1996} for a more modern exposition of the material.

\subsection{Sufficient jump condition and algorithmic section estimate}

As demonstrated in the previous section, gluing local sections to global sections is a non-trivial task. The exact details depend, among other things, on the relative position of the line bundle divisor and the intersection points of the curve components: the results change when some of these intersection points coincide and when the bundle divisor is special on some curve components.

In the following, we will propose a counting mechanism with the following key properties:
\begin{itemize}
 \item It relies mostly on topological data.
 \item It provides a lower bound on the number of global sections.
\end{itemize}

Of course, such a simplified counting procedure will fail to predict intricate geometries as discussed in \cite{Bacharach1886, Cayley_Arthur_On_1889, Eisenbud1996}. Still, it has two distinct advantages. First, since it relies mostly on topological data, it is very fast. Given a curve $C$ and a line bundle $\mathcal{L}$ on $C$, we can apply the strategy to place a lower bound on $h^0 \left( C( \mathbf{c} ), \mathcal{L} ( \mathbf{c} ) \right)$ for many different choices of parameters $\mathbf{c}$ of $C$. The collection of these lower bounds can then serve as an estimate of the vector-like spectrum of $(C,\mathcal{L})$ over the parameter space. Note that obtaining such an estimate is unfeasible with existing exact algorithms, e.g., those implemented in \cite{SheafCohomologyOnToricVarieties}, since these algorithms require extensive computational resources and often take a long time to finish.
The second advantage results from the fact that our counting procedure systematically underestimates the actual number of global sections. Therefore, it allows us to formulate sufficient conditions for a jump in the vector-like spectrum to happen.

\subsubsection{Counting procedure} \label{subsubsec:ProposalForCounting}

Let us consider a curve $C$ with
\begin{align}
C = \bigcup_{i = 1}^{N}{C_i} \, ,
\end{align}
i.e., $C$ has $N$ components $C_i$. For our counting procedure to be as simple and reliable as possible, let us avoid setups of the type discussed in \cref{subsubsec:NonTrivialBC} and \cref{subsubsec:NonTrivialBC2}. Hence, let us consider a line bundle $\mathcal{L}$ on $C$ such that neighboring curve components do not support non-trivial sections simultaneously. 
Put different, we only consider setups where for all curve components $C_i$ the following holds true:
\begin{align}
h^0 \left( C_i, \left. \mathcal{L} \right|_{C_i} \right) > 0 \quad \Rightarrow \quad h^0 \left( C_j, \left. \mathcal{L} \right|_{C_j} \right) = 0 \quad \forall \; C_j \text{ with } C_i \cap C_j \neq \emptyset \, .
\end{align}
Let us denote by $b_i$ the number of intersection points of $C_i$ with the other curve components. Generically, we then impose $b_i$ conditions on the ``local'' sections in $H^0( C_i, \left. \mathcal{L} \right|_{C_i} )$. Consequently,
\begin{align}
n_i ( C_i ) = \begin{cases} h^0( C_i, \left. \mathcal{L} \right|_{C_i} ) - b_i &\qquad \text{if } h^0( C_i, \left. \mathcal{L} \right|_{C_i} ) \geq b_i \\ 0 &\qquad \text{else} \end{cases}
\end{align}
is a lower bound to the number of sections on $C_i$ which satisfy the gluing boundary conditions. The sum of these contributions over all curve components places a lower bound on $h^0( C, \mathcal{L} )$:
\begin{align}
   \sum_{i = 1}^{N}{n_i( C_i )} \leq h^0( C, \mathcal{L} ) \, .
\end{align}
We expect that equality holds in generic situations and that only fairly tuned geometries, in the spirit of \cite{Bacharach1886, Cayley_Arthur_On_1889, Eisenbud1996}, will lead to a proper inequality.

As simple demonstration, let us apply this procedure to the geometry discussed in \cref{subsubsec:TrivialBCExample}:
\cref{subsubsec:TrivialBCExample}:
\begin{center}
\begin{tabular}{cccc}
\toprule
component $C_i$ & $h^0( C_i, \left. \mathcal{L} \right|_{C_i} )$ & $b_i$ & $n_i$ \\
\midrule
$V( x_1 )$ & 0 & 2 & 0 \\
$V( x_3^2 )$ & 9 & 6 & 3 \\
$V( x_5 )$ & 0 & 2 & 0 \\
$A$ & 0 & 2 & 0 \\
\bottomrule
\end{tabular}
\end{center}
Indeed, $\sum_{i = 1}^{3}{n_i} = 3$ in agreement with our discussion in \cref{subsubsec:GapExample}. However, if we apply this counting to $A_3 \cup D_2$, as discussed in \cref{subsubsec:NonTrivialBC2}, then we find the inequality
\begin{align}
n_1 + n_2 = (5-2) + 0 = 3 < 4 = h^0( A_3 \cup D_2, \mathcal{L} ) \, .
\end{align}
This shows that, if we are interested in the exact number rather than a lower bound, we should restrict our counting procedure to curve geometries where neighboring curve components do not support non-trivial sections simultaneously. Furthermore, the geometry studied in \cref{subsubsec:OvercountingBCs} shows that even under this assumption, there are exceptions to this counting procedure. In this case, this can be attributed to a special alignment of the line bundle divisor and the intersection points, such that one of the sections automatically satisfies all of the boundary conditions.

\subsubsection{Accuracy on our database}

Let us now apply this counting procedure to our database \cite{Database} to obtain an estimate of how often the inequality is satisfied. To this end, we need to identify the number of local sections, which can be challenging for complicated curve geometries and could call for an application of, e.g., the exact methods implemented in \cite{SheafCohomologyOnToricVarieties}. However, given the vast number of curve components in our database, we find it more appealing to focus on those curves for which we can identify the number of local sections quicker. To this end, we focus on the following two types of curves:
\begin{itemize}
 \item Smooth curves: \\
      We consider the line bundle degree $d = \mathrm{deg} ( \left. \mathcal{L} \right|_{C_i} )$. Provided that $d < 0$, we know that $\left. \mathcal{L} \right|_{C_i}$ does not admit non-trivial sections. Conversely, if $d > 2 g( C_i ) - 2$, then it follows from application of the Kodaira vanishing theorem, that $h^0( C_i, \left. \mathcal{L} \right|_{C_i} ) = d - g + 1$. If none of these conditions is satisfied, we discard the curve for this test.
 \item Non-split curves: \\
      For these curves, we can simply read off the number of local sections from our database.
\end{itemize}
Based on these local section counts, we have then applied the counting procedure presented in \cref{subsubsec:ProposalForCounting}. Recall that a large number of curves in our database do neither consist of smooth curve components nor are non-split. Furthermore, recall that we subject the curve geometry to the condition that neighboring components do not support non-trivial sections simultaneously. Let us emphasize that the latter is a simplifying assumption to simplify our counting procedure. Whilst we leave extensions in this direction to future work, we can still apply our (restricted) counting procedure to roughly 60\% of the cases in our database. For these, we predict the correct number of global sections with an accuracy of more than 99\%, i.e.\ our counting procedure works remarkably well. We list the detailed results in \cref{subsubsec:Accuracy}.

\subsubsection{Sufficient conditions for jumps in cohomology}

These insights of gluing local sections to form global sections, imply sufficient conditions for jumps in cohomology. First, we have the following

\begin{lemma} \label{Lemma1}

Let $S$ be a smooth surface, $\mathcal{L} \in \mathrm{Pic}(S)$ a line bundle, and $\left| C \right|$ a linear system of curves on $S$. Consider a special member $C_1 \cup C_2$ such that the curves $C_1$, $C_2$ meeting transversely in $C_1 \cdot C_2 > 0$ distinct points. Let $N_1 = h^0( C_1, \left. \mathcal{L} \right|_{C_1} )$ and $N_2 = h^0( C_2, \left. \mathcal{L} \right|_{C_2} )$. Then
\begin{align}
h^0 \left( C_1 \cup C_2, \mathcal{L} \right) \geq N_1 + N_2 - C_1 \cdot C_2 \, .
\end{align}

\end{lemma}

\paragraph{Proof}

We consider the short exact sequence $0 \to \left. \mathcal{L} \right|_{C_1 \cup C_2} \to \left. \mathcal{L} \right|_{C_1 \sqcup C_2} \to \left. \mathcal{L} \right|_{C_1 \cap C_2} \to 0$. The associated long exact sequence in sheaf cohomology begins with
\begin{align}
0 \to h^0 \left( C_1 \cup C_2, \left. \mathcal{L} \right|_{C_1 \cup C_2} \right) \to h^0 \left( C_1 \sqcup C_2, \left. \mathcal{L} \right|_{C_1 \sqcup C_2} \right) \to h^0 \left( C_1 \cap C_2, \left. \mathcal{L} \right|_{C_1 \cap C_2} \right) \to \dots
\end{align}
Now, since $h^0( C_1 \sqcup C_2, \left. \mathcal{L} \right|_{C_1 \sqcup C_2} ) = N_1 + N_2$ and $h^0( C_1 \cap C_2, \left. \mathcal{L} \right|_{C_1 \cap C_2} ) = C_1 \cdot C_2$, the statement follows. \qed

We can use this result, together with the insights on gluing local sections to global sections, to derive the following

\begin{corollary} \label{Corollary1}

Let $S$ be a smooth surface, $\mathcal{L} \in \mathrm{Pic}(S)$ a line bundle, and $\left| C \right|$ a linear system of curves on $S$ with smooth general member $C$ and special member $C_1 \cup C_2$ where $C_1$, $C_2$ are smooth curves of genera $g_1$, $g_2$ meeting transversely in $C_1 \cdot C_2 > 0$ distinct points. We assume $h^1( C, \left. \mathcal{L} \right|_C ) = 0$, $\mathrm{deg} \left( \left. \mathcal{L} \right|_{C_2} \right) > 2 g_2 - 2$ and $\mathrm{deg} \left( \left. \mathcal{L} \right|_{C_1} \right) < \mathrm{min} \left\{ 0, g_1 -1 \right\}$. Then
\begin{align}
h^0 \left( C_1 \cup C_2, \mathcal{L} \right) - h^0 \left( C, \mathcal{L} \right) \geq g_1 - 1 - \mathrm{deg} \left( \left. \mathcal{L} \right|_{C_1} \right) \, .
\end{align}

\end{corollary}

\paragraph{Proof}
Since $\mathrm{deg} \left( \left. \mathcal{L} \right|_{C_1} \right) < 0$, there are no sections on $C_1$. Hence, from \cref{Lemma1} we obtain the inequality
\begin{align}
h^0 \left( C_1 \cup C_2, \mathcal{L} \right) \geq h^0 \left( C_2, \left. \mathcal{L} \right|_{C_2} \right) - C_1 \cdot C_2 \, .
\end{align}
Note that $g_C = g_1 + g_2 + C_1 \cdot C_2 - 1$. Consequently, since $\mathrm{deg} \left( \left. \mathcal{L} \right|_{C_2} \right) > 2 g_2 - 2$, we can write
\begin{align}
\begin{split}
        h^0 ( C_2, \left. \mathcal{L} \right|_{C_2} ) &= \mathrm{deg} \left( \left. \mathcal{L} \right|_{C_2} \right) - g_2 + 1 \\
                                   &= \mathrm{deg} \left( \left. \mathcal{L} \right|_{C_2} \right) - ( g_C - g_1 - C_1 \cdot C_2 + 1 ) + 1 \\
                                   &= \left( \mathrm{deg} \left( \left. \mathcal{L} \right|_{C} \right) - g_C + 1 \right) + C_1 \cdot C_2 + g_1 - 1 - \mathrm{deg} \left( \left. \mathcal{L} \right|_{C_1} \right) \\
                                   &= h^0 ( C, \left. \mathcal{L} \right|_C ) + C_1 \cdot C_2 + g_1 - 1 - \mathrm{deg} \left( \left. \mathcal{L} \right|_{C_1} \right) \, .
\end{split}
\end{align}
Hence, we conclude
\begin{align}
\begin{split}
&h^0 ( C_1 \cup C_2, \mathcal{L} ) \geq h^0 \left( C_2, \left. \mathcal{L} \right|_{C_2} \right) - C_1 \cdot C_2 = h^0 ( C, \left. \mathcal{L} \right|_C ) + g_1 -1 - \mathrm{deg} \left( \left. \mathcal{L} \right|_{C_1} \right) \, , \\
\Leftrightarrow \quad &h^0 ( C_1 \cup C_2, \mathcal{L} ) - h^0 ( C, \left. \mathcal{L} \right|_C ) \geq g_1 - 1 - \mathrm{deg} \left( \left. \mathcal{L} \right|_{C_1} \right) \, .
\end{split}
\end{align}
Finally, since we assume $\mathrm{deg} \left( \left. \mathcal{L} \right|_{C_1} \right) < \mathrm{min} \left\{ 0, g_1 -1 \right\}$, the number of additional sections on $C_1 \cup C_2$ is bounded from below by the positive integer $g_1 - \mathrm{deg} \left( \left. \mathcal{L} \right|_{C_1} \right) - 1$. \hfill \qed

We expect that equality holds in generic situations and that only special setups in the spirit of \cite{Bacharach1886, Cayley_Arthur_On_1889} lead to a proper inequality. Still, our result is powerful enough to give a sufficient condition for a jump. Let us demonstrate this in the geometries discussed in \cref{subsec:JumpsInF-theoryExample}. Recall that we are looking at $S = dP_3$ and
\begin{align}
D_C = (10;-3,-3,-4) \, , \qquad D_L = (5;-4,-4,3) \, .
\end{align}
We found that on the genus $g = 24$ curve $C$ it holds $h^1( C, \left. \mathcal{L} \right|_C ) = 0$. Moreover, let us consider the splitting $C \to C_1 \cup C_2$ where $C_1 = V( x_6 )$. These two curves have the following properties:
\begin{align}
\begin{tabular}{ccccc}
\toprule
curve & class & degree & genus & $h^0$ \\
\midrule
$C_1$ & (0;0,0,1) & -3 & 0 & 0\\
$C_2$ & (10;-3,-3,-5) & 41 & 20 & 22 \\
\bottomrule
\end{tabular}
\end{align}
From this we see that \cref{Corollary1} applies to this geometry and implies
\begin{align}
h^0 \left( C_1 \cup C_2, \mathcal{L} \right) - h^0 \left( C, \mathcal{L} \right) \geq g_1 - 1 - \mathrm{deg} \left( \left. \mathcal{L} \right|_{C_1} \right) = 0 - 1 - (-3) = 2 \, ,
\end{align}
This is in agreement with our discussion in \cref{subsec:JumpsInF-theoryExample}.

In many string theory constructions, it is important to engineer exactly one additional vector-like pair. This is particularly true when generating exactly one Higgs pair in MSSM constructions. It is intuitive, that such a minimal change in the vector-like spectrum, requires only mild changes in the geometry. As long as \cref{Corollary1} applies, a necessary condition for such a mild change is to merely split off either a $\mathbb{P}^1$ or a torus --- $g_1 \geq 2$ implies $h^0 \left( C_1 \cup C_2, \mathcal{L} \right) - h^0 \left( C, \mathcal{L} \right) \geq 2$.

More generally, it is of interest to identify the allowed numbers of global sections on a given curve. Therefore, we will now describe an estimate for these values, which is based on the counting procedure presented in \cref{subsubsec:ProposalForCounting}, \cref{Lemma1} and \cref{Corollary1}.

\subsubsection{Algorithmic spectrum estimates}

We can use our results to formulate an algorithmic estimate for the vector-like spectrum over the parameter space of a given setup $(D_C, D_L)$ in a global model. For the time being, our algorithm is focused on the case of a curve in $dP_3$ defined by $\{P=0\}$ and pullback line bundles on these curves. We have implemented this algorithm in the package \emph{H0Approximator} \cite{H0Approximator} as part of \cite{SheafCohomologyOnToricVarieties}. Our algorithm proceeds as follows:
\begin{enumerate}
 \item Input: Curve class $D_C$ and line bundle class $D_L$
 \item Identify all combinations of toric $\mathbb{P}^1$s that can be split off from the curve $D_C$.
 \item Identify the \emph{generic} number of sections of $D_L$ on each curve component.
 \item Use the counting procedure presented in \cref{subsubsec:ProposalForCounting} as well as \cref{Lemma1} and \cref{Corollary1} to place a lower bound on the number of global sections.
 \item[$\Rightarrow$] The collection of all these global section estimates forms an estimate for $h^0$ of $D_L$ on the parameter space of the curve $D_C$.
\end{enumerate}

Let us emphasize a couple of important points of this counting procedure. First, in the second step we do not apply exact methods, such as \cite{SheafCohomologyOnToricVarieties}, to find the exact number of local sections. Rather, we identify the generic number of sections, by which we mean $h^0( C, \mathcal{L} ) = \chi( \mathcal{L} )$ if $\chi( \mathcal{L} ) \geq 0$ and $h^0( C, \mathcal{L} ) = 0$ otherwise. The advantage of this is, that the chiral index can be obtained from topology only. Hence, the number of global sections can be estimated very quickly. Furthermore, this strategy does not violate our \emph{lower bound philosophy}, since the generic number of sections is never larger than the actual number of sections. Consequently, this strategy allows us to quickly identify a lower bound to the actual number of global sections. 

Secondly, let us point out that one disadvantage of our approach of \emph{generic local sections} is that we are unable to identify Brill--Noether jumps on the curve components in this way. However, since such a quick spectrum estimate over the entire parameter space of the curve is currently unfeasible or impossible to obtain with the fully accurate methods, we accept this minor drawback.

Finally, note that upon splitting off $\mathbb{P}^1$s from the curve, the curve could (accidentally) factor further. Computing these further factorizations requires a primary ideal decomposition of the corresponding principal ideal. Currently, this is the most time consuming operation in our algorithm. We reserve optimizations for future work.

This algorithm correctly predicts all the possible values of $h^0$ for 67 of the 83 pairs $(D_C, D_L)$ in our database \cite{Database}. Only for one pair $(D_C, D_L)$, our prediction misses more than 2 values of the exact spectrum. Given the simplicity of our approximation, which means that we cannot detect intricate Brill--Noether jumps and effects discussed in~\cite{Bacharach1886, Cayley_Arthur_On_1889}, we consider this a very positive result. We list the details in \cref{subsubsec:SpectrumEstimate}.

Let us complete this section by applying our procedure to estimate the vector-like spectrum of the F-theory setup discussed in \cref{sec:F-theoryModel}. Recall that in this case we are looking at $D_C = (10;-3,-3,-4)$, i.e., a complicated genus 24 curve. The line bundle in this case is $D_L = (5;-4,-4,3)$. Even though this geometry is fairly involved, our approximator can estimate the spectrum in a couple of minutes\footnote{In this case, this long run time is mostly attributed to the primary decomposition, which we perform to check irreducibility of the curve components.}:
\begin{gapConsole}
!gapprompt@gap>& !gapinput@LoadPackage( "H0Approximator" );&
true
!gapprompt@gap>& !gapinput@FineApproximation( [10,-3,-3,-4],[5,-4,-4,3] );&
(*) 56 rough approximations
(*) Rough spectrum estimate: [ 15, 17, 18, 19, 20, 21 ]
     (x) h0 = 15: 9
     (x) h0 = 17: 18
     (x) h0 = 18: 4
     (x) h0 = 19: 9
     (x) h0 = 20: 12
     (x) h0 = 21: 4
(*) Checking irreducibility of curves...
(*) 26 fine approximations
(*) Fine spectrum estimate: [ 15, 17, 18, 19, 20, 21 ]
     (x) h0 = 15: 3
     (x) h0 = 17: 6
     (x) h0 = 18: 4
     (x) h0 = 19: 3
     (x) h0 = 20: 6
     (x) h0 = 21: 4
[ 15, 17, 18, 19, 20, 21 ]
\end{gapConsole}
Hence, we have identified 26 curve splittings into irreducible components, for which our counting procedure can estimate the spectrum. Based on this, we expect $h^0 \in \{ 15, 17, 18, 19, 20, 21 \}$. As we know from our analysis in \cref{sec:F-theoryModel}, indeed $15 \leq h^0 \leq 21$ and $h^0 = 16$ is only possible by a Brill--Noether jump. 
The latter cannot be predicted by this method.
More information on this implementation can be found in \cite{SheafCohomologyOnToricVarieties}.

\section{Conclusion and Outlook}

Motivated by a better understanding of the exact massless spectra of 4d F-theory compactifications, we have analyzed in this work families of curves $C({\bf c})$ in a complex surface and line bundles ${\cal L}({\bf c})$ on these. Our focus has been on the interplay between changes in the cohomology $h^0(C({\bf c}), {\cal L})$ and variations of the parameters ${\bf c}$, which play the role of complex structure moduli in the context of global F-theory models. To gain insights on how these two are related, we have used two approaches.

To begin with, we first used ideas from Big data and machine learning to gain some intuitions, based on computationally simpler examples, under what circumstances the cohomology may jump, leading to additional vector-like pairs in the F-theory interpretation.
To this end we have generated, in \cref{sec:MachineLearning}, a database \cite{Database} of cohomologies for pairs $(C({\bf c}), {\cal L}(\bf c))$ by varying the parameters ${\bf c}$, where the curves are of genus $1 \leq g \leq 6$, and the line bundles were pullback bundles from a $dP_3$ surface.
For these less complex examples, the cohomologies can be computed using the computer implementations in \cite{SheafCohomologyOnToricVarieties}.
We then use supervised learning on decision trees to predict jumps in the value of $h^0$.
Using different features for training, we find that, while not performing perfectly, topological criteria are surprisingly well-suited (reaching about 95\% accuracy) for distinguishing cases with generic vs.~enhanced $h^0$.
In particular, the algorithm learns from the data a strong correlation between jumps and curves $C({\bf c})$ which split into various components.
This intuition can be applied, without any detailed understanding of the origin of the jumps, directly to find complex structure tunings targeted at generating additional vector-like pairs in F-theory model building.
We demonstrate this in \cref{sec:F-theoryModel} with an F-theory toy model containing a curve of genus 24, for which a scan over the relevant parameter space would be computationally infeasible.
Nevertheless, we found that we can use curve splittings alone to easily engineer 2 to 5 additional vector-like pairs.
This highlights the effectiveness of the machine learning approach to learn certain features from simpler examples, and without any previous knowledge.
However, we also saw there that by curve splitting alone, a spectrum with just one vector-like pair is impossible to achieve.

To overcome this obstacle, we have employed well-known techniques in algebraic geometry, such as the Koszul resolution and {\v C}ech cohomology, which also helps to explain our findings from the machine learning approach in more detail.
We conclude that deformations of the parameters ${\bf c}$ leading to a jump in cohomology can be largely classified as either the curve $C({\bf c})$ or the line bundle ${\cal L}(\bf c)$ becoming non-generic.
While the former comes from curve splittings and is thus topological\footnote{More generally, a curve can also remain smooth while being non-generic, e.g., if it becomes hyperelliptic.
Such transitions are of non-topological nature, and therefore more subtle to detect.
We have neglected them for simplicity in our discussions.
}, the latter is due to special alignments of the points on $C({\bf c})$ defining ${\cal L}({\bf c})$, and not visible just from topological criteria.
The fact that the learner performed so well with the topological criteria is due to a bias in the dataset, which contains only a small number of instances with non-generic line bundles.
Such jumps can never be predicted by the learner based just on split type and intersection numbers.
However, as we discussed in \cref{sec:DataMeetsAnalysis}, we find in general ``equally likely'' jumps due to non-generic line bundles.
The likeliness can be quantified by comparing the dimension of the corresponding subspace of the parameter space on which the jumps occur, which for non-generic line bundles is the subject of \emph{Brill--Noether} theory.
This is generalized in the F-theoretic setup, where complex structure deformations affect genericity of the curve and line bundle democratically.
This leads to a \emph{stratification} of the parameter space by the values of $h^0$.
That is, the complex structure moduli space of global F-theory models decomposes into disjoint subspaces labelled by the vector-like spectrum.
The relationship between the strata can be represented by a Hasse-type diagram, which we term \emph{$h^0$-stratification diagrams}.

The connection between decision trees and the stratification diagrams, which are also Hasse diagrams, is rather intriguing. While they bear some resemblance with decision trees, a key difference is that, unlike in decision trees, nodes can have more than one \emph{incoming} edge. It would be interesting to investigate whether other graph-based machine learning techniques, such as Graph NNs, can be used to train algorithms that can predict the presence of jumps more accurately than the decision trees. Furthermore, recall that global F-theory models typically contain more than one matter curve. The complex structures of these curves are determined by the global moduli of the elliptic fibration, and it is in general not possible to tune the complex structures of all of these curves independently. Therefore, it would be important to extend our analysis to a simultaneous $h^0$-stratification of the moduli space by all the matter curves in a global F-theory model.

In \cref{sec:LocalToGlobalSectionCounting}, we have then investigated the ``microscopic'' origins of jumps due to curve-splittings. It follows a simple counting procedure of local sections on individual curve components, which we then glue to global contributions to $h^0$ on the whole curve. Depending on the boundary conditions imposed by the intersection patterns of the components, this can lead to a net-increase of global sections on the reducible curve compared to the generic case. We have used this understanding to formulate sufficient conditions for a jump in the vector-like spectrum to occur as a result of a curve splitting. These criteria are purely topological, and combine the gluing arguments with vanishing theorems on individual components. Let us stress that this in general provides only a lower bound for $h^0$ for the split curve, because it does not take into account alignments of the intersection points of the components and divisors on the individual components. It will be interesting to investigate, if these bounds can be further improved by topological considerations.

Despite these simplifications, we found these criteria extremely useful to provide a rough estimate of the possible spectrum of $h^0$ on the moduli space of F-theory compactifications, and implemented the algorithm in \cite{H0Approximator}. To fully appreciate this implementation, let us mention that to the best knowledge of the authors, the exact algorithms implemented in\cite{M2,MR1484478,SheafCohomologyOnToricVarieties} do not allow for a parametric cohomology computation. Rather, they will focus on one particular point in the complex structure moduli space and provide the exact answer at this very point. Since each of these computations requires huge amounts of computational resources and runtime, it is impractical to repeat such computations for many points in the complex structure moduli space. In contrast, the new algorithm yields an approximate, but oftentimes sufficiently accurate, estimate --- even for complicated examples such as the genus 24 curve discussed in \cref{sec:F-theoryModel} --- within minutes. We leave generalizations of this counting algorithm, as well as extensions to other toric surfaces, for future work.

Another limitation of our approach is that we have only considered pullback line bundles so far.
However, as already alluded to in the introduction, vector-like spectra in F-theory are oftentimes encoded in line bundles described by a formal weighted sum of points.  
Such a description is computationally harder for two main reasons.
First, it takes much longer to compute line bundle cohomologies of non-pullback bundles with the technologies of \cite{SheafCohomologyOnToricVarieties}. This makes it more challenging to generate a sufficiently large database to apply ideas from Big data and machine learning.
The second obstacle is the parametrization of the line bundles. 
Namely, distinct point configuration can encode equivalent line bundles if their difference is the divisor of a meromorphic function.
To have a better handle on tracking how these equivalences change with complex structure deformations, we need a better understanding of meromorphic functions on higher genus curves.
The crucial tool in this direction is the \emph{Abel--Jacobi map}, which also plays a similar role in the hyperelliptic curve cryptography.
It would be interesting to see to what extent machine learning ideas can be beneficial here.

A related issue arises for \emph{fractional bundles} or \emph{root bundles}.
These appear frequently in explicit global F-theory constructions that engineer a three-generation Standard-Model-like particle physics sector \cite{Krause:2011xj,Cvetic:2015txa,Lin:2016vus,Cvetic:2018ryq,Cvetic:2019gnh}.
The constraint to have chiral indices with $|\chi| = 3$ in these models lead to line bundles $\mathcal{L}$ on curves $C$ which satisfy ${\cal L}^{\otimes n} = L|_C$, where $L$ is a line bundle on the base ${\cal B}_3$ of the elliptic fibration.
In case $n=2$ and $L = K_{{\cal B}_3}$ is the canonical bundle of the base, the bundle ${\cal L}$ can be understood as the pullback of the spin bundle of ${\cal B}_3$ to $C$.
However, for general F-theory constructions, also 3rd and higher roots of bundles ${\cal L} \neq K_{{\cal B}_3}$ appear.
An understanding of which line bundles $\mathcal{L}$ on $C$ satisfy such an equation again requires a detailed understanding of which points --- in this case the intersection points of $C$ with the divisor on ${\cal B}_3$ dual to $L$ --- on the curve define equivalent divisors.
We expect that this will also be intimately related to satisfying the quantization condition \cite{Witten:1996md} for the gauge flux background.

Finally, it is important to point out that the complex structure parameters of the elliptic fibration are not the only parameters of the physical theory.
Rather, a large part of this parameter space which we have not touched upon is in the parametrization of all possible gauge backgrounds. This includes in particular backgrounds with so-called \emph{non-vertical} $G_4$-flux \cite{Greene:1993vm,Braun:2014xka}, for which explicit construction methods in global models are largely unknown.
While these typically do not contribute to the chiral index, it is not clear at the moment if they could modify the flux-induced line bundles on the matter curves.
However, since non-vertical fluxes contribute prominently to a superpotential for the moduli, their presence will dynamically select points in the moduli space that can be a vacuum for the theory, thus have a very different, but direct influence on the vector-like spectrum.
We will therefore need a much better handle on these gauge backgrounds first before we can develop a full understanding for the space of 4d F-theory vacua.

\paragraph{Acknowledgements}
We thank \texttt{Plesken}, \texttt{google cloud} and the \texttt{Oxford Hydra cluster} for their reliable computations. The work of M.B.~is supported by the Wiener--Anspach foundation. M.C.~and M.L.~are supported by DOE Award DE-SC0013528Y. M.C.~further acknowledges the support by the Simons Foundation Collaboration grant \#724069 on ``Special Holonomy in Geometry, Analysis and Physics'', the Slovenian Research Agency No.~P1-0306, and the Fay R.~and Eugene L.~Langberg Chair funds. R.D.~was supported in part by NSF grant DMS 2001673 and by the Simons Foundation Collaboration grant \#390287 on ``Homological Mirror Symmetry''. M.C., L.L. and M.L.~thank String Phenomenology 2019 for hospitality during the early stages of this project. M.B., M.C., and L.L.~thank the ITP Heidelberg for hospitality. M.B~and L.L.~further thank the University of Pennsylvania for hospitality.

\appendix

\section{Tools: Koszul resolution, Brill--Noether theory and fat points} \label{subsec:Tools}
\label{sec:Tools}

The purpose of this appendix is to cover some of the necessary mathematical backgrounds, and also provide more details of computations carried out throughout the paper.

\subsection{Brill--Noether theory} \label{subsec:BNTheory}

Our exposition of Brill--Noether theory is based on \cite{mumford1975curves,Griffiths:433962}. We refer the interested reader to these references for more details.

\subsubsection{The Jacobian of Riemann surfaces}

To each smooth Riemann surface $C_g$ one can associate a Jacobian variety $\mathrm{Jac} ( C_g )$. This variety is of dimension $g$ and classifies equivalence classes of line bundle divisors of degree $0$:
\begin{align}
\mathrm{Jac}( C_g ) = \mathrm{Div}_0( C_g ) / \mathrm{Prin}( C_g ) \, . \label{equ:Div0Jacobian}
\end{align}
In this expression $\mathrm{Div}^0 ( C_g )$ is the group of all divisors of degree $0$ and $\mathrm{Prin}( C_g )$ the group of all principal divisors on $C_g$. Line bundles on $C_g$ are isomorphic iff their divisors differ by a divisor in $\mathrm{Prin}( C_g )$. Hence, sheaf cohomologies of line bundles can only differ if the line bundles are not isomorphic, or equivalently if their divisors differ by more than elements of $\mathrm{Prin}( C_g )$. Consequently, the Jacobian of $C_g$ plays an important role for our analysis and in Brill--Noether theory. Let us therefore introduce the Jacobian in more detail.

Historically, the Jacobian of a curve $C_g$ of genus $g$ was discovered by investigating integrals $\int_{\mathcal{P}}{\omega}$ where $\mathcal{P} \subset C_g$ is a (not necessarily closed) path and $\omega$ a holomorphic differential. More generally, mark a point $p_0 \in C_g$, let $(\omega_1, \dots, \omega_g)$ be a basis of the holomorphic differentials on $C_g$ and consider the map
\begin{align}
\phi \colon C_g \to \mathbb{C}^g \, , \, p \mapsto \left( \int_{p_0}^{p}{\omega_1} \, , \, \dots \, , \, \int_{p_0}^{p}{\omega_g} \right) \, .
\end{align}
The value of this map strongly depends on the path $\mathcal{P} \subset C_g$ which we choose to connect $p_0$ and $p$. This redundancy can be removed by taking the period lattice of $C_g$ into account. To this end, recall that there are $2g$ homologically distinct closed 1-cycles in $C_g$, i.e., $H_1( C_g, \mathbb{Z} )$ is a $2g$-dimensional vector space over $\mathbb{Z}$.\footnote{See e.g. \cite{mumford1975curves} for an explicit construction of the $2g$-generators $A_i$, $B_i$ of $H_1( C_g, \mathbb{Z} )$.} We now consider the map
\begin{align}
\phi \colon H_1( C_g, \mathbb{Z} ) \to \mathbb{C}^g \, , \, \alpha \mapsto \left( \int_\alpha{\omega_1} \, , \, \dots \, , \, \int_\alpha{\omega_g} \right) \, ,
\end{align}
where $\omega_i$ denote the above basis of holomorphic differentials on $C_g$. Hence, for every of the $2g$-basis elements of $H_1( C_g, \mathbb{Z} )$, we obtain an element $\phi( \alpha ) \in \mathbb{C}^g$. It turns out that these $2g$ elements span a full-dimensional lattice $\Lambda$ in $\mathbb{C}^g$ --- the period lattice of $C_g$. By virtue of this lattice, we obtain a well-defined map
\begin{align}
\phi \colon C_g \to \mathbb{C}^g / \Lambda \, , \, p \mapsto \left( \int_{p_0}^{p}{\omega_1} \, , \, \dots \, , \, \int_{p_0}^{p}{\omega_g} \right) \label{equ:AbelJacobi} \, .
\end{align}
This map is known as the \emph{Abel--Jacobi map}. It can easily be extended to divisors in $C_g$. Namely, for a divisor
\begin{align}
D = \sum_{i = 1}^{N}{\lambda_i \cdot p_i} \, , \, \lambda_i \in \mathbb{Z}, \qquad p_i \in C_g \, ,
\end{align}
we define
\begin{align}
\phi \colon \mathrm{Div} ( C_g ) \to \mathbb{C}^g / \Lambda \, , \, D \mapsto \sum_{i = 1}^{N}{\lambda_i \cdot \phi \left( p_i \right)} \, . \label{equ:AbelJacobiForDivisorClasses}
\end{align}
The theorem of Abel (see \cite{freitag2011complex} and references therein) states that two effective divisors $D$ and $E$ satisfy $\phi( D ) = \phi( E )$ iff $D$ and $E$ are linearly equivalent. Consequently, we obtain an injective group homomorphism
\begin{align}
\Phi \colon \mathrm{Div}_0 ( C_g ) / \mathrm{Prin} ( C_g ) \to \mathbb{C}^g / \Lambda \, , \, [ D ] \mapsto \sum_{i = 1}^{N}{\lambda_i \cdot \phi \left( p_i \right)} \, ,
\end{align}
of divisor classes of degree $0$. It turns out that this map is also surjective (see \cite{freitag2011complex} for a proof). Hence, there is a natural isomorphism
\begin{align}
\mathrm{Jac} ( C_g ) = \mathrm{Div}_0 ( C_g ) / \mathrm{Prin} ( C_g ) \cong \mathbb{C}^g / \Lambda \, .
\end{align}

\subsubsection{Central results}

For ease of notation let $\mathrm{Div}( C_g )_d$ denote all divisors of degree $d$. Then, let us consider the restriction of \cref{equ:AbelJacobiForDivisorClasses} to $\mathrm{Div}( C_g )_d$, i.e.
\begin{align}
\Phi_d \colon \mathrm{Div} ( C_g )_d / \mathrm{Prim} ( C_g ) \to \mathbb{C}^g / \Lambda \, , \, D \mapsto \sum_{i = 1}^{N}{\lambda_i \cdot \phi \left( p_i \right)} \, .
\end{align}
Let us pick an integer $r \geq -1$ and study the subvariety of $\mathrm{Jac} ( C_g )$
\begin{align}
G^r_d = \left\{ p \in \mathrm{im} \left( \Phi_d \right) \, , \, h^0 \left( C_g, \mathcal{O}_{C_g} ( \Phi^{-1}_d ( p ) ) \right) = r+1 \right\} \, .
\end{align}
Then, the central result of Brill--Noether theory states \cite{Brill1874}
\begin{align}
\mathrm{dim}( G^r_d ) \geq \rho \left( r, d, g \right) \equiv g - \left( r + 1 \right) \cdot \left( ( r + 1 ) - ( d - g + 1 ) \right) \, .
\end{align}
By use of the Riemann--Roch theorem
\begin{align}
h^0 \left( C_g, \mathcal{O}_{C_g} ( D ) \right) - h^1 \left( C_g, \mathcal{O}_{C_g} ( D ) \right) = \mathrm{deg} \left( \mathcal{O}_{C_g} ( D ) \right) - g + 1 = d - g +1 \, ,
\end{align}
we can rewrite this results in the suggestive form
\begin{align}
\mathrm{dim}( G^{r}_d ) \geq \rho \left( r, d, g \right) \equiv g - n^0 \cdot n^1 \, ,
\end{align}
with $n^0 \equiv r+1$ and $n^1 = r+1 - ( d - g + 1 )$. 
We may thus use $\rho \left( r, d, g \right)$ as a measure for how likely it is that a line bundle of degree $d$ on a genus $g$ curve $C_g$ has $n^0 = r+1$ global sections.

Let us demonstrate this for degree $d = 2$ bundles on a genus-3 curve. By general theory, the number of section of a line bundle on a curve $C_g$ with $g \geq 1$ can never exceed its degree. Hence $n^0 \in \{ 0,1,2 \}$. 
With this information, let us compute $\rho( r, d, g )$ for the admissible values of $r$:
\begin{equation}
\begin{array}{cc|c}
\toprule
r & (n^0, n^1) & \rho( r, d, g ) \\
\midrule
-1 & (0,0) & 3 \\
0  & (1,1) & 2 \\
1  & (2,2) & -1\\
\bottomrule
\end{array} \label{equ:BNSimpleExample2}
\end{equation}
From this we learn, that most line bundles $\mathcal{L}$ of degree $2$ on a genus-3 curve $C_3$ satisfy $h^0 \left( C_3, \mathcal{L} \right) = 0$. Since for these bundles $\rho$ matches the dimension of the Jacobian of $C_3$, we can say that these line bundles are associated to generic points of the Jacobian. Furthermore, we learn that there are such line bundles with $h^0 \left( C_3, \mathcal{L} \right) = 1$. However, these are special in the sense that they are associated to a codimension-1 locus in the Jacobian $\mathrm{Jac} ( C_3 )$.

Finally, $\rho = -1$ for $r = 1$ begs for an explanation. This explanation follows from work of Griffiths and Harris \cite{griffiths1980}:
\begin{center}
On \textbf{generic} curves, $\mathrm{dim} ( G^{r}_d ) = \rho \left( r, d, g \right)$.
\end{center}
So in particular, on generic curves it holds $G^{r}_d = \emptyset$ if and only if $\rho \left( r, d, g \right) < 0$. Consequently, we conclude from \cref{equ:BNSimpleExample2}, that on generic genus $g = 3$ curve, there is no line bundle $\mathcal{L}$ of degree $2$ such that $h^0 ( C_3, \mathcal{L} ) = 2$.

Note however, that this does not rule out the possibility that non-generic curves may host such line bundles. In the case at hand, it follows from the theorem of Cliffford \cite{griffiths1980} that hyperelliptic curves $H_3$ of genus $g = 3$ admit line bundles $\mathcal{L}$ of degree $2$ and $h^0 ( H_3, \mathcal{L} ) = 2$.

\subsubsection{Brill--Noether jump}

As we see from \cref{equ:BNSimpleExample2}, we can in general modify a line bundle on a generic curve such that it admits additional sections. A jump from $r = r_{\text{generic}}$ to $r_{\text{generic}} + 1$ is equivalent to saying that the Serre-dual bundle admits a section, i.e., becomes effective:
\begin{align}
K_C - D > 0 \quad \Leftrightarrow \quad \exists p_i \colon K_C - D \sim \sum_{i}{p_i} \, .
\end{align}
where $\sim$ represents linear equivalence of divisors. Obviously, this requires the line bundle divisor $D$ to move into special alignment relative to $K_C$. Such a divisor is termed a \emph{special} divisor. We term a change in $h^0$, which is solely attributed to a special alignment of the line bundle divisor, a \emph{Brill--Noether jump}.

\subsection{Koszul resolution}

\subsubsection{Generalities}

Given a curve $C$ and a line bundle $\mathcal{L}$ on $C$, we wish to identify which deformations of the curve lead to an increased number of global sections for $\mathcal{L}$. For hypersurface curves in $dP_3$, the answer follows from a study of the Koszul resolution. In this case $C ( \mathbf{c} ) = V( P ( \mathbf{c} ) )$ for a polynomial $P ( \mathbf{c} )$. The coefficients $\mathbf{c}$ model the complex structure moduli of a global F-theory setting.

For such a setup, the Koszul resolution is given by the short-exact sequence
\begin{align}
0 &\to \mathcal{O}_{dP_3} \left( D_L - D_C \right) \xrightarrow{\alpha} \mathcal{O}_{dP_3} \left( D_L \right) \to \mathcal{L} ( \mathbf{c} ) \to 0 \, .
\end{align}
The map $\alpha$ is induced by the polynomial $P( \mathbf{c} )$. Namely, for $U \subseteq dP_3$ open, $\alpha$ is given by
\begin{align}
s \in \mathcal{O}_{dP_3} \left( D_L - D_C \right)( U ) \mapsto s \cdot P( \mathbf{c} ) \in \mathcal{O}_{dP_3} \left( D_L \right)( U ) \, . \label{equ:LocalMaps}
\end{align}
The Koszul resolution then induces the following long exact sequence in sheaf cohomology:
\begin{equation}
\begin{tikzpicture}[scale=2, baseline=(current  bounding  box.center)]
          \matrix(m)[matrix of math nodes,column sep=15pt,row sep=15pt]{
                0 & H^0 \left( dP_3, D_L - D_C \right)
                  & H^0 \left( dP_3, D_L \right)
                  & H^0 \left( C ( \mathbf{c} ), \mathcal{L} ( \mathbf{c} ) \right) \\
                  & H^1 \left( dP_3, D_L - D_C \right)
                  & H^1 \left( dP_3, D_L \right)
                  & H^1 \left( C ( \mathbf{c} ), \mathcal{L} ( \mathbf{c} ) \right) \\
                  & H^2 \left( dP_3, D_L - D_C \right)
                  & H^2 \left( dP_3, D_L \right)
                  & 0 & 0 \, . \\
                     };
               \draw[->,font=\scriptsize,every node/.style={above},rounded corners]
                     (m-1-1) edge (m-1-2)
                     (m-1-2) edge node {$\varphi_0$} (m-1-3)
                     (m-1-3) edge (m-1-4)
                     (m-1-4.east) --+(5pt,0)|-+(0,-7.5pt)-|([xshift=-5pt]m-2-2.west)--(m-2-2.west);
                \draw[->,font=\scriptsize,every node/.style={above},rounded corners]
                     (m-3-2) edge node {$\varphi_2$} (m-3-3)
                     (m-3-3) edge (m-3-4)
                     (m-3-4) edge (m-3-5);
                \draw[->,font=\scriptsize] (m-2-2) edge node [above] {$\varphi_1$} (m-2-3);
                \draw[->,font=\scriptsize] (m-2-3) edge node [above] {} (m-2-4);
                \draw[->,font=\scriptsize,every node/.style={above},rounded corners] (m-2-4.east) --+(5pt,0)|-+(0,-7.5pt)-|([xshift=-5pt]m-3-2.west)-- (m-3-2.west);
\end{tikzpicture}
\end{equation}
The maps $\varphi_i = \varphi_i ( \mathbf{c} )$ are induced from multiplication with $P( \mathbf{c} )$. Therefore, these maps are sensitive to the choice of parameters $\mathbf{c}$ for the curve $C( \mathbf{c} )$. Explicitly, the maps $\varphi_i$ are vector-space morphisms and the entries of their defining matrices are functions of the parameters $c_i$. Provided that we know these mapping matrices, we may thus use the exactness of the Koszul resolution of infer $h^i \left( C( \mathbf{c} ), \mathcal{L} ( \mathbf{c} ) \right)$ as a function of the coefficients $c_i$ in $P ( \mathbf{c} )$.

For example, in \cref{subsec:OneAdditionalSection}, we consider $D_C = (4;-1,-2,-1)$ and $D_L = (3;-3,-1,-2)$. In this case, the Koszul resolution simplifies and takes the form
\begin{equation}
\begin{tikzpicture}[scale=2, baseline=(current  bounding  box.center)]
          \matrix(m)[matrix of math nodes,column sep=15pt,row sep=15pt]{
                0 & 0
                  & 0
                  & H^0 \left( C ( \mathbf{c} ), \mathcal{L} ( \mathbf{c} ) \right) \\
                  & H^1 \left( dP_3, D_L - D_C \right) \cong \mathbb{C}^4
                  & H^1 \left( dP_3, D_L \right) \cong \mathbb{C}^1
                  & H^1 \left( C ( \mathbf{c} ), \mathcal{L} ( \mathbf{c} ) \right) \\
                  & 0
                  & 0
                  & 0 & 0 \, .  \\
                     };
               \draw[->,font=\scriptsize,every node/.style={above},rounded corners]
                     (m-1-1) edge (m-1-2)
                     (m-1-2) edge (m-1-3)
                     (m-1-3) edge (m-1-4)
                     (m-1-4.east) --+(5pt,0)|-+(0,-7.5pt)-|([xshift=-5pt]m-2-2.west)--(m-2-2.west);
                \draw[->,font=\scriptsize,every node/.style={above},rounded corners]
                     (m-3-2) edge (m-3-3)
                     (m-3-3) edge (m-3-4)
                     (m-3-4) edge (m-3-5);
                \draw[->,font=\scriptsize] (m-2-2) edge node [above] {$\varphi$} (m-2-3);
                \draw[->,font=\scriptsize] (m-2-3) edge node [above] {} (m-2-4);
                \draw[->,font=\scriptsize,every node/.style={above},rounded corners] (m-2-4.east) --+(5pt,0)|-+(0,-7.5pt)-|([xshift=-5pt]m-3-2.west)-- (m-3-2.west);
\end{tikzpicture}
\end{equation}
Then it follows
\begin{align}
\begin{split}
H^1 \left( C ( \mathbf{c} ), \mathcal{L} ( \mathbf{c} ) \right) &\cong \mathrm{coker} \varphi \, , \\
h^1 ( C ( \mathbf{c} ), \mathcal{L} ( \mathbf{c} ) ) &= 1 - \mathrm{dim} \left( \mathrm{im} \varphi \right) \, .
\end{split}
\end{align}
A detailed study of \v{C}ech cohomology \cite{cox2011toric} shows that in this geometry we have $M_\varphi = \left( c_3, c_6, c_9, 0 \right)$. Hence, $h^1 ( C ( \mathbf{c} ), \mathcal{L} ( \mathbf{c} ) ) = 1$ on curves with $c_3 = c_6 = c_9 = 0$ and otherwise $h^1 ( C ( \mathbf{c} ), \mathcal{L} ( \mathbf{c} ) ) = 0$. Along these lines, we classify the curve geometries according to their admitted number of global sections.

Recall that \v{C}ech cohomology expresses $H^i( dP_3, \mathcal{O}_{dP_3} ( D_L - D_C ) )$ and $H^i( dP_3, \mathcal{O}_{dP_3} ( D_L ) )$ as collections of local sections. The mappings of these local sections follow from \cref{equ:LocalMaps}, i.e., are given by multiplication with the polynomial $P( \mathbf{c} )$ which defines the curve $C ( \mathbf{c} )$. Importantly, these bases are expressed modulo equivalence relations induced from \v{C}ech coboundaries. Therefore, these computations are typically fairly tedious.

Oftentimes, \emph{cohomCalg} \cite{Blumenhagen:2010pv, 2011JMP....52c3506J, Blumenhagen:2010ed, Blumenhagen:2011xn, KoszulExtensionManual, cohomCalg:Implementation, Rahn:2010fm} can help to simplify this task. Namely, it identifies bases of $H^i( dP_3, \mathcal{O}_{dP_3} ( D_L - D_C ) )$ and $H^i( dP_3, \mathcal{O}_{dP_3} ( D_L ) )$ in terms of rationoms --- quotients of monomials in the homogeneous coordinates --- and therefore simplifies the task to find the bases in \v{C}ech cohomology. Even more, we may be tempted to simply multiply the basis elements identifed by \emph{cohomCalg} \cite{Blumenhagen:2010pv, 2011JMP....52c3506J, Blumenhagen:2010ed, Blumenhagen:2011xn, KoszulExtensionManual, cohomCalg:Implementation, Rahn:2010fm} with the polynomial $P( \mathbf{c} )$ and ignore all image rationoms that have not been identified as bases for $H^i( dP_3, \mathcal{O}_{dP_3} ( D_L ) )$ by \emph{cohomCalg} under the assumption that they correspond to \v{C}ech coboundaries.

This procedure fails whenever \emph{\v{C}ech cohomology chamber factors} greater than $1$ appear. In this case, \emph{cohomCalg} finds that one rationom $R$ spans a vector space of dimension greater than $1$ in sheaf cohomology. The interpretation of this is, that there are at least two distinct \v{C}ech cochains, i.e., collections of local sections, in which the rationom $R$ is the only non-trivial entry. Hence, these distinct \v{C}ech cochains are both canonically isomorphic to $R$. However, to identify the mapping matrices of the line bundle cohomologies correctly, the information about $R$ is insufficient. Rather, the corresponding \v{C}ech cochains need to be identified explicitly.

Given these insights, we have taken extra care, to work out the mappings presented in this work carefully with \v{C}ech cohomology. We present such a computation in large detail in the following section.

Before we come to this, let us mentioned that a detailed study of the Koszul resolution is not original to this work. For example, in the context of heterotic compactifications, these resolutions --- including the mappings in the induced long exact sequence --- have been studied extensively \cite{Anderson:2007nc, Anderson:2008ex, Anderson:2009mh,Anderson:2011ns,Anderson:2012yf}. However, to the best of our knowledge, chamber factor greater than 1 do not show in products of projective spaces. Hence, this complication does not arise in heterotic compactifications with CICYs.

\subsubsection{{\v C}ech cohomologies for \texorpdfstring{\cref{subsubsec:BN1}}{section 4.2.1}} \label{subsec:CechForExample3}

Here, we present a more detailed computation of the example discussed in \cref{subsubsec:BN1}.
Recall that the curve and line bundle in question are given by
\begin{align}
D_C = ( 4; -1, -1, -1 ) \, , \qquad D_L = ( 1; 2, -2, -1 ) \, .
\end{align}
Moreover, recall that in this case $h^0 \left( C ( \mathbf{c} ), \mathcal{L} ( \mathbf{c} ) \right)$ is uniquely determined by the mapping
\begin{align}
\varphi \colon H^1 \left( dP_3, \mathcal{O}_{dP_3} \left( D_L - D_C \right) \right) \xrightarrow{\cdot P ( \mathbf{c} )} H^1 \left( dP_3, \mathcal{O}_{dP_3} \left( D_L \right) \right) \, ,
\end{align}
where
\begin{align}
\begin{split}
P ( \mathbf{c} ) &= c_1 x_1^3 x_2^3 x_3^2 x_4 + c_2 x_1^2 x_2^3 x_3 x_4^2 x_6 + c_3 x_1 x_2^3 x_4^3 x_6^2 + c_4 x_1^3 x_2^2 x_3^3 x_5 + c_5 x_1^2 x_2^2 x_3^2 x_4 x_5 x_6 \\
                 &\qquad + c_6 x_1 x_2^2 x_3 x_4^2 x_5 x_6^2 + c_7 x_2^2 x_4^3 x_5 x_6^3 + c_8 x_1^2 x_2 x_3^3 x_5^2 x_6 + c_9 x_1 x_2 x_3^2 x_4 x_5^2 x_6^2 \\
                 &\qquad + c_{10} x_2 x_3 x_4^2 x_5^2 x_6^3 + c_{11} x_1 x_3^3 x_5^3 x_6^2 + c_{12} x_3^2 x_4 x_5^3 x_6^3 \, . \label{equ:DefiningEquation2}
\end{split}
\end{align}
Namely, $h^0 ( C( \mathbf{c} ), \mathcal{L} ( \mathbf{c} ) ) = 3 - \mathrm{rk} \left( M_\varphi \right)$. With \emph{cohomCalg}  \cite{Blumenhagen:2010pv, Blumenhagen:2010ed, Blumenhagen:2011xn, KoszulExtensionManual, cohomCalg:Implementation, 2011JMP....52c3506J, Rahn:2010fm}, we obtain basis of the line bundle cohomologies as follows:
\begin{align}
H^1( D_L - D_C ) &\cong \text{Span}_{\mathbb{C}} \left\{ \frac{1}{x_3 x_4^3 x_6^3}, \frac{1}{x_1 x_3^2 x_4^2 x_6^2}, \frac{1}{x_1^2 x_3^3 x_4 x_6} \right\} \cong \mathbb{C}^3 \, ,
\label{equ:H1ForDL-DCByCohomCalg} \\
H^1( D_L ) &\cong \text{Span}_{\mathbb{C}} \left\{ \frac{x_5^3 x_6}{x_1 x_4}, \frac{x_1 x_2^3}{x_3 x_6} \right\} \cong \mathbb{C}^2 \, . \label{equ:BasisForCohomologies-Example3}
\end{align}
By polynomial multiplication we then have
\begin{align}
\begin{split}
\frac{1}{x_3 x_4^3 x_6^3} \cdot P( \mathbf{c} ) &= c_3 \frac{x_1 x_2^3}{x_3 x_6} + \dots \, , \\
\frac{1}{x_1 x_3^2 x_4^2 x_6^2} \cdot P( \mathbf{c} ) &= c_2 \frac{x_1 x_2^3}{x_3 x_6} + c_{12} \frac{x_5^3 x_6}{x_1 x_4} + \dots \, ,\\
\frac{1}{x_1^2 x_3^3 x_4 x_6} \cdot P( \mathbf{c} ) &= c_1 \frac{x_1 x_2^3}{x_3 x_6} + c_{11} \frac{x_5^3 x_6}{x_1 x_4} + \dots \, .
\label{equ:LazyMultiplication}
\end{split}
\end{align}
On the RHS of these equations, we have omitted all rationoms which cannot be expressed as linear combinations of \cref{equ:BasisForCohomologies-Example3}. The remainder of this section will justify that we can indeed omit these terms. For the time being, note that this leads to
\begin{align}
M_\varphi = \begin{psmallmatrix} c_3 & c_2 & c_1 \\ 0 & c_{12} & c_{11} \end{psmallmatrix} \, , \label{equ:ExplicitMatrixVerification}
\end{align}
which is the matrix analyzed in \cref{subsubsec:BN1}.

\paragraph{Strategy}

In order to justify that all omitted terms in \cref{equ:LazyMultiplication} can be ignored, we will now analyse $H^1( dP_3, \mathcal{O}_{dP_3} ( D_L ) )$ and $H^1( dP_3, \mathcal{O}_{dP_3} ( D_L - D_C ) )$ from the perspective of \v{C}ech cohomology. For additional background we refer the interested reader to \cite{cox2011toric}. Recall that for $H^1( dP_3, \mathcal{O}_{dP_3} ( D_L ) )$ it holds
\begin{align}
H^1( dP_3, \mathcal{O}_{dP_3} ( D_L ) ) \cong \check{H}^1( \mathcal{U}, \mathcal{O}_{dP_3} ( D_L ) ) = \text{ker} \left( \delta_1 \right) / \text{im} \left( \delta_0 \right) \, .
\end{align}
In this expression, $\mathcal{U}$ is the affine open cover of the $dP_3$ surface --- we will discuss this momentarily --- and the maps $\delta_i$ are the boundary morphisms in the \v{C}ech complex
\begin{align}
0 \to \check{C}^0( \mathcal{U}, \mathcal{O}_{dP_3} ( D_L ) ) \xrightarrow{\delta_0} \check{C}^1( \mathcal{U}, \mathcal{O}_{dP_3} ( D_L ) ) \xrightarrow{\delta_1} \dots \, .
\end{align}
Thereby, let us specify our statement regarding the RHS of \cref{equ:LazyMultiplication}. We claim that all omitted terms are in $\mathrm{im} ( \delta_0 )$, i.e., are \v{C}ech coboundaries. To justify this statement, we proceed by investigating the following objects:
\begin{enumerate}
 \item $\mathrm{im} \left( \delta_0 ( D_L ) \right)$.
 \item $\mathrm{ker} \left( \delta_1 ( D_L ) \right)$,
 \item $\mathrm{ker} \left( \delta_1 ( D_L - D_C ) \right)$,
 \item the map $\mathrm{ker} \left( \delta_1 ( D_L - D_C ) \right) \to \mathrm{ker} \left( \delta_1 ( D_L ) \right)$.
\end{enumerate}

\paragraph{\v{C}ech 0-cocycles of \boldmath{${D_L}$}}

To understand $\check{C}^0( \mathcal{U}, \mathcal{O}_{dP_3} ( D_L ) )$, recall that $dP_3$ has 6 homogeneous variables $x_i$. These correspond to the ray generators
\begin{align}
& u_1 = (0,-1) \, , \qquad u_2 = (-1,0) \, , \qquad u_3 = (1,-1) \, , \\
& u_4 = (-1,1) \, , \qquad u_5 = (1,0) \, , \qquad u_6 = (0,1) \, .
\end{align}
In terms of these, the maximal cones in the fan of $dP_3$ are given by
\begin{align}
\begin{split}
U_1 = \text{Span}_{\geq 0} \left\{ u_1, u_3 \right\} \, , \qquad U_2 = \text{Span}_{\geq 0} \left\{ u_3, u_5 \right\} \, , \qquad U_3 = \text{Span}_{\geq 0} \left\{ u_5, u_6 \right\} \, , \\
U_4 = \text{Span}_{\geq 0} \left\{ u_6, u_4 \right\} \, , \qquad U_5 = \text{Span}_{\geq 0} \left\{ u_4, u_2 \right\} \, , \qquad U_6 = \text{Span}_{\geq 0} \left\{ u_2, u_1 \right\} \, .
\end{split}
\end{align}
These cones correspond to open affine subsets of the $dP_3$, namely the subsets of the form $\{ x_i \neq 0 \}$. Collectively, $\mathcal{U} = \{ U_i \}_{1 \leq i \leq 6}$ is the open affine cover  of $dP_3$. To compute $\check{C}^0( \mathcal{U}, \mathcal{O}_{X_\Sigma} ( D_L ) )$ with respect to this open affine cover $\mathcal{U}$, we note
\begin{align}
D_L = ( 1; 2, -2, -1 ) = H + 2 E_1 - 2 E_2 - E_3 = \sum_{i = 1}^{6}{a_i V( x_i )} \, ,
\end{align}
with $a_1 = a_4 = a_6 = 0$ and $a_2 = 2$, $a_3 = -1$, $a_5 = 1$. Now, we can quote from \cite{cox2011toric} that
\begin{align}
\check{C}^0( \mathcal{U}, \mathcal{O}_{X_\Sigma} ( D_L ) ) &= \bigoplus_{1 \leq i \leq 6}{H^0( U_i, \left. \mathcal{O}_{X_\Sigma}( D_L ) \right|_{U_i}} \, , \\
H^0( U_i, \left. \mathcal{O}_{X_\Sigma}( D_L ) \right|_{U_i} ) &\cong \left( \prod_{j = 1}^{6}{x_j^{a_j}} \right) \cdot \bigoplus_{m \in P_D ( U_i )}{\mathbb{C} \cdot \left( \prod_{j = 1}^{6}{x_j^{\langle m, u_j \rangle}} \right) } \, , \label{equ:Convenient} \\
P_D( U_i ) &= \{ m \in \mathbb{Z}^2 \, , \, \langle m, u_\rho \rangle \geq - a_\rho \; \forall \rho \in \sigma( 1 ) \} \, .
\end{align}
The normalization in \cref{equ:Convenient} ensures that we are looking at rationoms of degree $D_L$, as analysed by \emph{cohomCalg}. Explicitly, it holds
\begin{align}
P_D( U_1 ) &= \{ m \in \mathbb{Z}^2 \, , -m_2 \geq 0 \text{ and } m_1 - m_2 \geq 1 \} \, , \\
P_D( U_2 ) &= \{ m \in \mathbb{Z}^2 \, , m_1 - m_2 \geq 1 \text{ and } m_1 \geq -1 \} \, , \\
P_D( U_3 ) &= \{ m \in \mathbb{Z}^2 \, , m_1 \geq -1 \text{ and } m_2 \geq 0 \} \, , \\
P_D( U_4 ) &= \{ m \in \mathbb{Z}^2 \, , m_2 \geq 0 \text{ and } - m_1 + m_2 \geq 0 \} \, , \\
P_D( U_5 ) &= \{ m \in \mathbb{Z}^2 \, , - m_1 + m_2 \geq 0 \text{ and } -m_1 \geq -2 \} \, , \\
P_D( U_6 ) &= \{ m \in \mathbb{Z}^2 \, , - m_1 \geq -2 \text{ and } -m_2 \geq 0 \} \, .
\end{align}
To express these polytopes in simpler terms, we define the regions $A$, $B$, $C$, $D$, $E$, $F$, $G$, $H$:
\begin{center}
\begin{tikzpicture}[scale=0.5]

    \draw [thick,->] (0,-8)--(0,8) node (yaxis) [above] {};
    \draw [thick,->] (-8,0)--(8,0) node (xaxis) [right] {};

	\draw { (-7,-8)--(-1,-2)--(-1,-8)--(-7,-8) } [opacity = 0.5, fill=green];
	\node at (-3.5,-6.5) {$H$};
	\draw { (-1,-2)--(1,0)--(2,0)--(2,-8)--(-1,-8)--(-1,-2) } [opacity = 0.5, fill=red];
	\node at (0.5,-6.5) {$A$};
	\draw { (2,0)--(8,0)--(8,-8)--(2,-8)--(2,0) } [opacity = 0.5, fill=orange];
	\node at (4.5,-6.5) {$B$};
	\draw { (1,0)--(8,0)--(8,7)--(1,0) } [opacity = 0.5, fill=yellow];
	\node at (4.5,2) {$C$};
	\draw { (2,2)--(2,8)--(8,8)--(2,2) } [opacity = 0.5, fill=blue];
	\node at (4.5,6) {$D$};
	\draw { (-1,0)--(-1,8)--(2,8)--(2,2)--(0,0)--(-1,0) } [opacity = 0.5, fill=cyan];
	\node at (0.5,6) {$E$};
	\draw { (-8,0)--(-8,8)--(-1,8)--(-1,0)--(-8,0) } [opacity = 0.5, fill=brown];
	\node at (-3.5,6) {$F$};
	\draw { (-8,0)--(0,0)--(-8,-8)--(-8,0) } [opacity = 0.5, fill=sacramentostategreen];
	\node at (-3.5,-2) {$G$};
	
\end{tikzpicture}
\end{center}
In an abuse of terminology, we use $A$ to denote all polynomials formed from linear combination of the Laurent monomials associated to the lattice points of the region $A$. Similarly, we use the names for the other regions. Thereby, we can write
\begin{align}
\check{C}^0( \mathcal{U}, \mathcal{O}_{X_\Sigma} ( D_L ) ) &= \frac{x_2^2 x_5}{x_3} \cdot \left( H+A+B, A+B+C, C+D+E, \right. \\
                                                           & \qquad \qquad \qquad \qquad \left. D+E+F, E+F+G, G+H+A \right) \, . \label{equ:C0-Cohomologies-Example3}
\end{align}
Finally note that the map $\delta_0 \colon \check{C}^0( \mathcal{U}, \mathcal{O}_{X_\Sigma} ( D_L ) ) \to \check{C}^1( \mathcal{U}, \mathcal{O}_{X_\Sigma} ( D_L ) )$ is given by multiplication with the following matrix:
\begin{align}
M_{\delta_0} = 
\begin{psmallmatrix}
-1 & 1 & 0 & 0 & 0 & 0 \\ 
-1 & 0 & 1 & 0 & 0 & 0 \\ 
-1 & 0 & 0 & 1 & 0 & 0 \\ 
-1 & 0 & 0 & 0 & 1 & 0 \\ 
-1 & 0 & 0 & 0 & 0 & 1 \\ 
0 & -1 & 1 & 0 & 0 & 0 \\ 
0 & -1 & 0 & 1 & 0 & 0 \\
0 & -1 & 0 & 0 & 1 & 0 \\
0 & -1 & 0 & 0 & 0 & 1 \\
0 & 0 & -1 & 1 & 0 & 0 \\
0 & 0 & -1 & 0 & 1 & 0 \\
0 & 0 & -1 & 0 & 0 & 1 \\
0 & 0 & 0 & -1 & 1 & 0 \\
0 & 0 & 0 & -1 & 0 & 1 \\
0 & 0 & 0 & 0 & -1 & 1 \\
\end{psmallmatrix} \, .
\end{align}

\paragraph{\v{C}ech 1-cocycles  of \boldmath{$D_L$}}

We repeat this analysis for $\check{C}^1 ( \mathcal{U}, \mathcal{O}_{dP_3} ( D_L ) )$. The elements in this \v{C}ech cohomology are given by local sections on pairwise intersections of the $U_i$ which form the affine open cover of $dP_3$. These pairwise intersections and the corresponding polytopes are as follows:
\begin{center}
\begin{tabular}{c|ccc}
\toprule
Intersection & Cone & $P_D( U_{ij} )$ & Presentation \\
\midrule
$U_1 \cap U_2$ & $\text{Span}_{\geq 0} ( u_3 )$ & $\{ m \in \mathbb{Z}^2 \, , \, m_1 \geq 1 + m_2 \}$ & $B,C,D,E,F,N$ \\
$U_1 \cap U_3$ & $\text{Span}_{\geq 0} ( 0 )$ & $\mathbb{Z}^2$ \\
$U_1 \cap U_4$ & $\text{Span}_{\geq 0} ( 0 )$ & $\mathbb{Z}^2$ \\
$U_1 \cap U_5$ & $\text{Span}_{\geq 0} ( 0 )$ & $\mathbb{Z}^2$ \\
$U_1 \cap U_6$ & $\text{Span}_{\geq 0} ( u_1 )$ & $\{ m \in \mathbb{Z}^2 \, , \, m_1 \leq 0 \}$ & $A,B,C,I,K,L$ \\
\midrule
$U_2 \cap U_3$ & $\text{Span}_{\geq 0} ( u_5 )$ & $\{ m \in \mathbb{Z}^2 \, , \, m_1 \geq -1 \}$ & $C,D,E,F,G,H,I,L,M,N$ \\
$U_2 \cap U_4$ & $\text{Span}_{\geq 0} ( 0 )$ & $\mathbb{Z}^2$ \\
$U_2 \cap U_5$ & $\text{Span}_{\geq 0} ( 0 )$ & $\mathbb{Z}^2$ \\
$U_2 \cap U_6$ & $\text{Span}_{\geq 0} ( 0 )$ & $\mathbb{Z}^2$ \\
\midrule
$U_3 \cap U_4$ & $\text{Span}_{\geq 0} ( u_6 )$ & $\{ m \in \mathbb{Z}^2 \, , \, m_2 \geq 0 \}$ & $A,B,C,D,E,L,$M\\
$U_3 \cap U_5$ & $\text{Span}_{\geq 0} ( 0 )$ & $\mathbb{Z}^2$ \\
$U_3 \cap U_6$ & $\text{Span}_{\geq 0} ( 0 )$ & $\mathbb{Z}^2$ \\
\midrule
$U_4 \cap U_5$ & $\text{Span}_{\geq 0} ( u_4 )$ & $\{ m \in \mathbb{Z}^2 \, , \, m_2 \geq m_1 \}$ & $A,G,H,I,K,L$ \\
$U_4 \cap U_6$ & $\text{Span}_{\geq 0} ( 0 )$ & $\mathbb{Z}^2$ \\
\midrule
$U_5 \cap U_6$ & $\text{Span}_{\geq 0} ( u_2 )$ & $\{ m \in \mathbb{Z}^2 \, , \, m_1 \leq 2 \}$ & $A,B,C,D,H,I,K,L,M,N$ \\
\bottomrule
\end{tabular}
\end{center}
In this table, we have use the following geometric loci to express the polytopes in question:
\begin{center}
\begin{tikzpicture}[scale=0.5]

    \draw [thick,->] (0,-8)--(0,8) node (yaxis) [above] {};
    \draw [thick,->] (-8,0)--(8,0) node (xaxis) [right] {};

	\draw { (2,0)--(2,1)--(8,7)--(8,0)--(2,0) } [opacity = 0.5, fill=green];
	\draw { (2,0)--(8,0)--(8,-8)--(2,-8)--(2,0) } [opacity = 0.5, fill=red];
	\draw { (2,0)--(1,0)--(0,-1)--(0,-8)--(2,-8)--(2,0) } [opacity = 0.5, fill=yellow];
	\draw { (1,0)--(2,0)--(2,1)--(1,0) } [opacity = 0.5, fill=blue];
	\draw { (0,-1)--(-1,-2)--(-1,-8)--(0,-8)--(0,-1) } [opacity = 0.5, fill=cyan];
	\draw { (0,0)--(1,0)--(0,-1)--(0,0) } [opacity = 0.5, fill=purple];
	\draw { (-1,-2)--(-1,-8)--(-7,-8)--(-1,-2) } [opacity = 0.5, fill=red];
	\draw { (-8,-8)--(-1,-1)--(-1,0)--(-8,0)--(-8,-8) } [opacity = 0.5, fill=yellow];
	\draw { (-1,-1)--(0,0)--(-1,0)--(-1,-1) } [opacity = 0.5, fill=green];
	\draw { (0,0)--(0,8)--(-1,8)--(-1,0)--(0,0) } [opacity = 0.5, fill=blue];
	\draw { (-1,8)--(-8,8)--(-8,0)--(-1,0)--(-1,8) } [opacity = 0.5, fill=purple];
	\draw { (0,0)--(2,2)--(2,8)--(0,8)--(0,0) } [opacity = 0.5, fill=cyan];
	\draw { (2,2)--(2,8)--(8,8)--(2,2) } [opacity = 0.5, fill=yellow];

    \node at (-6,-5) {$A$};
    \node at (-2.5,-5) {$B$};
    \node at (-0.5,-5) {$C$};
    \node at (1,-5) {$D$};
    \node at (5,-5) {$E$};
    \node at (5,2) {$F$};
    \node at (5,6) {$G$};
    \node at (1,6) {$H$};
    \node at (-0.5,6) {$I$};
    \node at (-6,6) {$K$};
    \node at (-0.5,-0.2) {$L$};
    \node at (0.5,-0.2) {$M$};
    \node at (1.5,0.2) {$N$};

\end{tikzpicture}
\end{center}
To identify a basis of $\mathrm{ker} ( \delta_1 )$, we look at the corresponding mapping matrix
\begin{align}
M_{\delta_1} = \begin{psmallmatrix}
-1 & 1 & 0 & 0 & 0 & -1 & 0 & 0 & 0 & 0 & 0 & 0 & 0 & 0 & 0  \\
-1 & 0 & 1 & 0 & 0 & 0 & -1 & 0 & 0 & 0 & 0 & 0 & 0 & 0 & 0  \\ 
-1 & 0 & 0 & 1 & 0 & 0 & 0 & -1 & 0 & 0 & 0 & 0 & 0 & 0 & 0  \\
-1 & 0 & 0 & 0 & 1 & 0 & 0 & 0 & -1 & 0 & 0 & 0 & 0 & 0 & 0  \\
 0 & -1 & 1 & 0 & 0 & 0 & 0 & 0 & 0 & -1 & 0 & 0 & 0 & 0 & 0 \\
 0 & -1 & 0 & 1 & 0 & 0 & 0 & 0 & 0 & 0 & -1 & 0 & 0 & 0 & 0 \\
 0 & -1 & 0 & 0 & 1 & 0 & 0 & 0 & 0 & 0 & 0 & -1 & 0 & 0 & 0 \\
 0 & 0 & -1 & 1 & 0 & 0 & 0 & 0 & 0 & 0 & 0 & 0 & -1 & 0 & 0 \\
 0 & 0 & -1 & 0 & 1 & 0 & 0 & 0 & 0 & 0 & 0 & 0 & 0 & -1 & 0 \\
 0 & 0 & 0 & -1 & 1 & 0 & 0 & 0 & 0 & 0 & 0 & 0 & 0 & 0 & -1 \\
 0 & 0 & 0 & 0 & 0 & -1 & 1 & 0 & 0 & -1 & 0 & 0 & 0 & 0 & 0 \\
0 & 0 & 0 & 0 & 0 & -1 & 0 & 1 & 0 & 0 & -1 & 0 & 0 & 0 & 0 \\
0 & 0 & 0 & 0 & 0 & -1 & 0 & 0 & 1 & 0 & 0 & -1 & 0 & 0 & 0 \\
0 & 0 & 0 & 0 & 0 & 0 & -1 & 1 & 0 & 0 & 0 & 0 & -1 & 0 & 0 \\
0 & 0 & 0 & 0 & 0 & 0 & -1 & 0 & 1 & 0 & 0 & 0 & 0 & -1 & 0 \\
0 & 0 & 0 & 0 & 0 & 0 & 0 & -1 & 1 & 0 & 0 & 0 & 0 & 0 & -1 \\
0 & 0 & 0 & 0 & 0 & 0 & 0 & 0 & 0 & -1 & 1 & 0 & -1 & 0 & 0 \\
0 & 0 & 0 & 0 & 0 & 0 & 0 & 0 & 0 & -1 & 0 & 1 & 0 & -1 & 0 \\
0 & 0 & 0 & 0 & 0 & 0 & 0 & 0 & 0 & 0 & -1 & 1 & 0 & 0 & -1 \\
0 & 0 & 0 & 0 & 0 & 0 & 0 & 0 & 0 & 0 & 0 & 0 & -1 & 1 & -1 \\
\end{psmallmatrix} \, .
\end{align}
Let us introduce the points
\begin{align}
\begin{split}
p_2 = (2,1) \, , \qquad p_9 = (-1,-1) \, . \label{equ:Points}
\end{split}
\end{align}
The corresponding Laurent monomials, once multiplied by $x^a \equiv \prod_{j = 1}^{6}{x_j^{a_j}}$, are $\frac{x_5^3 x_6}{x_1 x_4}$, $\frac{x_1 x_2^3}{x_3 x_6}$, i.e., exactly those rationoms which \emph{cohomCalg} identified in 
\cref{equ:BasisForCohomologies-Example3} as basis of the cohomology:
\begin{align}
H^1( D_L ) &\cong \text{Span}_{\mathbb{C}} \left\{ \frac{x_5^3 x_6}{x_1 x_4}, \frac{x_1 x_2^3}{x_3 x_6} \right\} \, .
\end{align}
However, here we can make this isomorphism explicit. In an abuse of terminology let $p_2$, $p_9$ denote their Laurent monomials. Then it is readily verified that the following \v{C}ech 1-cocycles furnish a basis of $\mathrm{ker} \left( \delta_1 \right)$:
\begin{align}
\left( 0, - p_2, - p_2, - p_2, 0, -p_2, -p_2, -p_2, 0, 0, 0, p_2, 0, p_2, p_2 \right) \cong \frac{x_5^3 x_6}{x_1 x_4} \, , \\
\left( 0, - p_9, - p_9, - p_9, 0, -p_9, -p_9, -p_9, 0, 0, 0, p_9, 0, p_9, p_9 \right) \cong \frac{x_1 x_2^3}{x_3 x_6} \, . \label{equ:C1-For-Cohomologies-In-Example3}
\end{align}

\paragraph{\v{C}ech 1-cocycles  of \boldmath{${D_L-D_C}$}}

Finally, let us identify $\check{C}^1 ( \mathcal{U}, \mathcal{O}_{dP_3} ( D_L - D_C ) )$.
We have
\begin{align}
D_L - D_C = (-3;3,-1,0) = 3 V( x_2 ) - 4 V( x_3 ) - 3 V( x_5 ) - 3 V( x_6 ) \, .
\end{align}
Thus $a_1 = a_4 = 0$, $a_2 = 3$, $a_3 = -4$ and $a_5 = a_6 = -3$. The points associated to the Laurent monomials identified by \emph{cohomCalg} in \cref{equ:H1ForDL-DCByCohomCalg} are:
\begin{align}
\begin{split}
\frac{1}{x_3 x_4^3 x_6^3} = \frac{x_2^3}{x_3^4 x_5^3 x_6^3} \cdot \frac{x_3^3 x_5^3}{x_2^3 x_4^3} &\qquad \leftrightarrow \qquad q_1 = (3,0) \, , \\
\frac{1}{x_1 x_3^2 x_4^2 x_6^2} = \frac{x_2^3}{x_3^4 x_5^3 x_6^3} \cdot \frac{x_3^2 x_5^3 x_6}{x_2^3 x_4^2 x_1} &\qquad \leftrightarrow \qquad q_2 = (3,1) \, , \\
\frac{1}{x_1^2 x_3^3 x_4 x_6} = \frac{x_2^3}{x_3^4 x_5^3 x_6^3} \cdot \frac{x_3 x_5^3 x_6^2}{x_4 x_2^3 x_1^2} &\qquad \leftrightarrow \qquad q_3 = (3,2) \, .
\end{split}
\end{align}
The relevant pairwise intersection and polytopes are as follows:
\begin{center}
\begin{tabular}{c|ccc}
\toprule
Intersection & Cone & $P_D( U_{ij} )$ & Points contained \\
\midrule
$U_1 \cap U_2$ & $\text{Span}_{\geq 0} ( u_3 )$ & $\{ m \in \mathbb{Z}^2 \, , \, m_1 - m_2 \geq 4 \}$ & $\emptyset$ \\
$U_1 \cap U_3$ & $\text{Span}_{\geq 0} ( 0 )$ & $\mathbb{Z}^2$ & $q_1, q_2, q_3$ \\
$U_1 \cap U_4$ & $\text{Span}_{\geq 0} ( 0 )$ & $\mathbb{Z}^2$ & $q_1, q_2, q_3$ \\
$U_1 \cap U_5$ & $\text{Span}_{\geq 0} ( 0 )$ & $\mathbb{Z}^2$ & $q_1, q_2, q_3$ \\
$U_1 \cap U_6$ & $\text{Span}_{\geq 0} ( u_1 )$ & $\{ m \in \mathbb{Z}^2 \, , \, - m_2 \geq 0 \}$ & $q_1$ \\
\midrule
$U_2 \cap U_3$ & $\text{Span}_{\geq 0} ( u_5 )$ & $\{ m \in \mathbb{Z}^2 \, , \, m_1 \geq 3 \}$ & $q_1, q_2, q_3$  \\
$U_2 \cap U_4$ & $\text{Span}_{\geq 0} ( 0 )$ & $\mathbb{Z}^2$ & $q_1, q_2, q_3$ \\
$U_2 \cap U_5$ & $\text{Span}_{\geq 0} ( 0 )$ & $\mathbb{Z}^2$ & $q_1, q_2, q_3$ \\
$U_2 \cap U_6$ & $\text{Span}_{\geq 0} ( 0 )$ & $\mathbb{Z}^2$ & $q_1, q_2, q_3$ \\
\midrule
$U_3 \cap U_4$ & $\text{Span}_{\geq 0} ( u_6 )$ & $\{ m \in \mathbb{Z}^2 \, , \, m_2 \geq 3 \}$ & $\emptyset$ \\
$U_3 \cap U_5$ & $\text{Span}_{\geq 0} ( 0 )$ & $\mathbb{Z}^2$ & $q_1, q_2, q_3$ \\
$U_3 \cap U_6$ & $\text{Span}_{\geq 0} ( 0 )$ & $\mathbb{Z}^2$ & $q_1, q_2, q_3$ \\
\midrule
$U_4 \cap U_5$ & $\text{Span}_{\geq 0} ( u_4 )$ & $\{ m \in \mathbb{Z}^2 \, , \, - m_1 + m_2 \geq 0 \}$ & $\emptyset$ \\
$U_4 \cap U_6$ & $\text{Span}_{\geq 0} ( 0 )$ & $\mathbb{Z}^2$ & $q_1, q_2, q_3$ \\
\midrule
$U_5 \cap U_6$ & $\text{Span}_{\geq 0} ( u_2 )$ & $\{ m \in \mathbb{Z}^2 \, , \, - m_1 \leq -3 \}$ & $q_1, q_2, q_3$ \\
\bottomrule
\end{tabular}
\end{center}
It is not hard to verify that $\mathrm{ker} \left( \delta_1 \right) = \mathrm{Span}_{\mathbb{Z}} \left\{ b_1, b_2, b_3 \right\}$ where
\begin{align}
\begin{split}
b_1 = \left( 0, q_1, q_1, q_1, 0, q_1, q_1, q_1, 0,0,0,-q_1,0,-q_1,-q_1 \right) \, , \\
b_2 = \left( 0, q_2, q_2, q_2, 0, q_2, q_2, q_2, 0,0,0,-q_2,0,-q_2,-q_2 \right) \, , \\
b_3 = \left( 0, q_3, q_3, q_3, 0, q_3, q_3, q_3, 0,0,0,-q_3,0,-q_3,-q_3 \right) \, .
\end{split}
\end{align}

\paragraph{Images of \boldmath{$b_1$}, \boldmath{$b_2$}, \boldmath{$b_3$} in \boldmath{$\check{C}^1( \mathcal{U}, D_L )$}}

The mapping between the \v{C}ech cocycles happens through the following mapping of complexes
\begin{equation}
\begin{tikzpicture}[baseline=(current bounding box.center)]

          \def\w{4.5};
          \def\h{1.8};

          \node (A) at (0.5*\w,0) {$0$};
          \node (B) at (\w,0) {$\check{C}^0( \mathcal{U}, D_L - D_C )$};
          \node (C) at (2*\w,0) {$\check{C}^1( \mathcal{U}, D_L - D_C )$};
          \node (D) at (3*\w,0) {$\check{C}^2( \mathcal{U}, D_L - D_C )$};
          \node (E) at (3.5*\w,0) {$\cdots$};
          \node (A2) at (0.5*\w,-\h) {$0$};
          \node (B2) at (\w,-\h) {$\check{C}^0( \mathcal{U}, D_L )$};
          \node (C2) at (2*\w,-\h) {$\check{C}^1( \mathcal{U}, D_L )$};
          \node (D2) at (3*\w,-\h) {$\check{C}^2( \mathcal{U}, D_L )$};
          \node (E2) at (3.5*\w,-\h) {$\cdots$};
          
          \draw[->,thick] (A) -- (B);
          \draw[->,thick] (B) --node[above] {$\delta_0$} (C);
          \draw[->,thick] (C) --node[above] {$\delta_1$} (D);
          \draw[->,thick] (D) -- (E);
          \draw[->,thick] (A2) -- (B2);
          \draw[->,thick] (B2) --node[above] {$\delta_0$} (C2);
          \draw[->,thick] (C2) --node[above] {$\delta_1$} (D2);
          \draw[->,thick] (D2) -- (E2);
          
          \draw[->,thick] (B) --node[left] {$\cdot P( \mathbf{c} )$} (B2);
          \draw[->,thick] (C) --node[left] {$\cdot P( \mathbf{c} )$} (C2);
          \draw[->,thick] (D) --node[left] {$\cdot P( \mathbf{c} )$} (D2);
          
\end{tikzpicture}
\end{equation}
where $P( \mathbf{c} )$ is the global section of $D_C$ in \cref{equ:DefiningEquation2}. From this it is now readily verified, that the terms omitted on the RHS of \cref{equ:LazyMultiplication} correspond to elements of ${\check{C}^1( \mathcal{U}, D_L )}$ of the form
\begin{align}
\varphi_i = \left( 0, r_i, r_i, r_i, 0, r_i, r_i, r_i, 0,0,0,-r_i,0,-r_i,-r_i \right) \, ,
\end{align}
where $r_i$ is the Laurent monomial associated --- upon multiplication by $x^a = \frac{x_2^2 x_5}{x_3}$ --- to
\begin{align}
\begin{tabular}{llll}
$r_1 = (-1,-3)$, & $r_2 = (-1,-2)$, & $r_3 = (2,-1)$, & $r_4 = (-1,0)$, \\
$r_5 = (2,0)$, & $r_6 = (-1,1)$, & $r_7 = (1,1)$.
\end{tabular}
\end{align}
From this we can verify that $\varphi_i = \delta_0 ( \mu_i )$ for $\mu_i \in \check{C}^0( \mathcal{U}, D_L )$ as follows:
\begin{center}
\begin{tabular}{cc}
\toprule
$\varphi_i$ & $\mu_i$ \\
\midrule
$\varphi_1$ & $(r_1, r_1, 0, 0, 0, r_1 )$ \\
$\varphi_2$ & $(r_2, r_2, 0, 0, 0, r_2 )$ \\
$\varphi_3$ & $(-r_3, -r_3, 0, 0, 0, -r_3 )$ \\
$\varphi_4$ & $(0, 0, r_4, r_4, r_4, 0 )$ \\
\midrule
$\varphi_5$ & $(-r_5, -r_5, 0, 0, 0, -r_5 )$ \\
$\varphi_6$ & $(0, 0, r_6, r_6, r_6, 0 )$ \\
$\varphi_7$ & $(0, 0, r_7, r_7, r_7, 0 )$ \\
\bottomrule
\end{tabular}
\end{center}
Hence, we conclude
\begin{align}
\begin{split}
\varphi \left( b_1 \right) \cong c_3 \frac{x_1 x_2^3}{x_3 x_6} \, , \quad
\varphi \left( b_2 \right) \cong c_2 \frac{x_1 x_2^3}{x_3 x_6} + c_{12} \frac{x_5^3 x_6}{x_1 x_4} \, , \quad
\varphi \left( b_3 \right) \cong c_1 \frac{x_1 x_2^3}{x_3 x_6} + c_{11} \frac{x_5^3 x_6}{x_1 x_4} \, .
\end{split}
\end{align}
This justifies our analysis based on the matrix in \cref{equ:ExplicitMatrixVerification}.

\subsubsection{Application to GUT-example}

In the example discussed in \cref{sec:F-theoryModel} we consider $D_C = (10;-3, -3, -4)$ and $D_L = (5;-4,-4,3)$. This curve $C_{\mathbf{5}_3}$ is cut-out by the following polynomial $a_{3,2}$:
\begin{align}\label{eq:explicit_form_a32}
     \begin{split}
      a_{3,2} &=c_{44} x_1^6 x_2^7 x_3^3 x_4^4+c_{43} x_1^6 x_2^6 x_3^4 x_4^3 x_5+c_{42} x_1^6 x_2^5 x_3^5 x_4^2 x_5^2+c_{41} x_1^6 x_2^4 x_3^6 x_4 x_5^3+c_{40} x_1^6 x_2^3 x_3^7 x_5^4\\
      &+c_{39} x_1^5 x_2^7 x_3^2 x_4^5 x_6+c_{38} x_1^5 x_2^6 x_3^3 x_4^4 x_5 x_6+c_{37} x_1^5 x_2^5 x_3^4 x_4^3 x_5^2 x_6+c_{36} x_1^5 x_2^4 x_3^5 x_4^2 x_5^3 x_6\\
      &+c_{35} x_1^5 x_2^3 x_3^6 x_4 x_5^4 x_6+c_{34} x_1^5 x_2^2 x_3^7 x_5^5 x_6+c_{33} x_1^4 x_2^7 x_3 x_4^6 x_6^2+c_{32} x_1^4 x_2^6 x_3^2 x_4^5 x_5 x_6^2\\
      &+c_{31} x_1^4 x_2^5 x_3^3 x_4^4 x_5^2 x_6^2+c_{30} x_1^4 x_2^4 x_3^4 x_4^3 x_5^3 x_6^2+c_{29} x_1^4 x_2^3 x_3^5 x_4^2 x_5^4 x_6^2+c_{28} x_1^4 x_2^2 x_3^6 x_4 x_5^5 x_6^2\\
      &+c_{27} x_1^4 x_2 x_3^7 x_5^6 x_6^2+c_{26} x_1^3 x_2^7 x_4^7 x_6^3+c_{25} x_1^3 x_2^6 x_3 x_4^6 x_5 x_6^3+c_{24} x_1^3 x_2^5 x_3^2 x_4^5 x_5^2 x_6^3\\
      &+c_{23} x_1^3 x_2^4 x_3^3 x_4^4 x_5^3 x_6^3+c_{22} x_1^3 x_2^3 x_3^4 x_4^3 x_5^4 x_6^3+c_{21} x_1^3 x_2^2 x_3^5 x_4^2 x_5^5 x_6^3+c_{20} x_1^3 x_2 x_3^6 x_4 x_5^6 x_6^3\\
      &+c_{19} x_1^3 x_3^7 x_5^7 x_6^3+c_{18} x_1^2 x_2^6 x_4^7 x_5 x_6^4+c_{17} x_1^2 x_2^5 x_3 x_4^6 x_5^2 x_6^4+c_{16} x_1^2 x_2^4 x_3^2 x_4^5 x_5^3 x_6^4\\
      &+c_{15} x_1^2 x_2^3 x_3^3 x_4^4 x_5^4 x_6^4+c_{14} x_1^2 x_2^2 x_3^4 x_4^3 x_5^5 x_6^4+c_{13} x_1^2 x_2 x_3^5 x_4^2 x_5^6 x_6^4+c_{12} x_1^2 x_3^6 x_4 x_5^7 x_6^4\\
      &+c_{11} x_1 x_2^5 x_4^7 x_5^2 x_6^5+c_{10} x_1 x_2^4 x_3 x_4^6 x_5^3 x_6^5+c_9 x_1 x_2^3 x_3^2 x_4^5 x_5^4 x_6^5+c_8 x_1 x_2^2 x_3^3 x_4^4 x_5^5 x_6^5\\
      &+c_7 x_1 x_2 x_3^4 x_4^3 x_5^6 x_6^5+c_6 x_1 x_3^5 x_4^2 x_5^7 x_6^5+c_5 x_2^4 x_4^7 x_5^3 x_6^6+c_4 x_2^3 x_3 x_4^6 x_5^4 x_6^6\\
      &+c_3 x_2^2 x_3^2 x_4^5 x_5^5 x_6^6+c_2 x_2 x_3^3 x_4^4 x_5^6 x_6^6+c_1 x_3^4 x_4^3 x_5^7 x_6^6
     \end{split}
\end{align}
Hence, the Koszul resolution of the line bundle ${\cal L} = \left. \mathcal{O}_{dP_3} \left( D_L \right) \right|_{C_{{\bf 5}_3}}$ is given by
\begin{align}
0 &\to \mathcal{O}_{dP_3} \left( D_L - D_C \right) \xrightarrow{\phi} \mathcal{O}_{dP_3} \left( D_L \right) \to \mathcal{L} \to 0 \, ,
\end{align}
and the map $\phi$ is induced from multiplication with $a_{3,2}$. The associated long exact sequence in sheaf cohomology is then:
\begin{equation}
\begin{tikzpicture}[scale=2, baseline=(current  bounding  box.center)]
          \matrix(m)[matrix of math nodes,column sep=15pt,row sep=15pt]{
                0 & 0
                  & H^0 \left( dP_3, D_L \right) \cong \mathbb{C}^{4}
                  & H^0 \left( D_C, \mathcal{L} \right) \\
                  & H^1 \left( dP_3, D_L - D_C \right) \cong \mathbb{C}^{4}
                  & H^1 \left( dP_3, D_L \right) \cong \mathbb{C}^6
                  & H^1 \left( C_{{\bf 5}_3}, \mathcal{L} \right) \\
                  & 0
                  & 0
                  & 0 & 0 \, . \\
                     };
               \draw[->,font=\scriptsize,every node/.style={above},rounded corners]
                     (m-1-1) edge (m-1-2)
                     (m-1-2) edge (m-1-3)
                     (m-1-3) edge (m-1-4)
                     (m-1-4.east) --+(5pt,0)|-+(0,-7.5pt)-|([xshift=-5pt]m-2-2.west)--(m-2-2.west);
                \draw[->,font=\scriptsize,every node/.style={above},rounded corners]
                     (m-3-2) edge (m-3-3)
                     (m-3-3) edge (m-3-4)
                     (m-3-4) edge (m-3-5);
                \draw[->] (m-2-2) edge node [above] {$\varphi$} (m-2-3);
                \draw[->] (m-2-3) edge node [above] {} (m-2-4);
                \draw[->,font=\scriptsize,every node/.style={above},rounded corners] (m-2-4.east) --+(5pt,0)|-+(0,-7.5pt)-|([xshift=-5pt]m-3-2.west)-- (m-3-2.west);
\end{tikzpicture}
\end{equation}
By exactness of this sequence, we have $h^1( C_{\mathbf{5}_3}, \mathcal{L} ) = 6 - \mathrm{rk} ( M_{\varphi} )$, where the mapping matrix $M_\varphi$ is determined by the coefficients of $a_{3,2}$:
\begin{equation}
\resizebox{0.9\textwidth}{!}{$
M_\varphi = \left(
\begin{array}{ccccccccccccccccc}
  0 & c_1 & 0 & 0 &c_2  &c_3  &c_4  &c_5& 0 &0  &0  &0& 0 &0& 0&  0&  0\\
  c_5 &0  &0  &0  &c_1& c_2&  c_3 &c_4  &0  &0  &0  &0  &0  &0  &0  &0& 0\\
  c_{11}  &c_6& 0 &0& c_7&  c_8&  c_9&  c_{10}& c_1&  c_2&  c_3&  c_4&  c_5&  0&  0&  0 &0\\
  0 &0  &c_{39}&  c_{34}  &0  &0  &0  &0  &c_{40} &c_{41}&  c_{42}  &c_{43}&  c_{44}& c_{35}  &c_{36}&  c_{37}& c_{38}\\
  0 &0  &c_{44} &0& 0 &0  &0  &0  &0  &0  &0  &0  &0  &c_{40} &c_{41}&  c_{42}& c_{43}\\
  0 &0  &0  &c_{40} &0  &0  &0  &0  &0  &0  &0  &0  &0  &c_{41} &c_{42} &c_{43} &c_{44}
\end{array} \right) \, .$}
\end{equation}
Some linear algebra yields that the rank of this map drops by one, if
\begin{align}
  \begin{split}
  &c_1=c_2=c_3 = c_4= c_5 =c_7=c_8=c_9 = c_{10}=c_{35}=c_{36}=c_{37}=c_{38}=1\\
  &c_{40}=c_{41}=c_{42}=c_{43}=c_{44}=1, \quad c_{11}=c_{34}=-1,\quad   c_6 = c_{39}=2 \, .
\end{split}
\end{align}
One can easily verify that the polynomial \eqref{eq:explicit_form_a32} does not factorize for generic other coefficients not tuned above. Hence the curve $C_{{\bf 5}_3}$ remains irreducible. By applying \emph{sagemath} \cite{sagemath}, one can further justify the smoothness of $C_{{\bf 5}_3}$. Therefore, this tuning condition leads to one additional section without topology change for  $C_{{\bf 5}_3}$. This is an example of jump from Brill--Noether theory.

\subsection{The fat point}

Finally, in our analysis, non-reduced curves feature prominently. Consequently, a basic understanding of such curves is required. Let us therefore briefly discuss the mother of all non-reduced varieties, the \emph{fat point}. This is an example in non-compact affine space $\mathbb{C}^2$ with coordinates $x,y$. Most of this intuition carries over to compact curves. More details can for example be found in \cite{Griffiths:433962, hartshorne1977algebraic}.

Let us consider $V( x ) \subseteq \mathbb{C}^2$. This is the complex (non-compact) curve with coordinate $y$. The difference between $V( x )$ and $V( x^2 )$ is not the collection of points, which these vanishing sets contain, but rather the allowed functions on these spaces. Namely, recall that in the modern language of algebraic geometry, a scheme (or equivalently in the analytic regime --- a \emph{geometric space}) is a pair of a topological space and a structure sheaf. The difference between $V( x )$ and $V( x^2 )$ is this very structure sheaf.

In staying within the regime of algebraic geometry, the structure sheaf of $\mathbb{C}^2$ is given by (the sheafification of) the total coordinate ring $\mathbb{C} [ x,y ]$ --- the ring of all polynomials in the variables $x$ and $y$. Likewise, we can understand the structure sheaf on $V( x )$ from its coordinate ring:
\begin{align}
R_{V( x )} = \mathbb{C} [ x,y ] / \left\langle x \right\rangle = \mathbb{C} [ y ] \, .
\end{align}
Hence, functions on the variety $V( x )$ correspond to polynomials in $y$. How about $V( x^2 )$? On this space it holds
\begin{align}
R_{V( x^2 )} = \mathbb{C} [ x,y ] / \left\langle x^2 \right\rangle = \mathbb{C} [ y ] \oplus \left\langle x \right\rangle \, .
\end{align}
Consequently, on $V( x^2 )$, the polynomial $x$ provides a \emph{non-trivial} function! This is the difference between $V( x )$ and $V( x^2 )$.

We can extend this example slightly by looking at $V( y, x^2 )$. For this space we find
\begin{align}
R_{V( y, x^2 )} = \mathbb{C} [ x,y ] / \left\langle y, x^2 \right\rangle = \left\langle x \right\rangle \, .
\end{align}
Hence, on this point in the affine plane $\mathbb{C}$, the set of non-trivial functions is 1-dimensional and is generated by the polynomial $x$. This lends $V( y, x^2 )$ its name --- as point set it is just a single point, yet this point is large enough to admit non-trivial functions --- it is a \emph{fat point}.

\section{Collection of data}

\subsection{Curve splittings and jumps} \label{sec:DetailsRigidDivisorJumps}
\label{sec:CurveSplittingsAndJumps}

Recall that the six toric $\mathbb{P}^1$s of $dP_3$ correspond to the exceptional divisors $E_1$, $E_2$, $E_3$ and the following three divisors:
\begin{align}
E_4 = H - E_1 - E_2 \, , \qquad E_5 = H - E_1 - E_3 \, , \qquad E_6 = H - E_2 - E_3 \, .
\end{align}

\subsubsection{\texorpdfstring{$\mathbf{D_C = (3;-1,-1,-1)}$}{DC=(3;-1,-1,-1)}}

For this genus-1 curve we find:
\begin{center}
\resizebox{\textwidth}{!}{
\begin{tabular}{cc|cc|cc|cc}
\toprule
bundle & $h^0$-values & $E_1$-splits & $E_2$-splits & $E_3$-splits & $E_4$-splits & $E_5$-splits & $E_6$-splits\\
\midrule
(2, 1, -4, 1) & (4, 5, 6) & (4, 5, 6) & (4, 5, 6) & (4, 5, 6) & (4, 5, 6) & (4, 5, 6) & (4, 5, 6) \\
(1, -3, -3, -2) & (0, 1, 2, 3, 4, 5) & (2, 3, 4, 5) & (2, 3, 4, 5) & (1, 3, 4, 5) & (0, 1, 2, 3, 4, 5) & (0, 1, 2, 3, 4, 5) & (0, 1, 2, 3, 4, 5) \\
(1, -1, -3, 0) & (0, 1, 2, 3) & (0, 1, 2, 3) & (2) & (0, 1, 2, 3) & (1, 2, 3) & (0, 1, 2) & (0, 2, 3) \\
(1, -2, -3, -2) & (0, 1, 2, 3, 4) & (1, 2, 3, 4) & (2, 3, 4) & (1, 2, 3, 4) & (0, 1, 2, 3, 4) & (0, 1, 2, 3, 4) & (0, 1, 2, 3, 4) \\
(1, -1, -3, -1) & (0, 2, 3) & (0, 2, 3) & (2) & (0, 2, 3) & (0, 2, 3) & (0, 2) & (0, 2, 3) \\
\midrule
\multirow{2}{*}{(1, -3, -4, -2)} & (0, 1, 2, 3, 4) & \multirow{2}{*}{(2, 3, 5, 6, 7)} & \multirow{2}{*}{(3, 4, 5, 6)} &\multirow{2}{*}{(1, 3, 4, 6)} & (0, 1, 2, 3, 4) & (0, 1, 2, 3, 4) & (0, 1, 2, 3, 4) \\
 & (5, 6, 7) &  &  & & (5, 6) & (5, 6) & (5, 6, 7) \\
 \midrule
(2, 1, -4, 2) & (5, 6, 7) & (5, 6, 7) & (5, 6, 7) & (6, 7) & (5, 6, 7) & (5, 6, 7) & (5, 6, 7) \\
(2, 2, -4, 2) & (6, 7, 8, 9) & (7, 8, 9) & (6, 7, 8) & (7, 8, 9) & (6, 7, 8, 9) & (6, 7, 8) & (6, 7, 8, 9) \\
(1, -1, -4, -1) & (0, 3, 5) & (0, 3, 5) & (3) & (0, 3, 5) & (0, 3, 5) & (0, 3) & (0, 3, 5) \\
(1, 1, -3, 1) & (2, 3, 4) & (2, 3, 4) & (2, 3, 4) & (2, 3, 4) & (2, 3, 4) & (2, 3, 4) & (2, 3, 4) \\
\midrule
(1, 1, -3, 0) & (1, 2, 3) & (1, 2, 3) & (2, 3) & (1, 2, 3) & (1, 2, 3) & (1, 2, 3) & (2, 3) \\
(1, -1, -2, 0) & (0, 1) & (0, 1) & (1) & (0, 1) & (1) & (0, 1) & (0, 1) \\
(1, 1, -3, 2) & (3, 4, 5) & (3, 4, 5) & (3, 4, 5) & (4, 5) & (3, 4, 5) & (3, 4, 5) & (3, 4, 5) \\
\bottomrule
\end{tabular}}
\end{center}

\subsubsection{\texorpdfstring{$\mathbf{D_C = (4;-1,-2,1)}$}{DC=(4;-1,-2,1)}}

For this (generically disjoint) union of a genus-0 and a genus-2 curve, we find:
\begin{center}
\resizebox{\textwidth}{!}{
\begin{tabular}{cc|cc|cc|cc}
\toprule
bundle & $h^0$-values & $E_1$-splits & $E_2$-splits & $E_3$-splits & $E_4$-splits & $E_5$-splits & $E_6$-splits\\
\midrule
(2, -1, -2, 5) & (2, 5, 7, 8)  & (2, 5, 7, 8) & (2, 5, 7, 8) & (2)  & (2, 5, 7, 8) & (5, 7, 8) & (5, 7, 8) \\
(1, -1, -2, -1) & (2, 3)  & (2, 3) & (2, 3) & (2)  & (2, 3) & (3) & (3) \\
(1, -2, -2, -2) & (3, 4, 5, 6, 7)  & (4, 6, 7) & (3, 4, 5, 6, 7) & (3, 4)  & (3, 4, 5, 6, 7) & (5, 6, 7) & (5, 6, 7) \\
(2, -3, -2, -1) & (2, 3, 4, 5, 6)  & (4, 5) & (2, 3, 4, 5, 6) & (2, 3, 4, 5)  & (4, 5, 6) & (3, 5, 6) & (3, 4, 5, 6) \\
(1, -2, -1, 4) & (0, 1, 2, 3)  & (1, 2, 3) & (0, 1, 2, 3) & (0, 1)  & (1, 2, 3) & (1, 2, 3) & (1, 2, 3) \\
\midrule
\multirow{2}{*}{(1, -2, -2, -3)}  & (4, 5, 7, 8, 9)  & \multirow{2}{*}{(5, 8, 10, 11)} & (4, 5, 7, 8, 9) & \multirow{2}{*}{(4,5)}  & (4, 5, 7, 8, 9) & \multirow{2}{*}{ (7, 8, 9, 10, 11)} &\multirow{2}{*}{ (7, 8, 9, 10, 11)} \\
  &  (10, 11) &  & (10, 11) &  &  (10, 11) &  &   \\
 \midrule
 \multirow{2}{*}{(2, -3, -2, -2)} & (3, 4, 5, 6, 7)  & \multirow{2}{*}{(5, 6, 7, 8)} & \multirow{2}{*}{(3, 5, 6, 7, 8, 9)} & \multirow{2}{*}{(3, 4, 5, 6)}  & \multirow{2}{*}{(5, 6, 7, 8, 9)} & \multirow{2}{*}{(5, 6, 7, 8, 9)} & \multirow{2}{*}{(5, 6, 7, 8, 9)} \\
 & (8, 9) & & & & & & \\
\midrule
(1, -2, 1, -1) & (5, 6)  & (5, 6) & (5, 6) & (5)  & (5, 6) & (6) & (5, 6) \\
(2, -2, -1, -2) & (6, 7)  & (6, 7) & (6, 7) & (6)  & (6, 7) & (7) & (6, 7) \\
\midrule
\multirow{2}{*}{(2, -2, -2, 7)} & (1, 2, 6, 7, 10)  & (2, 6, 7, 10, 11) & (1, 2, 6, 7, 10) & \multirow{2}{*}{(1, 2)}  & \multirow{2}{*}{(2, 7, 11, 14)} & (6, 7, 10, 11, 13) & {(6, 7, 10, 11, 13)} \\
& (11, 13, 14, 15) &  (11, 13, 14, 15) & (13, 14, 15) &   &  & (13, 14, 15) & (14) \\
\midrule
(3, -1, -2, 10) & (6, 14, 21, 27, 32)  & (6, 14, 21, 27, 32) & (6, 14, 21, 27, 32) & (6)  & (6, 14, 21, 27) & (14, 21, 27, 32) & (14, 21, 27) \\
(1, -3, 1, -1) & (4, 5, 6, 7)  & (4, 5, 6, 7) & (4, 5, 6, 7) & (4, 5, 6)  & (4, 5, 6, 7) & (6, 7) & (4, 5, 6, 7) \\
\bottomrule
\end{tabular}}
\end{center}

\subsubsection{\texorpdfstring{$\mathbf{D_C = (4;-1,-2,-1)}$}{DC=(4;-1,-2,-1)}}

For this genus-2 curve we find:
\begin{center}
\resizebox{\textwidth}{!}{
\begin{tabular}{cc|cc|cc|cc}
\toprule
bundle & $h^0$-values & $E_1$-splits & $E_2$-splits & $E_3$-splits & $E_4$-splits & $E_5$-splits & $E_6$-splits\\
\midrule
(2, 3, -3, 1) &  (5, 7, 8) & (7) & (5, 7, 8) & (5, 7, 8) & (5, 7, 8) & (5, 7, 8) & (5, 7, 8) \\
(3, 1, -4, -1) &  (3, 4) & (3, 4) & (3, 4) & (3, 4) & (3, 4) & (3, 4) & (4) \\
(2, 2, -4, 0) &  (1, 2, 3, 4) & (2, 3, 4) & (2, 3, 4) & (1, 2, 3, 4) & (1, 2, 3, 4) & (2, 3, 4) & (2, 3, 4) \\
\midrule
\multirow{2}{*}{(2, 1, -4, -3)} &  (0, 1, 2, 3, 4) & (0, 1, 2, 3, 4) & (2, 3, 4, 5, 6) & (2, 3, 4, 5) & (0, 1, 2, 3, 4) & (0, 1, 2, 3, 4) & \multirow{2}{*}{(1, 2, 3, 4)} \\
 &  (5, 6) & (5) &  &  & (5, 6) & (5, 6) &  \\
\midrule
(1, -1, -3, -2) &  (0, 1, 2) & (0, 1, 2) & (1, 2) & (1, 2) & (0, 1, 2) & (0, 1, 2) & (0, 1, 2) \\
(1, -2, -4, 2) &  (0, 1, 2, 3, 4) & (1, 2, 3, 4) & (2, 3, 4) & (0, 1, 2, 3, 4) & (0, 1, 2, 3) & (0, 1, 2, 3, 4) & (0, 1, 2, 3, 4) \\
\midrule
\multirow{3}{*}{(4, 3, -3, -8)} &  (4, 5, 6, 7, 8) & (6, 7, 8, 9, 10) & (4, 5, 6, 7, 8) & (7, 10, 12, 13, 15) & (4, 5, 6, 7, 8) & (4, 7, 9, 10, 12) & \multirow{3}{*}{(10, 12, 13, 15, 16)} \\
&  (10, 12, 13, 15) &  (12, 13, 15, 16)& ( 9, 10, 12, 13) & ( 17) & (9, 10, 12, 13, 15)  & (13, 15, 16, 18, 19) &  \\
&  (16, 17, 18, 19) &  (17, 18)& (15, 17, 18, 19) &  & ( 16, 17, 18, 19)  &  &  \\
\midrule
\multirow{2}{*}{(1, 3, -4, -5)} &  (0, 1, 2, 4, 6) & (0, 2, 4, 6, 7) & (2, 4, 6, 8, 9) & (4, 6, 8) & (0, 1, 2, 4, 6) & (0, 2, 4, 6, 7) & (0, 1, 2, 4, 6) \\
&  (7, 8, 9, 11) &  (8, 9)& ( 11) &  & ( 7, 8, 9, 11)  & (9,11) & (7) \\
\midrule
\multirow{2}{*}{(3, 1, -4, -5)}  &  (0, 1, 2, 4, 5) & (0, 1, 2, 4, 5) & (2, 4, 6, 8, 9) & (4, 5, 6, 8) & (0, 1, 2, 4, 5) & (0, 1, 2, 4, 5) & (4, 5, 6, 7) \\
&  (6, 7, 8, 9, 11) & (6, 7, 8, 9) & (11) &  & (6, 7, 8, 9, 11) & (6, 7, 8, 9, 11) & \\
\midrule
\multirow{3}{*}{(3, 2, -3, -7)} &  (0, 1, 2, 3, 4) & (1, 2, 3, 4, 6) & (1, 3, 4, 6, 7) & (6, 7, 9, 10, 11) & (1, 2, 3, 4, 6) & (1, 3, 4, 6, 7) & (6, 7, 9, 10, 11) \\
&  (6, 7, 9, 10, 11) &  (7, 9, 10, 11)& ( 7, 9, 10, 11) & ( 12) & ( 7, 9, 10, 11)  & (9, 10, 11, 12) & ( 12) \\
&  (12, 14, 15, 16) &  (12, 14, 15)& (12, 14, 15, 16) &  & (12, 14, 15)  & (14, 15, 16) &  \\
\midrule
\multirow{2}{*}{(3, 2, -3, -5)} &  (2, 3, 4, 5, 6) & (3, 4, 5, 7, 8) & (2, 3, 4, 6, 7) & (4, 5, 6, 7, 8) & (2, 3, 4, 5, 6) & (2, 3, 4, 5, 6) & (6, 7, 8) \\
&  (7, 8, 9, 10, 11) &  (10)& ( 8, 9, 10, 11) &( 9,10)   & ( 7, 8, 9, 10, 11)  & (7, 8, 9, 10, 11) &  \\
\midrule
(1, 1, -4, 2) &  (0, 1, 2, 3, 4) & (0, 1, 2, 3, 4) & (2, 3, 4) & (0, 1, 2, 3, 4) & (0, 1, 2, 3, 4) & (2, 3, 4) & (0, 1, 2, 3, 4) \\
(1, 0, -4, -1) &  (0, 2, 3) & (0, 2, 3) & (2) & (0, 2, 3) & (0, 2, 3) & (0, 2) & (0, 2, 3) \\
(3, -3, -1, -2) & (4, 5) & (4, 5) & (4, 5) & (4, 5) & (4, 5) & (4, 5) & (4, 5) \\
\midrule
\multirow{3}{*}{(4, -7, -1, -3) }& (3, 4, 5, 6, 7) & (3, 4, 5, 6, 7) & (3, 4, 5, 6, 7) & (3, 4, 5, 6, 8) & (3, 4, 5, 6, 7) & (3, 4, 5, 6, 8) & (3, 4, 5, 6, 7) \\
&  ( 8, 10, 11, 12) &  (8, 10, 11, 12)& (8, 10, 11, 13, 15) & ( 10, 11, 12, 13) & ( 8, 11, 13, 15)  & (10, 11, 13, 15) & (10, 11, 12, 13, 15) \\
&  (13, 15, 17) &  &  & (15, 17) &   &  &  \\
\bottomrule
\end{tabular}}
\end{center}

\subsubsection{\texorpdfstring{$\mathbf{D_C = (4;-1,-2,0)}$}{DC=(4;-1,-2,0)}}

For this genus-2 curve we find:
\begin{center}
\resizebox{0.99\textwidth}{!}{
\begin{tabular}{cc|cc|cc|cc}
\toprule
bundle & $h^0$-values & $E_1$-splits & $E_2$-splits & $E_3$-splits & $E_4$-splits & $E_5$-splits & $E_6$-splits\\
\midrule
(1, -2, -1, 4) & (0, 1, 2, 3, 4, 5, 6)  & (1, 2, 3, 4, 5, 6) & (0, 1, 2, 3, 4, 5, 6) & (2, 3)  & (1, 3, 5, 6) & (2, 3, 4, 5, 6) & (2, 3, 4, 5, 6) \\
\bottomrule
\end{tabular}}
\end{center}

\subsubsection{\texorpdfstring{$\mathbf{D_C = (4;-1,-1,-1)}$}{DC=(4;-1,-1,-1)}}

On this genus-3 curve we find:
\label{sec:Genus3Details}
\begin{center}
\resizebox{0.99\textwidth}{!}{
\begin{tabular}{cc|cc|cc|cc}
\toprule
bundle & $h^0$-values & $E_1$-splits & $E_2$-splits & $E_3$-splits & $E_4$-splits & $E_5$-splits & $E_6$-splits\\
\midrule
(1, -2, -3, -1) &  (0, 1, 2, 3, 4) &  (1, 3, 4) &  (2, 3) &  (0, 1, 2, 3, 4) &  (0, 1, 2, 3, 4) &  (0, 1, 2, 3, 4) &  (0, 1, 2, 3, 4) \\
\midrule
\multirow{2}{*}{(1, -3, -4, -3)} &  (0, 2, 3, 4, 5) &  (2, 4, 5, 7, 8) &  (3, 5, 6, 7, 8) &  (2, 4, 5, 7, 8) &  (0, 2, 3, 4, 5) &  (0, 2, 3, 4, 5) &  (0, 2, 3, 4, 5) \\
&  (6, 7, 8, 9, 10) &  (9, 10) &   (9)& (9, 10) &  (6, 7, 8, 9, 10) & (6, 7, 8, 9, 10) &(6, 7, 8, 9, 10)  \\
\midrule
(1, 1, -3, 0) &  (0, 1, 2, 3) &  (0, 1, 2, 3) &  (2) &  (0, 1, 2, 3) &  (1, 2, 3) &  (0, 1, 2, 3) &  (1, 2, 3) \\
\midrule
\multirow{2}{*}{(1, -3, -3, -3)} &  (0, 2, 3, 4, 5) &  (2, 4, 5, 6, 7) &  (2, 4, 5, 6, 7) &  (2, 4, 5, 6, 7) &  (0, 2, 3, 4, 5) &  (0, 2, 3, 4, 5) &  (0, 2, 3, 4, 5) \\
&  (6, 7, 8) &  (8) &   (8)& (8) &  (6, 7, 8) & (6, 7, 8) &(6, 7, 8)  \\
\midrule
\multirow{2}{*}{(1, -3, -2, -3)} &  (0, 1, 2, 3, 4) &  (2, 3, 4, 5, 6) &  (1, 3, 4, 5, 6) &  (2, 3, 4, 5, 6) &  (0, 1, 2, 3, 4) &  (0, 1, 2, 3, 4) &  (0, 1, 2, 3, 4) \\
&  (5, 6, 7) &  & (7)& &  (5, 6, 7) & (5, 6) &(5, 6, 7)  \\
\midrule
(1, 2, -2, -1) &  (1, 2, 3) &  (2, 3) &  (1, 2, 3) &  (1, 2, 3) &  (1, 2, 3) &  (1, 2, 3) &  (2, 3) \\
\midrule
\multirow{2}{*}{(1, 1, -3, -3)} &  (0, 1, 2, 3, 4) &  (0, 1, 2, 3, 4) &  (2, 4, 5) &  (2, 4, 5) &  (0, 2, 3, 4, 5) &  (0, 2, 3, 4, 5) &  (1, 2, 3, 4, 5) \\
&  (5, 6) &  (5, 6) &  &  &  (6) & (6) &  \\
\midrule
(2, 3, -4, -1) &  (4, 5, 6, 7, 8, 9) &  (6, 7, 8, 9) &  (4, 5, 6, 7, 8, 9) &  (4, 5, 6, 7, 8, 9) &  (4, 5, 6, 7, 8, 9) &  (4, 5, 6, 7, 8, 9) &  (6, 7, 8, 9) \\
\midrule
\multirow{2}{*}{(1, 2, -4, 2)} &  (2, 3, 4, 5, 6) &  (3, 4, 5, 6, 7) &  (3, 4, 5, 6, 7) &  (3, 4, 5, 6, 7) &  (2, 3, 4, 5, 6) &  (3, 4, 5, 6, 7) &  (2, 3, 4, 5, 6) \\
&  (7, 8) &  (8) & (8) & (8) &  (7, 8) & (8) &   (7, 8)\\
\midrule
(1, -2, -3, -2) &  (0, 1, 2, 3, 4, 5) &  (1, 2, 3, 4, 5) &  (2, 3, 4) &  (1, 2, 3, 4, 5) &  (0, 1, 2, 3, 4, 5) &  (0, 1, 2, 3, 4, 5) &  (0, 1, 2, 3, 4, 5) \\
(1, 3, -3, 1) &  (3, 4, 5, 6, 7) &  (5, 6, 7) &  (3, 4, 5, 6, 7) &  (3, 4, 5, 6, 7) &  (3, 4, 5, 6, 7) &  (3, 4, 5, 6, 7) &  (3, 4, 5, 6, 7) \\
(1, -1, -3, 0) &  (0, 1, 2, 3) &  (0, 1, 2, 3) &  (2) &  (0, 1, 2, 3) &  (1, 2, 3) &  (0, 1, 2, 3) &  (0, 2, 3) \\
\bottomrule
\end{tabular}}
\end{center}

\subsubsection{\texorpdfstring{$\mathbf{D_C = (5;-2,-2,-1)}$}{DC=(5;-2,-2,-1)}}

On this genus-4 curve we find:
\begin{center}
\resizebox{\textwidth}{!}{
\begin{tabular}{cc|cc|cc|cc}
\toprule
bundle & $h^0$-values & $E_1$-splits & $E_2$-splits & $E_3$-splits & $E_4$-splits & $E_5$-splits & $E_6$-splits\\
\midrule
(2, -2, -4, -2) &  (0, 1, 2, 3, 4) &  (0, 1, 2, 3, 4) &  (2, 3) & (1, 3, 4) & (0, 1, 2, 3, 4) &  (0, 1, 2, 3, 4) &  (0, 1, 2, 3, 4) \\
(1, -1, -3, 0) &  (0, 1) &  (0, 1) &  (1) & (0, 1) & (0, 1) &  (0, 1) &  (0, 1) \\
(1, 2, -2, 0) &  (2, 3) &  (3) &  (2, 3) & (2, 3) & (2, 3) &  (2, 3) &  (2, 3) \\
(1, 2, -2, 1) &  (3, 4) &  (4) &  (3, 4) & (3, 4) & (3, 4) &  (3, 4) &  (3, 4) \\
(1, 1, -4, -1) &  (0, 2, 3) &  (0, 2, 3) &  (2) & (0, 2, 3) & (0, 2, 3) &  (0, 2, 3) &  (0, 2, 3) \\
(1, -1, -4, -1) &  (0, 2, 3) &  (0, 2, 3) &  (2) & (0, 2, 3) & (0, 2, 3) &  (0, 2, 3) &  (0, 2, 3) \\
(1, -2, -4, 2) &  (0, 2, 3) &  (0, 2, 3) &  (2) & (0, 2, 3) & (0, 2, 3) &  (0, 2, 3) &  (0, 2, 3) \\
(1, 1, -4, 1) &  (0, 1, 2, 3, 4) &  (0, 1, 2, 3, 4) &  (2, 3) & (0, 1, 2, 3, 4) & (0, 1, 2, 3, 4) &  (1, 3, 4) &  (0, 1, 2, 3, 4) \\
(1, -1, -2, 3) &  (0, 1, 2, 3) &  (0, 1, 2, 3) &  (0, 1, 2, 3) & (1, 2) & (0, 1, 2, 3) &  (1, 2, 3) &  (0, 1, 2, 3) \\
(2, -1, -4, 1) &  (0, 1, 2, 3) &  (0, 1, 2, 3) &  (2) & (0, 1, 2, 3) & (0, 1, 2, 3) &  (0, 1, 2, 3) &  (0, 1, 2, 3) \\
(1, 2, -3, 1) &  (1, 2, 3, 4) &  (2, 3, 4) &  (1, 2, 3, 4) & (1, 2, 3, 4) & (1, 2, 3, 4) &  (2, 3, 4) &  (1, 2, 3, 4) \\
(1, 1, -4, 0) &  (0, 2, 3) &  (0, 2, 3) &  (2) & (0, 2, 3) & (0, 2, 3) &  (0, 2, 3) &  (0, 2, 3) \\
(1, 2, -2, -2) &  (0, 1, 2, 3, 4) &  (1, 2, 3, 4) &  (0, 1, 2, 3, 4) & (1, 2, 3, 4) & (0, 1, 2, 3, 4) &  (0, 1, 2, 3, 4) &  (2, 3) \\
\bottomrule
\end{tabular}}
\end{center}

\subsubsection{\texorpdfstring{$\mathbf{D_C = (5;-1,-1,-2)}$}{DC=(5;-1,-1,-2)}}

On this genus-5 curve we find:
\begin{center}
\resizebox{\textwidth}{!}{
\begin{tabular}{cc|cc|cc|cc}
\toprule
bundle & $h^0$-values & $E_1$-splits & $E_2$-splits & $E_3$-splits & $E_4$-splits & $E_5$-splits & $E_6$-splits\\
\midrule
(1, -2, -2, -3) & (0, 1, 2, 3) & (1, 2, 3) & (1, 2, 3) & (1, 2, 3) & (0, 1, 2, 3) & (0, 1, 2, 3) & (0, 1, 2, 3) \\
\midrule
\multirow{2}{*}{(1, 1, -4, 2)} & (2, 3, 4, 5, 6) & (2, 3, 4, 5, 6) & (3, 4, 5, 6, 7) & (3, 4, 5, 6, 7) & (3, 4, 5, 6, 7) & (2, 3, 4, 5, 6) & (2, 3, 4, 5, 6) \\
 & (7, 8) & (7, 8) & & (8) & (8) & (7, 8) & (7, 8) \\
\midrule
\multirow{2}{*}{(1, 1, -4, 1)} & (0, 1, 2, 3, 4) & (0, 1, 2, 3, 4) & (3, 4) & (0, 1, 2, 3, 4) & (1, 2, 3, 4, 5) & (1, 2, 4, 6, 7) & (1, 2, 3, 4, 5) \\
& (5, 6, 7) & (5, 6, 7) & & (5, 6, 7) & (5, 6, 7) & & (6, 7) \\
\midrule
(1, -1, -3, -2) & (0, 2, 3) & (0, 2, 3) & (2) & (0, 2, 3) & (0, 2, 3) & (0, 2, 3) & (0, 2, 3) \\
(1, 1, -3, -1) & (0, 1, 2, 3) & (0, 1, 2, 3) & (2) & (0, 1, 2, 3) & (0, 2, 3) & (0, 1, 2, 3) & (1, 2, 3) \\
(1, 1, -3, -2) & (0, 2, 3) & (0, 2, 3) & (2) & (0, 2, 3) & (0, 2, 3) & (0, 2, 3) & (0, 2, 3) \\
(1, 2, -2, -1) & (0, 1) & (0, 1) & (1) & (0, 1) & (0, 1) & (0, 1) & (1) \\
(1, 1, -4, 0) & (0, 1, 3, 5, 6) & (0, 1, 3, 5, 6) & (3) & (0, 1, 3, 5, 6) & (1, 3, 5, 6) & (0, 1, 3, 5, 6) & (1, 3, 5, 6) \\
(1, -2, -1, -3) & (0, 1, 2) & (1, 2) & (0, 1, 2) & (1, 2) & (0, 1, 2) & (0, 1, 2) & (0, 1, 2) \\
(1, 1, -3, 1) & (1, 2, 3, 4) & (1, 2, 3, 4) & (2, 3) & (1, 2, 3, 4) & (1, 2, 3, 4) & (1, 2, 3, 4) & (1, 2, 3, 4) \\
(1, -1, -2, -2) & (0, 1) & (0, 1) & (1) & (0, 1) & (0, 1) & (0, 1) & (0, 1) \\
(1, -2, -3, -3) & (0, 1, 2, 3, 4, 5) & (1, 2, 3, 4, 5) & (2, 3, 4) & (1, 2, 3, 4, 5) & (0, 1, 2, 3, 4, 5) & (0, 1, 2, 3, 4, 5) & (0, 1, 2, 3, 4, 5) \\
(1, 1, -4, -1) & (0, 1, 3, 5, 6) & (0, 1, 3, 5, 6) & (3) & (0, 1, 3, 5, 6) & (0, 3, 5, 6) & (0, 1, 3, 5, 6) & (1, 3, 5, 6) \\
\bottomrule
\end{tabular}}
\end{center}

\subsubsection{\texorpdfstring{$\mathbf{D_C = (6;-3,-2,-1)}$}{DC=(6;-3,-2,-1)}}

On this genus-6 curve we find:
\begin{center}
\resizebox{\textwidth}{!}{
\begin{tabular}{cc|cc|cc|cc}
\toprule
bundle & $h^0$-values & $E_1$-splits & $E_2$-splits & $E_3$-splits & $E_4$-splits & $E_5$-splits & $E_6$-splits\\
\midrule
(1, 1, -4, 1)  & (0, 1, 2, 3, 4) & (0, 1, 2, 3, 4) & (2, 3) & (0, 1, 2, 3, 4) & (0, 1, 2, 3, 4) & (1, 3, 4) & (0, 1, 2, 3, 4) \\
(1, 0, -3, 1)  & (0, 1) & (0, 1) & (1) & (0, 1) & (0, 1) & (0, 1) & (0, 1) \\
\bottomrule
\end{tabular}}
\end{center}

\subsection{Local to global section counting applied to our database} \label{sec:DetailsApplicationToDatabase}

In this section, we list results which quantify how good the counting procedure proposed in \cref{subsubsec:ProposalForCounting} works, when applied to our database. We have preformed two tests:
\begin{enumerate}
 \item We consider those curves in our data, for which we can quickly identify the exact number of sections on all curve components. This can be done quickly for non-split curves and for curves with only smooth components. For the latter curves, we have read-off the genus $g$ and the line bundle degree $d$ from our database. If $d < 0$, we know that there are no non-trivial sections on this curve component. However, if $d > 2g - 2$, then $h^0 (C, \mathcal{L} ) = d - g + 1$. Based on these exact local section counts, we have then tried to predict the number of global sections. The accuracy for this is listed in \cref{subsubsec:Accuracy}.
 \item Our second test is based on our \emph{H0Approximator}-program \cite{H0Approximator}, which is part of \cite{ToricVarietiesProject}. This program considers curve degeneration, which split-off combinations of the 6 toric $\mathbb{P}^1$s in $dP_3$. For each such curve splitting, the program assumes that the number of local sections on each curve component is generic. Since this generic value is a lower bound to the actual number of local sections, we can use these estimates to derive a lower bound on the number of global sections. By repeating this strategy for many curve splittings, we obtain an estimate for the allowed $h^0$-values over the parameter space of the curve in question. We list the so-obtained results for all pairs $(D_C, D_L)$ in our database \cite{Database} in \cref{subsubsec:SpectrumEstimate}.
\end{enumerate}

\subsubsection{Accuracy} \label{subsubsec:Accuracy}

\begin{center}
\begin{longtable}{cc|cc}
\caption{Accuracy of counting procedure for exact numbers of local sections} \\
\hline
$\mathbf{D_C}$ & $\mathbf{D_L}$ & \textbf{Applicable data sets [\%]} & \textbf{Accuracy [\%]} \\
\hline
\endfirsthead
\multicolumn{4}{c}%
{\tablename\ \thetable\ -- \textit{Continued from previous page}} \\
\hline
$\mathbf{D_C}$ & $\mathbf{D_L}$ & \textbf{Applicable data sets [\%]} & \textbf{Accuracy [\%]} \\
\hline
\endhead
\hline \multicolumn{4}{r}{\textit{Continued on next page}} \\
\endfoot
\hline
\endlastfoot
(3, -1, -1, -1) & (1, 1, -3, 0) & 62.2 & 100 \\
(3, -1, -1, -1) & (1, 1, -3, 1) & 71.6 & 100 \\
(3, -1, -1, -1) & (1, 1, -3, 2) & 52.7 & 100 \\
(3, -1, -1, -1) & (1, -1, -2, 0) & 52.7 & 100 \\
(3, -1, -1, -1) & (1, -1, -3, 0) & 66.9 & 100 \\
(3, -1, -1, -1) & (1, -1, -3, -1) & 76.4 & 100 \\
(3, -1, -1, -1) & (1, -1, -4, -1) & 76.4 & 100 \\
(3, -1, -1, -1) & (1, -2, -3, -2) & 90.5 & 100 \\
(3, -1, -1, -1) & (1, -3, -3, -2) & 90.5 & 100 \\
(3, -1, -1, -1) & (1, -3, -4, -2) & 90.5 & 100 \\
(3, -1, -1, -1) & (2, 1, -4, 1) & 62.2 & 100 \\
(3, -1, -1, -1) & (2, 1, -4, 2) & 48.0 & 100 \\
(3, -1, -1, -1) & (2, 2, -4, 2) & 37.0 & 100 \\
\hline
(4, -1, -2, 1) & (1, -1, -2, 0) & 38.7 & 100 \\
(4, -1, -2, 1) & (1, -1, -2, -1) & 38.7 & 100 \\
(4, -1, -2, 1) & (1, -2, 1, -1) & 26.9 & 100 \\
(4, -1, -2, 1) & (1, -2, -1, 4) & 12.6 & 65.1 \\
(4, -1, -2, 1) & (1, -2, -2, -2) & 43.4 & 100 \\
(4, -1, -2, 1) & (1, -2, -2, -3) & 43.4 & 100 \\
(4, -1, -2, 1) & (1, -3, 1, -1) & 9.2 & 100 \\
(4, -1, -2, 1) & (2, -1, -2, 5) & 4.3 & 100 \\
(4, -1, -2, 1) & (2, -2, -1, -2) & 28.3 & 100 \\
(4, -1, -2, 1) & (2, -2, -2, 7) & 4.4 & 100 \\
(4, -1, -2, 1) & (2, -3, -2, -1) & 12.6 & 100 \\
(4, -1, -2, 1) & (2, -3, -2, -2) & 12.6 & 100 \\
(4, -1, -2, 1) & (3, -1, -2, 10) & 23.9 & 100 \\
\hline
(4, -1, -2, -1) & (1, 0, -4, -1) & 80.4 & 100 \\
(4, -1, -2, -1) & (1, 3, -4, -5) & 83.4 & 99 \\
(4, -1, -2, -1) & (1, -1, -3, -2) & 88.3 & 100 \\
(4, -1, -2, -1) & (1, -2, -4, 2) & 84.2 & 100 \\
(4, -1, -2, -1) & (3, 2, -3, -7) & 71.8 & 100 \\
(4, -1, -2, -1) & (2, 1, -4, -3) & 76.4 & 100 \\
(4, -1, -2, -1) & (2, 2, -4, 0) & 50.6 & 100 \\
(4, -1, -2, -1) & (2, 3, -3, 1) & 44.8 & 100 \\
(4, -1, -2, -1) & (3, 1, -4, -1) & 45.4 & 100 \\
(4, -1, -2, -1) & (3, 1, -4, -5) & 69.4 & 100 \\
(4, -1, -2, -1) & (3, 2, -3, -5) & 54.3 & 100 \\
(4, -1, -2, -1) & (1, 1, -4, 2) & 76.3 & 98.6 \\
(4, -1, -2, -1) & (4, 3, -3, -8) & 60.6 & 100 \\
\hline
(4, -1, -2, -1) & (3, -3, -1, -2) & 66.5 & 98.7 \\
(4, -1, -2, -1) & (4, -7, -1, -3) & 74.1 & 92.6 \\
\hline
(4, -1, -2, 0) & (1, -2, -1, 4) & 58.7 & 92.5 \\
\hline
(4, -1, -1, -1) & (1, 1, -3, 0) & 52.2 & 95.8 \\
(4, -1, -1, -1) & (1, 1, -3, -1) & 56.5 & 100 \\
(4, -1, -1, -1) & (1, 1, -3, -3) & 73.8 & 100 \\
(4, -1, -1, -1) & (1, 2, -2, -1) & 45.6 & 100 \\
(4, -1, -1, -1) & (1, 2, -4, 2) & 65.3 & 100 \\
(4, -1, -1, -1) & (1, 3, -3, 1) & 56.9 & 100 \\
(4, -1, -1, -1) & (1, -1, -3, 0) & 64.3 & 96.6 \\
(4, -1, -1, -1) & (1, -2, -3, -1) & 82.4 & 100 \\
(4, -1, -1, -1) & (1, -2, -3, -2) & 87.5 & 100 \\
(4, -1, -1, -1) & (1, -3, -2, -3) & 85.8 & 100 \\
(4, -1, -1, -1) & (1, -3, -3, -3) & 84.0 & 100 \\
(4, -1, -1, -1) & (1, -3, -4, -3) & 86.2 & 100 \\
(4, -1, -1, -1) & (2, 3, -4, -1) & 45.5 & 100 \\
\hline
(5, -2, -2, -1) & (1, 1, -4, 0) & 62.0 & 100 \\
(5, -2, -2, -1) & (1, 1, -4, 1) & 58.4 & 99.7 \\
(5, -2, -2, -1) & (1, 1, -4, -1) & 67.9 & 100 \\
(5, -2, -2, -1) & (1, 2, -2, 0) & 45.2 & 100 \\
(5, -2, -2, -1) & (1, 2, -2, 1) & 50.8 & 100 \\
(5, -2, -2, -1) & (1, 2, -2, -2) & 46.7 & 98.9 \\
(5, -2, -2, -1) & (1, 2, -3, 1) & 48.9 & 99.0 \\
(5, -2, -2, -1) & (1, -1, -2, 3) & 45.1 & 99.9 \\
(5, -2, -2, -1) & (1, -1, -3, 0) & 72.7 & 100 \\
(5, -2, -2, -1) & (1, -1, -4, -1) & 88.6 & 100 \\
(5, -2, -2, -1) & (1, -2, -4, 2) & 77.3 & 100 \\
(5, -2, -2, -1) & (2, -1, -4, 1) & 51.4 & 97.9 \\
(5, -2, -2, -1) & (2, -2, -4, -2) & 75.2 & 100 \\
\hline
(5, -1, -1, -2) & (1, -2, -2, -3) & 88.1 & 100 \\
(5, -1, -1, -2) & (1, -2, -1, -3) & 84.4 & 100 \\
(5, -1, -1, -2) & (1, -1, -3, -2) & 82.7 & 100 \\
(5, -1, -1, -2) & (1, -1, -2, -2) & 79.3 & 100 \\
(5, -1, -1, -2) & (1, 2, -2, -1) & 42.1 & 100 \\
(5, -1, -1, -2) & (1, 1, -4, 2) & 54.3 & 99.2 \\
(5, -1, -1, -2) & (1, 1, -4, 1) & 47.5 & 99.2 \\
(5, -1, -1, -2) & (1, 1, -4, 0) & 56.8 & 95.0 \\
(5, -1, -1, -2) & (1, 1, -3, -2) & 65.9 & 100 \\
(5, -1, -1, -2) & (1, 1, -3, -1) & 55.5 & 98.6 \\
(5, -1, -1, -2) & (1, 1, -3, 1) & 46.1 & 99.4 \\
(5, -1, -1, -2) & (1, -2, -3, -3) & 88.1 & 100 \\
(5, -1, -1, -2) & (1, 1, -4, -1) & 64.4 & 98.8 \\
\hline
(6, -3, -2, -1) & (1, 0, -3, 1) & 51.8 & 100 \\
(6, -3, -2, -1) & (1, 1, -4, 1) & 52.4 & 99.7 \\
\end{longtable}
\end{center}

\subsubsection{Spectrum estimate} \label{subsubsec:SpectrumEstimate}

\begin{center}
\begin{longtable}{cc|cc}
\caption{Spectrum estimates from the \emph{H0Approximator}} \\
\hline
$\mathbf{D_C}$ & $\mathbf{D_L}$ & \textbf{Predicted spectrum} & \textbf{Missing values} \\
\hline
\endfirsthead
\multicolumn{4}{c}%
{\tablename\ \thetable\ -- \textit{Continued from previous page}} \\
\hline
$\mathbf{D_C}$ & $\mathbf{D_L}$ & \textbf{Predicted spectrum} & \textbf{Missing values} \\
\hline
\endhead
\hline \multicolumn{4}{r}{\textit{Continued on next page}} \\
\endfoot
\hline
\endlastfoot
(5, -2, -2, -1) & (2, -2, -4, -2) & ( 0, 1, 2, 3, 4 ) & -- \\
(5, -2, -2, -1) & (1, -1, -3, 0) & ( 0, 1 ) & --\\
(5, -2, -2, -1) & (1, 2, -2, 0) & ( 2, 3 ) & --\\
(5, -2, -2, -1) & (1, 2, -2, 1) & ( 3, 4 ) & --\\
(5, -2, -2, -1) & (1, 1, -4, -1) & ( 0, 2, 3 ) & --\\
(5, -2, -2, -1) & (1, -1, -4, -1) & ( 0, 2, 3 ) & --\\
(5, -2, -2, -1) & (1, -2, -4, 2) & ( 0, 2, 3 ) & --\\
(5, -2, -2, -1) & (1, 1, -4, 1) & ( 0, 1, 2, 3, 4 ) & --\\
(5, -2, -2, -1) & (1, -1, -2, 3) & ( 0, 1, 2, 3 ) & --\\
(5, -2, -2, -1) & (2, -1, -4, 1) & ( 0, 1, 2, 3 ) & --\\
(5, -2, -2, -1) & (1, 2, -3, 1) & ( 1, 2, 3, 4 ) & --\\
(5, -2, -2, -1) & (1, 1, -4, 0) & ( 0, 2, 3 ) & --\\
(5, -2, -2, -1) & (1, 2, -2, -2) & ( 0, 1, 2, 3, 4 ) & --\\
\hline
(3, -1, -1, -1) & (2, 1, -4, 1) & ( 4, 5, 6 ) & --\\
(3, -1, -1, -1) & (1, -3, -3, -2) & ( 0, 1, 2, 3, 4 ) & ( 5 ) \\
(3, -1, -1, -1) & (1, -1, -3, 0) & ( 0, 1, 2, 3 ) & --\\
(3, -1, -1, -1) & (1, -2, -3, -2) & ( 0, 1, 2, 3 ) & ( 4 ) \\
(3, -1, -1, -1) & (1, -1, -3, -1) & ( 0, 2, 3 ) & --\\
(3, -1, -1, -1) & (1, -3, -4, -2) & ( 0, 1, 2, 3, 4, 5 ) & ( 6, 7 ) \\
(3, -1, -1, -1) & (2, 1, -4, 2) & ( 5, 6, 7 ) & --\\
(3, -1, -1, -1) & (2, 2, -4, 2) & ( 6, 7, 8, 9 ) & --\\
(3, -1, -1, -1) & (1, -1, -4, -1) & ( 0, 3, 5 ) & -- \\
(3, -1, -1, -1) & (1, 1, -3, 1) & ( 2, 3, 4 ) & --\\
(3, -1, -1, -1) & (1, 1, -3, 0) & ( 1, 2, 3 ) & --\\
(3, -1, -1, -1) & (1, -1, -2, 0) & ( 0, 1 ) & --\\
(3, -1, -1, -1) & (1, 1, -3, 2) & ( 3, 4, 5 ) & --\\
\hline
(5, -1, -1, -2) & (1, -2, -2, -3) & ( 0, 1, 2, 3 ) & --\\
(5, -1, -1, -2) & (1, 1, -4, 2) & ( 2, 3, 4, 5, 6, 7, 8 ) & --\\
(5, -1, -1, -2) & (1, 1, -4, 1) & ( 0, 1, 2, 3, 4, 5, 6, 7 ) & --\\
(5, -1, -1, -2) & (1, -1, -3, -2) & ( 0, 2, 3 ) & --\\
(5, -1, -1, -2) & (1, 1, -3, -1) & ( 0, 2, 3 ) & ( 1 ) \\
(5, -1, -1, -2) & (1, 1, -3, -2) & ( 0, 2, 3 ) & --\\
(5, -1, -1, -2) & (1, 2, -2, -1) & ( 0, 1 ) & --\\
(5, -1, -1, -2) & (1, 1, -4, 0) & ( 0, 3, 5, 6 ) & ( 1 ) \\
(5, -1, -1, -2) & (1, -2, -1, -3) & ( 0, 1, 2 ) & --\\
(5, -1, -1, -2) & (1, 1, -3, 1) & ( 1, 2, 3, 4 ) & --\\
(5, -1, -1, -2) & (1, -1, -2, -2) & ( 0, 1 ) & --\\
(5, -1, -1, -2) & (1, -2, -3, -3) & ( 0, 1, 2, 3, 4, 5 ) & --\\
(5, -1, -1, -2) & (1, 1, -4, -1) & ( 0, 3, 5, 6 ) & ( 1 ) \\
\hline
(4, -1, -1, -1) & (1, -2, -3, -1) & ( 0, 1, 2, 3, 4 ) & --\\
(4, -1, -1, -1) & (1, -3, -4, -3) & ( 0, 2, 3, 4, 5, 6, 7, 8, 9 ) & ( 10 ) \\
(4, -1, -1, -1) & (1, 1, -3, 0) & ( 0, 1, 2, 3 ) & --\\
(4, -1, -1, -1) & (1, -3, -3, -3) & ( 0, 2, 3, 4, 5, 6, 7 ) & ( 8 ) \\
(4, -1, -1, -1) & (1, -3, -2, -3) & ( 0, 1, 2, 3, 4, 5, 6 ) & ( 7 ) \\
(4, -1, -1, -1) & (1, 2, -2, -1) & ( 1, 2, 3 ) & --\\
(4, -1, -1, -1) & (1, 1, -3, -3) & ( 0, 1, 2, 3, 4, 5, 6 ) & --\\
(4, -1, -1, -1) & (1, 1, -3, -3) & ( 0, 1, 2, 3, 4, 5, 6 ) & --\\
(4, -1, -1, -1) & (2, 3, -4, -1) & ( 4, 5, 6, 7, 8, 9 ) & --\\
(4, -1, -1, -1) & (1, 2, -4, 2) & ( 2, 3, 4, 5, 6, 7, 8 ) & --\\
(4, -1, -1, -1) & (1, -2, -3, -2) & ( 0, 1, 2, 3, 4, 5 ) & --\\
(4, -1, -1, -1) & (1, 3, -3, 1) & ( 3, 4, 5, 6, 7 ) & --\\
(4, -1, -1, -1) & (1, -1, -3, 0) & ( 0, 2, 3 ) & ( 1 ) \\
\hline
(4, -1, -2, 0) & (1, -2, -1, 4) & ( 0, 1, 2, 3, 4, 5, 6 ) & --\\
\hline
(4, -1, -2, 1) & (2, -1, -2, 5) & ( 2, 5, 7, 8 ) & --\\
(4, -1, -2, 1) & (1, -1, -2, -1) & ( 2, 3 ) & --\\
(4, -1, -2, 1) & (1, -2, -2, -2) & ( 3, 4, 5, 6, 7 ) & --\\
(4, -1, -2, 1) & (2, -3, -2, -1) & ( 2, 3, 4, 5, 6 ) & --\\
(4, -1, -2, 1) & (1, -2, -1, 4) & ( 0, 1, 2, 3 ) & --\\
(4, -1, -2, 1) & (1, -2, -2, -3) & ( 4, 5, 7, 8, 9, 10, 11 ) & --\\
(4, -1, -2, 1) & (2, -3, -2, -2) & ( 3, 5, 6, 7, 8, 9 ) & (4) \\
(4, -1, -2, 1) & (1, -2, 1, -1) & ( 5, 6 ) & --\\
(4, -1, -2, 1) & (2, -2, -1, -2) & ( 6, 7 ) & --\\
(4, -1, -2, 1) & (2, -2, -2, 7) & ( 1, 2, 6, 7, 10, 11, 13, 14, 15 ) & --\\
(4, -1, -2, 1) & (3, -1, -2, 10) & ( 6, 14, 21, 27, 32 ) & --\\
(4, -1, -2, 1) & (1, -3, 1, -1) & ( 4, 5, 6, 7 ) & --\\
\hline
(6, -3, -2, -1) & (1, 1, -4, 1) & ( 0, 1, 2, 3, 4 ) & --\\
(6, -3, -2, -1) & (1, 0, -3, 1) & ( 0, 1 ) & --\\
\hline
(4, -1, -2, -1) & (3, -3, -1, -2) & ( 4, 5 ) & --\\
(4, -1, -2, -1) & (4, -7, -1, -3) & ( 3, 6, 8, 11, 12, 13, 15 ) & ( 4, 5, 7, 10, 17 ) \\
(4, -1, -2, -1) & (2, 3, -3, 1) & ( 5, 7, 8 ) & --\\
(4, -1, -2, -1) & (3, 1, -4, -1) & ( 3, 4 ) & --\\
(4, -1, -2, -1) & (2, 2, -4, 0) & ( 1, 2, 3, 4 ) & --\\
(4, -1, -2, -1) & (2, 1, -4, -3) & ( 0, 1, 2, 3, 4, 5, 6 ) & --\\
(4, -1, -2, -1) & (1, -1, -3, -2) & ( 0, 1, 2 ) & --\\
(4, -1, -2, -1) & (1, -2, -4, 2) & ( 0, 1, 2, 3, 4 ) & --\\
(4, -1, -2, -1) & (4, 3, -3, -8) & ( 4, 6, 7, 9, 10, 12, 13, 15, 16, 17, 18, 19) & ( 5, 8 ) \\
(4, -1, -2, -1) & (1, 3, -4, -5) & ( 0, 2, 4, 6, 7, 8, 9, 11 ) & ( 1 ) \\
(4, -1, -2, -1) & (3, 1, -4, -5) & ( 0, 2, 4, 5, 6, 7, 8, 9, 11 ) & ( 1 ) \\
(4, -1, -2, -1) & (3, 2, -3, -7) & ( 0, 1, 3, 4, 6, 7, 9, 10, 11, 12, 14, 15 ) & ( 2, 16 ) \\
(4, -1, -2, -1) & (3, 2, -3, -5) & ( 2, 3, 4, 5, 6, 7, 8, 9, 10, 11 ) & --\\
(4, -1, -2, -1) & (1, 1, -4, 2) & ( 0, 1, 2, 3, 4 ) & --\\
(4, -1, -2, -1) & (1, 0, -4, -1) & ( 0, 2, 3 ) & --\\
\bottomrule
\end{longtable}
\end{center}

\bibliography{papers}{}
\bibliographystyle{JHEP} 

\end{document}